\documentclass[twocolumn]{article}
\usepackage[english]{babel}
\usepackage{my_packages}
\usepackage{my_style}
\usepackage{stellar_symbols}
\usepackage{url}

\usepackage[left=1.5cm, right=1.5cm, top=1.75cm, bottom=1.75cm]{geometry}

\def\docTitle{The Stellar decomposition: A compact representation for simplicial complexes and beyond}
\title{\docTitle}

\author{
	Riccardo Fellegara, 
	{\emph{German Aerospace Center (DLR), Braunschweig, Germany}}
	\and 
	Kenneth Weiss, 
	{\emph{Lawrence Livermore National Laboratory, Livermore, CA, USA}}
	\and 
	Leila De Floriani, 
	{\emph{University of Maryland at College Park, College Park, MD, USA}}
}
\date{}

\begin{document}
	\linespread{0.5}
	
	\maketitle

\begin{abstract}
	We introduce the \emph{Stellar decomposition}, a model for efficient topological data structures over a broad range of simplicial and cell complexes.
	A Stellar decomposition of a complex is a collection of \emph{regions} indexing the complex's vertices and cells such that each region has sufficient information 
	to locally reconstruct the \emph{star} of its vertices, i.e., the cells incident in the region's vertices. 
	Stellar decompositions are general in that they can compactly represent and efficiently traverse arbitrary complexes with a manifold or non-manifold domain.
	They are scalable to complexes in high dimension and of large size, and
	they enable users to easily construct tailored application-dependent data structures using a fraction 
	of the memory required by a corresponding global topological data structure on the complex.
	
	As a concrete realization of this model for spatially embedded complexes, we introduce the \emph{Stellar tree}, 
	which combines a nested spatial tree with a simple tuning parameter to control the number of vertices in a region.
	Stellar trees exploit the complex's spatial locality by reordering vertex and cell indices according to the spatial 
	decomposition and by compressing sequential ranges of indices.
	%
	%
	Stellar trees are competitive with state-of-the-art topological data structures for manifold simplicial complexes 
	and offer significant improvements for cell complexes and non-manifold simplicial complexes.
	%
	We conclude with a high-level description of several mesh processing and analysis applications that utilize Stellar trees to process large datasets.
\end{abstract}

\section{Introduction}
Efficient mesh data structures play a fundamental role in a broad range of mesh processing applications in computer graphics, geometric modeling, scientific visualization, geospatial data science and finite element analysis. 
Although simple problems can be easily modeled on small low dimensional meshes, phenomena of interest might occur only on much larger meshes and in higher dimensions.
Thus, we often require flexibility to deal with increasingly complex meshes including those defined by irregularly connected heterogeneous and/or multidimensional cell types discretizing spaces with complicated topology.
Moreover, as advances in computing capabilities continue to outpace those in memory, it becomes increasingly important to optimize and exploit locality within the mesh as we process and locally query it. Such queries are the primary means of interacting with the mesh and have  traditionally been posed in terms of a few spatial and topological primitives.
However, while there are simple, intuitive models for representing polygonal surfaces, there are numerous challenges in generalizing these structures to higher dimensions and in scaling to very large meshes.

In this paper, we introduce the \emph{Stellar decomposition}, a model for topological data structures that supports  efficient navigation of the topological connectivity of simplicial complexes and of certain classes of cell complexes, e.g., those composed of quadrilaterals, polygons, hexahedra, prisms and pyramids. 
We refer to this class of complexes as \emph{Canonical Polytope complexes (CP complexes)}.
The defining property of a Stellar decomposition is that the complex is broken up into \emph{regions} indexing a collection of vertices 
such that each vertex within a region has sufficient information to locally reconstruct its \emph{star}, 
i.e., the set of cells from the complex incident in that vertex.

A Stellar decomposition is
  \emph{general}, in that it can easily represent arbitrary complexes with a manifold or non-manifold domain, 
  it is \emph{scalable} to complexes both in high dimensions and with a large number of cells,
  and it is \emph{flexible}, in that it enables users to defer decisions about which topological connectivity relations to encode.
It, therefore, supports the generation of optimal application-dependent local data structures at runtime.
Due to the locality of successive queries in typical mesh processing applications, the construction costs of these local topological data structures are amortized over multiple mesh operations while processing a local region. 

We also formally define and analyze the \emph{Stellar tree} as a concrete instance of the Stellar decomposition model for spatially embedded complexes. Stellar trees utilize a hierarchical $n$-dimensional quadtree, or kD-tree, as vertex decomposition, and are easily \emph{tunable} using a single parameter 
that defines the maximum number of vertices allowed in each local region. 
%

While Stellar trees have been previously utilized in several mesh processing applications ranging from mesh simplification~\cite{Fellegara2020Efficient}
to morphological feature extraction~\cite{Weiss2013primaldual,Fellegara2017Efficient}, 
they have not been formally defined 
and their performance has not yet been characterized in relation to existing topological data structures for simplicial and cell complexes.
This paper presents a careful study of the storage requirements, generation algorithms and timings and query performance for Stellar trees over a wide range of CP complexes.
As we demonstrate in Section~\ref{sec:storage}, Stellar trees are competitive with dimension-specific state-of-the-art topological data structures for (pseudo)-manifold triangle and tetrahedral complexes 
and offer significant improvements for other CP complexes, especially over data structures for general simplicial complexes in 3D and higher dimensions.
The source code for our Stellar tree implementation 
will be released in the public domain.

\paragraph{Contributions} The contributions of this work include:
\begin{compactitem}
  \item The formal theoretical definition of a Stellar decomposition over \emph{Canonical Polytope (CP) complexes}, 
    a class of cell complexes that includes simplicial and cubical complexes of arbitrary dimension,
    as well as cells in the finite element `zoo', such as polygons, pyramids and prisms.
  \item The definition of the \emph{Stellar tree} as a concrete realization of the Stellar decomposition for spatially embedded complexes. 
    The decomposition in a Stellar tree is based on a hierarchical spatial index
    with a simple tuning parameter to facilitate balancing storage and performance needs.
  \item The definition of \emph{Sequential Range Encoding (SRE)}, a compact encoding for the entities indexed by each region of the decomposition.
    When applied to CP complexes reindexed by the spatial decomposition of a Stellar tree, SRE yields compressed Stellar trees 
    with only a small overhead relative to the original CP complex's cells. 
    %
    %
\end{compactitem}


\paragraph{Outline}
The remainder of this paper is organized as follows.
In Sections~\ref{background} and~\ref{related}, we review background notions and related work, respectively. 
In Section~\ref{sec:stellar_decomposition}, we define Stellar decompositions,
 describe our compact encoding, 
 and provide a high-level description of the procedure for generating a Stellar decomposition.
In Section~\ref{sec:stellar_tree}, we define the Stellar tree, a spatio-topological realization of the Stellar decomposition. 
%
%
%
In Section~\ref{sec:general_strategy}, we describe a general mesh processing paradigm that can be followed by applications defined on a Stellar tree.
In Section~\ref{sec:threshold_calibration}, we discuss our experimental setup and evaluate how our tuning parameter affects the quality of a Stellar tree's decomposition and its performance in extracting topological features.
We then compare Stellar trees to several state-of-the-art topological data structures in Section~\ref{sec:storage}.
%
In Section~\ref{sec:local_topo_rels}, we describe how to extract local connectivity information from the Stellar tree and evaluate the performance of these algorithms. 
%
%
We provide a high-level overview of several mesh processing and analysis applications 
that have benefited from Stellar trees to process large datasets in Section~\ref{sec:stellar_ecosystem}
and conclude in Section~\ref{sec:stellar_conclusions} with some remarks and directions for future work.


\begin{figure*}
	\centering	
	\subfloat[A pure complex]{
		\resizebox{.24\textwidth}{!}{
			\includegraphics{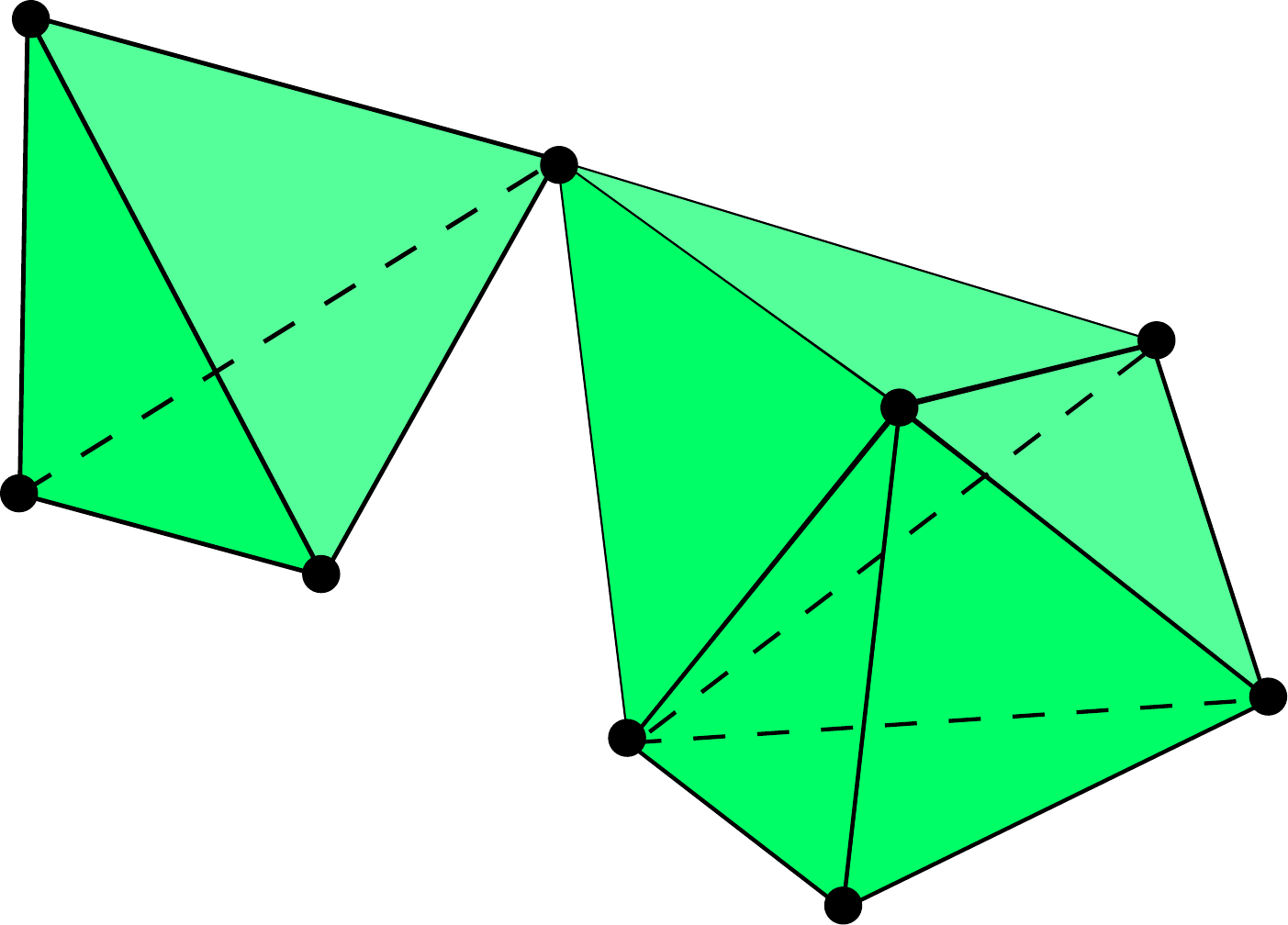}
		}
	}
	\hfil
	\subfloat[A CP complex]{
		\def\svgwidth{.22\textwidth}
		{\scriptsize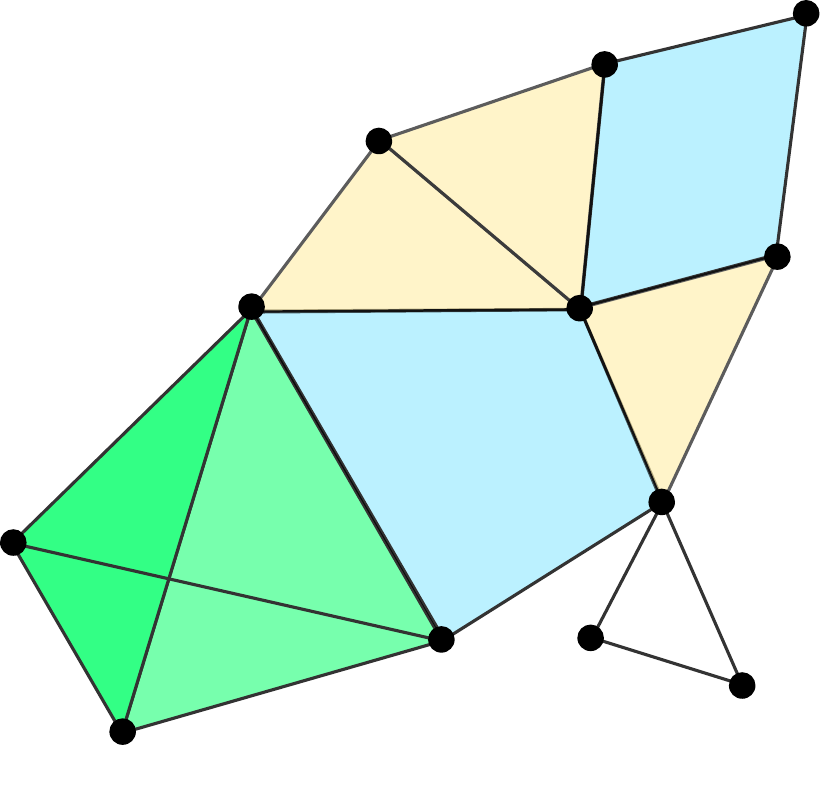}
		}
	\hfil
	\subfloat[A pseudo-manifold]{
		\resizebox{.22\textwidth}{!}{
			\includegraphics{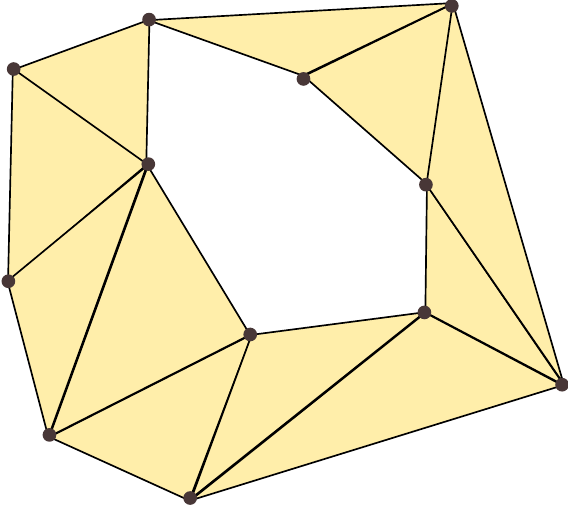}
		}
	}
	\caption{Examples of CP complexes. 		
		(a) A pure simplicial 3-complex with four tetrahedra. 
		(b) A CP complex with three top edges, three top triangles, two top quads and a top tetrahedron. 
		(c) A 2-dimensional \emph{pseudo-manifold} with eleven triangles.} 
	\label{fig:complexes_examples}
\end{figure*}

\begin{figure}
	\centering	
	\subfloat[Star]{
		\def\svgwidth{.4\columnwidth}
		{\scriptsize
\begingroup%
  \makeatletter%
  \providecommand\color[2][]{%
    \errmessage{(Inkscape) Color is used for the text in Inkscape, but the package 'color.sty' is not loaded}%
    \renewcommand\color[2][]{}%
  }%
  \providecommand\transparent[1]{%
    \errmessage{(Inkscape) Transparency is used (non-zero) for the text in Inkscape, but the package 'transparent.sty' is not loaded}%
    \renewcommand\transparent[1]{}%
  }%
  \providecommand\rotatebox[2]{#2}%
  \newcommand*\fsize{\dimexpr\f@size pt\relax}%
  \newcommand*\lineheight[1]{\fontsize{\fsize}{#1\fsize}\selectfont}%
  \ifx\svgwidth\undefined%
    \setlength{\unitlength}{323.99997711bp}%
    \ifx\svgscale\undefined%
      \relax%
    \else%
      \setlength{\unitlength}{\unitlength * \real{\svgscale}}%
    \fi%
  \else%
    \setlength{\unitlength}{\svgwidth}%
  \fi%
  \global\let\svgwidth\undefined%
  \global\let\svgscale\undefined%
  \makeatother%
  \begin{picture}(1,0.91435566)%
    \lineheight{1}%
    \setlength\tabcolsep{0pt}%
    \put(0,0){\includegraphics[width=\unitlength,page=1]{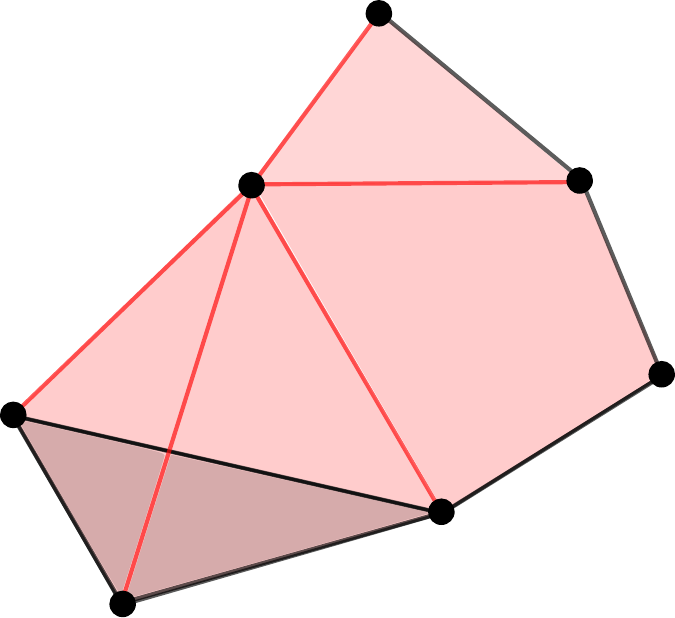}}%
    \put(0.29051605,0.67233013){\color[rgb]{0,0,0}\makebox(0,0)[lt]{\lineheight{0}\smash{\begin{tabular}[t]{l}$v_0$\end{tabular}}}}%
  \end{picture}%
\endgroup%
}
	}
	\hfil
	\subfloat[Link]{
		\def\svgwidth{.4\columnwidth}
		{\scriptsize
\begingroup%
  \makeatletter%
  \providecommand\color[2][]{%
    \errmessage{(Inkscape) Color is used for the text in Inkscape, but the package 'color.sty' is not loaded}%
    \renewcommand\color[2][]{}%
  }%
  \providecommand\transparent[1]{%
    \errmessage{(Inkscape) Transparency is used (non-zero) for the text in Inkscape, but the package 'transparent.sty' is not loaded}%
    \renewcommand\transparent[1]{}%
  }%
  \providecommand\rotatebox[2]{#2}%
  \newcommand*\fsize{\dimexpr\f@size pt\relax}%
  \newcommand*\lineheight[1]{\fontsize{\fsize}{#1\fsize}\selectfont}%
  \ifx\svgwidth\undefined%
    \setlength{\unitlength}{323.99997711bp}%
    \ifx\svgscale\undefined%
      \relax%
    \else%
      \setlength{\unitlength}{\unitlength * \real{\svgscale}}%
    \fi%
  \else%
    \setlength{\unitlength}{\svgwidth}%
  \fi%
  \global\let\svgwidth\undefined%
  \global\let\svgscale\undefined%
  \makeatother%
  \begin{picture}(1,0.91435566)%
    \lineheight{1}%
    \setlength\tabcolsep{0pt}%
    \put(0,0){\includegraphics[width=\unitlength,page=1]{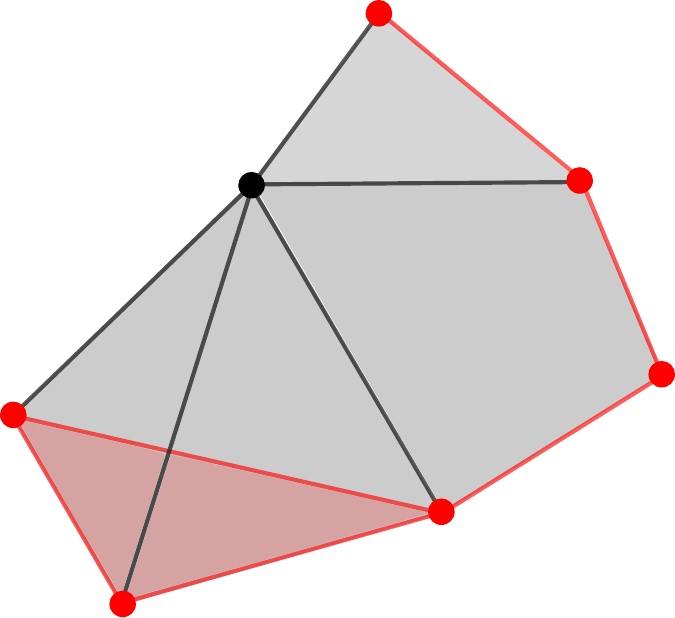}}%
    \put(0.29051605,0.67233013){\color[rgb]{0,0,0}\makebox(0,0)[lt]{\lineheight{0}\smash{\begin{tabular}[t]{l}$v_0$\end{tabular}}}}%
  \end{picture}%
\endgroup%
}
	}
	\caption{		
		The star and the link of 0-cell (vertex) $v_0$ from the complex in Figure \ref{fig:complexes_examples}(b).
		Cells belonging to either the star (a) or link (b) of $v_0$ are highlighted in red.
	} 
	\label{fig:relation_examples}
\end{figure}

\section{Background notions}
\label{background}


	In this section, we review notions related to cell and simplicial complexes, 
	which are the basic combinatorial structures for representing discretized shapes.
	Throughout the paper, 
		we use \sDim\ to denote the dimension of the ambient space in which the complex is embedded, 
		\cDim\ to represent the dimension of the  complex
		and \tDim\ to denote the dimension of a cell from the complex, where $0 \leq \tDim \leq \cDim$. 
	
	
	\NOTA{
	\felleComment{
			\textbf{R}: both cell and simplicial complexes are homeomorphic to a close k-dim ball. 
			We can state that as we are in a \emph{discrete} space (discussed it with Ulderico)}
	}
	
A $\tDim$-dimensional \emph{cell} in the $\sDim$-dimensional Euclidean space $\eSpace$ is a subset of $\eSpace$ homeomorphic to a closed $\tDim$-dimensional ball $ B^\tDim = \{ x \in \etSpace : \|x\| \leq 1 \}$.  
A $\cDim$-dimensional \emph{cell complex} $\cC$ in $\eSpace$ is a finite set of cells with disjoint interiors and of dimension at most $\cDim$ 
such that the boundary of each $\tDim$-cell $\cell$ in $\cC$ consists of the union of  other cells of $\cC$ with dimension less than $\tDim$. 
Such cells are referred to as the \emph{faces} of $\cell$.
%
%
A cell which does not belong to the boundary of any other cell in $\cC$ is called a \emph{top cell}. 
$\cC$ is a \emph{pure} cell complex when all top cells have dimension $\cDim$. 
The subset of $\eSpace$ spanned by the cells of $\cC$ is called the \emph{domain} of $\cC$. 
An example of a pure cell 3-complex is shown in Figure \ref{fig:complexes_examples}(a): all its top cells are 3-cells (tetrahedra).

Throughout this paper, we are concerned with a restricted class of cell complexes whose cells can be fully reconstructed by their set of vertices, 
e.g., via a canonical ordering~\cite{Scho94,Poir98,Rema03,Celes2005,Tautges2010Canonical}.
We refer to this class of complexes as \emph{Canonical Polytope complexes (CP complexes)},
and note that it includes simplicial complexes, cubical complexes, polygonal cell complexes
and heterogeneous meshes with cells from the finite element `zoo' (e.g., simplices, hexahedra, pyramids, and prisms).
In what follows, we denote a CP complex as $\sC$.
%
An example of a CP complex is shown in Figure~\ref{fig:complexes_examples}(b), which contains top edges, triangles, quadrilaterals, and tetrahedra.
		
A pair of cells in a CP complex $\sC$ are mutually \emph{incident} if one is a face of the other. They are 
 \emph{$h$-adjacent} if they have the same dimension $\tDim > h$ and are incident in a common $h$-face.
We informally refer to vertices (0-cells) as \emph{adjacent} if they are both incident in a common edge (1-cell)
and, similarly, for \tDim-cells that are incident in a common $(\tDimMinusOne)$-cell (i.e., they are (\tDimMinusOne)-adjacent).
The \emph{(combinatorial) boundary} of a CP cell $\simplex$ is defined by the set of its faces.
The \emph{star} of a CP cell $\simplex$ 
 is the set of its \emph{co-faces}, i.e., CP cells in $\sC$ that have $\simplex$ as a face. 
An example of star for a 0-cell (vertex) is shown in Figure \ref{fig:relation_examples}(a). 
In this example, the star of vertex $v_0$ is formed by five edges, four triangles, a quad, and a tetrahedron. 
Of these CP cells, tetrahedron $\sigma_5$, quad $\sigma_1$ and triangle $\sigma_4$ are \emph{top} cells.
The \emph{link} of a CP cell $\simplex$ 
is the set of all the faces of cells in the star 
that are not incident in $\simplex$.
An example of link for a 0-cell (vertex) is shown in Figure \ref{fig:relation_examples}(b). In this example, the link of $v_0$ is composed of six vertices, six edges, and a triangle. 

Two $h$-cells $\simplex$ and $\simplex^{\prime}$ in \sC\ are \emph{$(h{-}1)$-connected} 
if there is a sequence, called an \emph{$h$-path}, of $(h{-}1)$-adjacent $h$-cells in \sC\ from $\simplex$ to $\simplex^{\prime}$.
A complex $\sC$ is \textit{$h$-connected}, if for every pair of $h$-cells $\sigma_1$ and $\sigma_2$, there is an $h$-path in \sC\ joining $\sigma_1$ and $\sigma_2$.

We can now define a $\cDim$-dimensional \emph{CP complex} $\sC$ as a set of CP-cells in $\eSpace$ of dimension at most $\cDim$ such that:
\begin{inparaenum}[(1)]
	\item $\sC$ contains all CP-cells in the boundary of the CP-cells in $\sC$;
	\item the intersection of any two CP-cells in $\sC$ is \emph{conforming}, i.e., it is either empty, or it consists of faces shared by both CP-cells.
\end{inparaenum}	
\emph{Simplicial complexes} are an important subset of CP complexes whose cells are  \emph{simplices}.
Let $\tDim$ be a non-negative integer. A \tDim-simplex $\simplex$ is the convex hull of $\tDim + 1$ independent points in $\eSpace$ (with $\tDim \leq \sDim$), called vertices of $\simplex$. 
A \emph{face} of a \tDim-simplex $\simplex$ is an $h$-simplex ($0 \leq h \leq \tDim$) generated by $h + 1$ vertices of $\simplex$.

\NOTA{
  \felleComment{I think we can remove the parts in red. These were needed when we were describing in details how to validate a complex with the Stellar}
  \kennyComment{I simplified it a bit, but left the pseudo-manifold, since we use that terminology in the paper}
}
%
Other important notions are those of  \emph{manifolds} 
and \emph{pseudo-manifolds}.
A subset $M$ of the Euclidean space $\eSpace$ is called a \emph{\cDim-manifold}, with $\cDim \leq \sDim$, if and only if every point of $M$ has a neighborhood homeomorphic to the open \cDim-dimensional ball. 
%
%
A more practical concept for the purpose of representing CP complexes is that of pseudo-manifold.
A pure $\cDim$-dimensional CP complex $\sC$ is said to be a \emph{pseudo-manifold} when it is $(\cDimMinusOne)$-connected and its $(\cDimMinusOne)$-cells are incident in at most two \cDim-cells.
Informally, we refer to the connected and compact subspace of \eSpace\ not satisfying the manifold conditions as \emph{non-manifold}.
%
\NOTA{ 
	A subset $M$ of the Euclidean space $\eSpace$ is called a \emph{\cDim-manifold with boundary} (with $\cDim \leq \sDim$) 
	if and only if every point of $M$ has a neighborhood homeomorphic either to the open \cDim-dimensional ball 
	or the open \cDim-dimensional ball intersected with a hyperplane in $\eSpace$.
}

Queries on a cell complex are often posed in terms of \emph{topological relations}, which are  defined by the adjacencies and incidences of its cells.
Let us consider a $\cDim$-dimensional CP complex $\sC$ and  a $\tDim$-cell $\simplex\in\sC$, with $0\!\le \tDim\!\le \cDim$:
\begin{compactitem}
	\item a \emph{boundary relation} $\relation{\tDim,p}(\simplex)$, with $0\le p<\tDim$, consists of the $p$-cells of $\sC$ in the boundary of $\simplex$;
	\item a \emph{co-boundary relation} $\relation{\tDim,q}(\simplex)$, with $\tDim < q \le \cDim$, consists of the $q$-cells of $\sC$ in the star of $\simplex$;
	\item an \emph{adjacency relation} $\relation{\tDim,\tDim}(\simplex)$ consists of  the set of $\tDim$-cells of $\sC$ that are (\tDimMinusOne)-adjacent to $\simplex$.
\end{compactitem}
For some examples of topological relations, consider the CP complex in Figure~\ref{fig:complexes_examples}(b):
Boundary relation \relation{3,0} for tetrahedron \simplex$_5$ is the list of its boundary vertices, 
  i.e., \relation{3,0}(\simplex$_5$) = $\{v_0,v_2,v_4,v_5\}$. 
Co-boundary relation \relation{0,2} for vertex $v_3$ is the list of its incident 2-cells (triangles and quads), 
  i.e., \relation{0,2}($v_3$) = $\{\simplex_0,\simplex_1,\simplex_2,\simplex_3,\simplex_4\}$. 
Adjacency relation \relation{0,0} for vertex $v_0$ is the list of its adjacent vertices, 
  i.e., \relation{0,0}($v_0$) = $\{v_1,v_2,v_3,v_4,v_5\}$.


\section{Related work}
\label{related}

\NOTA{
  \leilaComment{for \textbf{Kenny} please add \\
    1. paper on cache mesh \cite{Nguyen2017Cache} \\
    2. comparison of Maximal Simplex tree and Simplex Array List \cite{Boissonnat2015Building} 
    }
  \kennyComment{ Done. }
}

In this section, we review the state of the art on topological mesh data structures, hierarchical spatial indexes, data layouts 
and distributed mesh data structures.


\subsection{Topological mesh data structures}
\label{sec:related_topological_ds}
	
There has been much research on efficient representations for manifold cell and simplicial complexes,  especially for the 2D case.
A comprehensive survey of topological data structures for manifold and non-manifold shapes can be found in~\cite{DeFloriani2005Data}.

A topological data structure over a cell complex encodes a subset of its topological relations and supports the efficient reconstruction of local topological connectivity over its cells. 
Topological data structures can be classified according to:
\begin{inparaenum}[(i)]
		\item the \emph{dimension} of the cell complex,
		\item the \emph{domain} to be approximated, i.e., manifolds versus
		non-manifold shapes,
		\item the subset of \emph{topological information} directly encoded, and 
		\item the \emph{organization} of topological information directly encoded, i.e., explicit or implicit data structures.
\end{inparaenum}

The explicit cells and topological relations can either be allocated on demand using small local data structures, such as linked lists,
or contiguously, e.g. using arrays.
In the former case, pointers are used to reference the elements, which can be useful when the data structure needs to support frequent updates 
	to the underlying cells or their connectivity.
In the latter case, indexes of the cells within the array can be used to efficiently reference the elements.
Recently, an approach has been proposed in \cite{Nguyen2017Cache} to reconstruct topological relations on demand and to cache them for later reuse.
	
Broadly speaking, topological data structures can be categorized as \emph{incidence-based} or \emph{adjacency-based}. 
Whereas incidence-based data structures primarily encode their topological connectivity through incidence relations over all the complex's cells,
adjacency-based data structures primarily encode their connectivity through adjacency relations over its top cells.

The \emph{Incidence Graph} ($IG$)~\cite{Edelsbrunner1987Algorithms} is the prototypical incidence-based data structure for cell complexes in arbitrary dimension.  The IG explicitly encodes all cells of a given cell complex $\cC$, 
and for each $p$-cell $\cell$, its immediate boundary and co-boundary relations (i.e., \relation{p,p{-}1} and \relation{p,p{+}1}).
	Several compact representations with the same expressive power as the IG have been developed for simplicial complexes~\cite{DeFloriani2004data,DeFloriani2010dimension},
	which typically require less than half the storage space as the IG~\cite{Canino2014Representing}.

Several incidence-based data structures have been developed for manifold 2-complexes, which encode the incidences among edges. 
The \emph{half-edge} data structure~\cite{Mantyla1988Introduction} is the most widely data structure of this type~\cite{CGAL,OML15}.
Design tradeoffs for data structures based on half-edges are discussed in~\cite{sieger2011design}.
%
\emph{Half-faces}~\cite{Kremer2013OpenVolumeMesh} generalize the notion of a half-edge to polyhedral complexes, 
while \emph{combinatorial maps}~\cite{Lienhardt1994N,Damiand2014Combinatorial} generalize this notion to higher dimensions.

\emph{Indexed data structures}~\cite{Lawson1977Software} provide a more compact alternative by explicitly encoding only vertices, top cells 
and the boundary relations from top cells to their vertices. Since the cells of a CP complex are entirely determined by their ordered list of vertices, 
this provides sufficient information to efficiently extract all boundary relations among the cells, but not the co-boundary or adjacency relations.
The \emph{Indexed data structure with Adjacencies ($IA$)}~\cite{Paoluzzi1993Dimension,Nielson1997Tools} 
extends the indexed representation to manifold simplicial complexes of arbitrary dimension by explicitly encoding adjacency relation \relation{\cDim,\cDim}, giving rise to an adjacency-based representation.
%
All remaining topological relations can be efficiently recovered if we also encode
a top simplex in the star of each vertex (i.e., a subset of relation \relation{0,\cDim}).

	
\NOTA{\ytodo{CoT and SOT}}
The \emph{Corner-Table (CoT)} data structure~\cite{Rossignac20013D} is also adjacency-based. 
It is defined only for 
triangle meshes, where it has the same representational power as the IA data structure. 
	It uses \emph{corners} as a conceptual abstraction to represent individual vertices of a triangle
	and encodes topological relations among corners and their incident vertices and triangles.
  Several efficient extensions of the Corner-Table data structure have been proposed 
  that exploit properties of manifold triangle meshes~\cite{Gurung2011SQuad,Luffel2014}.
	The \emph{Sorted Opposite Table (SOT)} data structure~\cite{Gurung2009SOT} extends the Corner-Table data structure to tetrahedral meshes 
	and introduces several storage optimizations.  
	Notably, it supports the reconstruction of boundary relation \relation{\cDim, 0} from co-boundary relations \relation{0,\cDim} (implicitly encoded) and \relation{\cDim,\cDim} relations (explicitly encoded), reducing its topological overhead by nearly a factor of two. 
	Since modifications to the mesh require non-local reconstructions of the associated data structures,
	this representation is suitable for applications on static meshes.
  %

\NOTA{\ytodo{IA*}}
The \emph{Generalized Indexed data structure with Adjacencies (\iastar\ data structure)}~\cite{Canino2011IA} extends the representational domain of the IA data structure to arbitrary non-manifold and mixed dimensional simplicial complexes.
The \iastar\ data structure is compact, in the sense that it gracefully degrades to the IA data structure in locally manifold neighborhoods of the mesh, and has been shown to be more compact than incidence-based data structures, especially as the dimension increases~\cite{Canino2014Representing}.
A similar data structure for non-manifold complexes was presented in~\cite{Dyedov2015}.
A detailed description can be found in Section \ref{sec:storage_other_structures}.

The \emph{Simplex tree} \cite{Boissonnat2014simplex} also encodes general simplicial complexes of arbitrary dimension. 
It explicitly stores all simplices of the complex within a \emph{trie}~\cite{Fredkin1960Trie} whose nodes are in bijection with the simplices of the complex. A public domain implementation is available  
in the \emph{GUDHI} library \cite{GUDHI}.
We provide a detailed description of this data structure in Section~\ref{sec:storage_other_structures}.
%
%
%
\NOTA{
\felleComment{removed the sentence with the comparison between IA* and Simplex tree done in Fugacci et al. as we are going to do the same type of comparison on much larger datasets, confirming practically what they are claiming \\ \textbf{notice}: I updated the reference of the Fugacci et al. paper \\ \textbf{notice2}: IA* is 10-30\% more compact than Simplex tree on low-dimensional datasets, on higher-dimensional they highlighted the trend that we identified in our comparisons.}
}
Boissonnat\etal~\cite{Boissonnat2017} also propose two top-based data structures targeting a compact Simplex tree representation. 
The \emph{Maximal Simplex Tree} ($MST$) is an induced subgraph of the Simplex tree, in which only the paths corresponding to top simplices are encoded,
but most operations require processing the entire complex.
The \emph{Simplex Array List} ($SAL$) is a hybrid data structure computed from the top simplices of a simplicial complex $\sC$
that improves processing efficiency by increasing the storage overhead.
Both the $MST$ and the $SAL$ are interesting structures from a theoretical point-of-view, 
but, as described in~\cite{Boissonnat2017}, the model does not currently scale to large datasets and results are limited to complexes with only a few thousand vertices. 
Moreover, to the best of our knowledge, there is no public domain implementation currently available.
\NOTA{KW: See beginning of Section 5.3 in \cite{Boissonnat2017} (page 544).
  > ``Unfortunately, [showing the compactness of MST] not been possible due to the lack of available libraries 
      able to handle very large automata''
}

\NOTA{skeleton blocker}
The \emph{Skeleton-Blocker} data structure~\cite{Attali2012Efficient} encodes simplicial complexes that are close to \emph{flag complexes}, simplicial complexes whose top simplices are entirely determined from the structure of their 1-skeleton, i.e., the vertices and edges of the complex, and has been successfully employed for executing edge contractions on such complexes. It encodes the 1-skeleton 
and the \emph{blockers}, simplices that are not in \sC, but whose faces are. 
Its generation procedure is computationally intensive for general simplicial complexes since 
identifying the \emph{blockers} requires inserting simplices of all dimensions.

We compare the Stellar tree representation with the IA, CoT, and SOT data structures as well as with the Simplex tree, 
and \iastar\ data structure in Section~\ref{sec:storage_other_structures}. 
	
\subsection{Hierarchical spatial indexes}
\label{sec:related_spatial_index}

A spatial index is a data structure used for indexing spatial information, such as points, lines or surfaces in the Euclidean space. 
Spatial indexes form a decomposition of the embedding space into \emph{regions}.  This can be driven by: 
\begin{inparaenum}[(i)]
	\item an \emph{object-based} or a \emph{space-based} criterion for generating the decomposition; and
	\item a \emph{hierarchical} or a \emph{non-hierarchical} (\emph{flat}) organization of the regions.
\end{inparaenum}
These properties are independent, and, thus, we can have hierarchical object-based decompositions as well as flat space-based ones. 

We now consider how the regions of a decomposition can intersect.
In an \emph{overlapping} decomposition the intersection between the regions can be non-empty on both the interiors and on the boundary of their domain, 
while, in a \emph{non-overlapping} decomposition, intersections can only occur on region boundaries.
We say that a region is \emph{nested} within another region if it is entirely contained within that region.
In the remainder of this section, we focus primarily on \emph{hierarchical spatial indexes}, 
which can be classified by the dimensionality of the underlying ambient space and by the types of entities indexed.

Hierarchical spatial indexes for point data are provided by \emph{Point Region (PR)} quadtrees/octrees and kD-trees~\cite{Samet2006Foundations}.
In these indexes, the shape of the tree is independent of the order in which the points are inserted, and the points are only indexed by leaf blocks. 
The storage requirements of these data structures can be reduced by allowing leaf blocks to index multiple points, as in the \emph{bucket PR} quadtree/octree~\cite{Samet2006Foundations}, whose \emph{bucketing threshold} determines the number of points that a leaf block can index before it is refined.

Several data structures have been proposed for spatial indexing of \emph{polygonal maps (PM)}, including graphs and planar triangle meshes.
\emph{PM quadtrees}~\cite{Samet1985Storing} extend the PR quadtrees to represent polygonal maps considered as a structured collection of edges.
While there are several variants (\emph{PM$_1$}, \emph{PM$_2$}, \emph{PM$_3$} and the randomized \emph{PMR)}, which differ in the criterion used to refine leaf blocks, all maintain within the leaf blocks a list of intersecting edges from the mesh. 
The \textit{PM$_2$-Triangle quadtree}~\cite{DeFloriani2008Hierarchical} specializes PM quadtrees over triangle meshes and has been applied to terrain models. 
The PM index family has also been extended to \emph{PM-octrees} encoding polyhedral objects in 3D~\cite{Carlbom1985hierarchical,Navazo1989Extended,Samet2006Foundations}, where the subdivision rules have been adjusted to handle edges and polygonal faces of the mesh elements.
Another proposal for triangulated terrain models are \emph{Terrain trees} \cite{Fellegara2017Efficient}, that are a spatial index family for the efficient representation and analysis of large-scale triangulated terrains generated from $LiDAR$ (\emph{Light Detection and Ranging}) point clouds.
%
A collection of spatial indexes for tetrahedral meshes called \emph{Tetrahedral trees} was developed in~\cite{DeFloriani2010Spatial,Fellegara2020Tetrahedral}. 
\NOTA{ In~\cite{DeFloriani2010Spatial}, we have developed a collection of spatial indexes for tetrahedral meshes, that we call \emph{Tetrahedral trees}. }

We note that data structures in the PM family are \emph{spatial data structures} optimized for efficient spatial queries on a complex (e.g., point location, containment and proximity queries) and are not equipped to reconstruct the connectivity of the complex.
In contrast, the \emph{PR-star octree}~\cite{Weiss2011PR} is a topological data structure for tetrahedral meshes embedded in 3D space. 
It augments the bucket PR octree with a list of tetrahedra incident in the vertices of its leaf blocks, i.e., those in the \emph{star} of its vertices. 
	
	In this paper, we have generalized the PR-star data structure to handle a broader class of complexes (CP complexes) 
  in arbitrary dimensions and with an arbitrary domain (i.e., non-manifold and non-pure complexes). 
	At the same time, our new leaf block encoding exploits the spatial coherence of the mesh, 
	yielding a significant storage saving compared to PR-star trees (see Section~\ref{sec:storage_encodings}).
	%
  %
	As we discuss in Section~\ref{sec:stellar_ecosystem}, Stellar trees have been shown to be effective in several 
  geometrical and topological applications including 
		local curvature estimation, mesh validation and simplification~\cite{Weiss2011PR}, 
		morphological feature extraction~\cite{Weiss2013primaldual} 
		and morphological simplification~\cite{Fellegara2014Efficient}, 
		among others.
		
\subsection{Optimized data layouts}
\label{sec:related_layouts}

  Considerable effort has been devoted to reindexing meshes to better exploit their underlying spatial locality, 
  for example, to support streamed processing~\cite{Isenburg2005}, better cache locality~\cite{Yoon05} or compression~\cite{Yoon2007}.
	Cignoni\etal~\cite{Cignoni2003External} introduce an external memory spatial data structure for triangle meshes embedded in $\eucl^3$.
	Whereas our aim is to enable efficient topological operations on the elements of general simplicial and CP complexes, 
	the objective of~\cite{Cignoni2003External} is to support compact out-of-core processing of massive triangle meshes.  
	Since the data structure in~\cite{Cignoni2003External} is dimension-specific, by exploiting geometric and topological properties of triangle meshes in $\eucl^3$, 
	it would be difficult to generalize to CP complexes or to higher dimensions.
	%
	Dey\etal~\cite{Dey2010Localized} use an octree to index a large triangle mesh for localized Delaunay remeshing.
	Due to the significant overhead associated with their computations, the octrees in~\cite{Dey2010Localized} are typically shallow, containing very few octree blocks. 
	In the context of interactive rendering and visualization of large triangulated terrains and polygonal models,
  Cignoni\etal~\cite{Cignoni2003BDAM,Cignoni2004Adaptive} associate patches of triangles with the simplices
	of a multiresolution diamond hierarchy~\cite{Weiss2011Simplex}.
	
	%

\subsection{Distributed mesh data structures}
\label{sec:related_distributed}

Stellar decompositions and Stellar trees are also related to distributed mesh data structures~\cite{Devine2009,Ibanez2016}, which partition large meshes 
across multiple processors for parallel processing e.g.\ in numerical simulations~\cite{Anderson2021mfem,Kirk2006,Edwards2010}.
In the latter, each computational \emph{domain} maintains a mapping between its boundary elements and their counterparts on neighboring domains.
To reduce inter-process communication during computation, each domain might also include one or more 
layers of elements from other domains surrounding its elements, 
typically referred to as \emph{ghost}, \emph{rind} or \emph{halo} layers~\cite{Poirier2000,Lawlor2006,Ollivier2010}.
Although each region of a Stellar decomposition (or tree) can be seen as a computational domain in a distributed data structure with a single ghost layer
(i.e., the elements in the star of its boundary vertices),
Stellar trees are aimed at providing efficient processing on coherent subsets of the mesh (regions),
where users can generate optimized local topological data structures.
In a distributed regime, we envision Stellar trees helping more with fine-grained
(intra-domain) parallelism than with coarse-grained multi-domain partitions.

\section{Stellar decomposition}
\label{sec:stellar_decomposition}

The \emph{Stellar decomposition} is a model for data structures representing \emph{Canonical Polytope (CP) complexes}.
%
%
We denote a CP complex as \sC, its ordered lists of vertices as \sCV\ and of \topcpcells\ as \sCT. 
We provide a definition of the Stellar decomposition in Section~\ref{sec:stellar_dec_def},
and describe its encoding in Section~\ref{sec:stellar_dec_enc}.

\subsection{Definition}
\label{sec:stellar_dec_def}



Given a CP complex \sC, a \emph{decomposition} \D\ of its vertices \sCV\ is a collection of subsets of \sCV\ such that 
every vertex $\vertex \in \sCV$ belongs to one of these subsets. 
We will refer to the elements of decomposition \D\ as \emph{regions}, which we will denote as \R.

A Stellar decomposition \sDec\ defines a map from the regions of a decomposition \D\ of its vertex set \sCV\ to the vertices and \topcpcells\ of complex \sC. 
%
Formally, a Stellar decomposition is defined by three components:
\begin{compactenum}
	\item a \emph{CP complex} \sC; 
	\item a \emph{decomposition} \D\ whose regions cover the vertices of \sC;
	\item a \emph{map} \PhiMap\ from regions of \D\ to entities of \sC.
\end{compactenum}
Thus, 
a Stellar decomposition is a triple $\sDec = (\sC,\D,\PhiMap)$.
Since \sC\ is entirely characterized by its vertices, and \topcpcells, we define map \PhiMap\ 
in terms of the two components: \PhiMapVert, the mapping to vertices and  \PhiMapTop, the mapping to \topcpcells.

\begin{figure}[t]
	\centering
	\subfloat[]{
		\resizebox{.4\columnwidth}{!}{
			\includegraphics{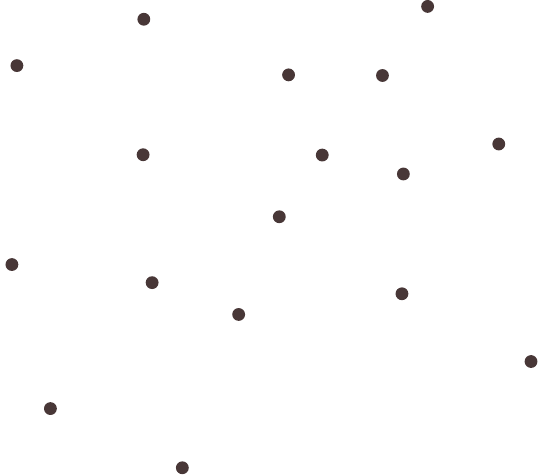}
		}
	}
	\hfil
	\subfloat[]{
		\resizebox{.4\columnwidth}{!}{
			\includegraphics{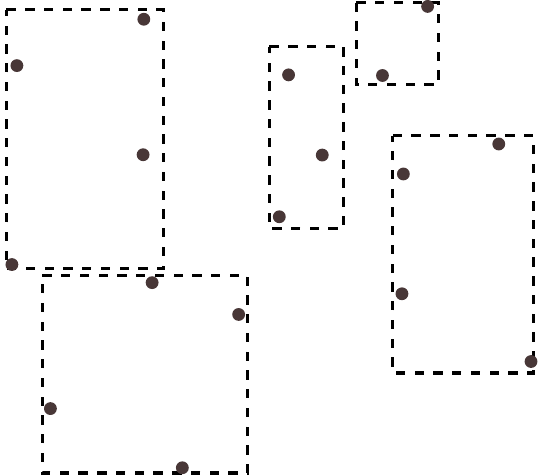}
		}
	}
	
	\caption{
		Example mapping function \PhiMapVert\ in 2D. 
		An initial set of points (a) is associated with the regions of a decomposition \D\ (b).
	} 
	\label{fig:mapping_verts_regions}
\end{figure}

\NOTA{\kennyComment{ Should we introduce the \vR\ and \tR\ notation here? Instead of in Section~\ref{sec:leaf_encodings}?}}
For the vertices, we have a map from \D\ to \sCV\ based on an application-dependent \emph{`belongs to'} property.
%
Formally, 
	$\PhiMapVert: \D \rightarrow \mathcal{P}(\sCV)$
is a map from \D\ to the powerset of \sCV\ where
\begin{equation*} 
	\forall \R \in \D, \PhiMapVert \left( \R \right) = \{\vertex \in \sCV : \vertex \text{ \emph{`belongs to'} } \R \}.
\end{equation*}
Figure~\ref{fig:mapping_verts_regions} illustrates an example decomposition \D\ over a point set where mapping function \PhiMapVert\ associates points with regions of \D. In this paper, we will assume that each vertex in \sCV\ is uniquely associated with a single region \R\ in \D.

\NOTA{
  While a region \R\ in \D\ is associated with a subset of vertices from \sCV, 
  the above definition does not limit a vertex $\vertex \in \sCV$ to be in a single region.
  However, we do require that for each vertex $\vertex \in \sCV$ there exists only one region \R\ containing \vertex, i.e., where $\vertex \in \PhiMapVert(\R)$. 
}

\begin{figure*}[t]
	\centering
	\subfloat[]{
		\resizebox{.25\textwidth}{!}{
			\includegraphics{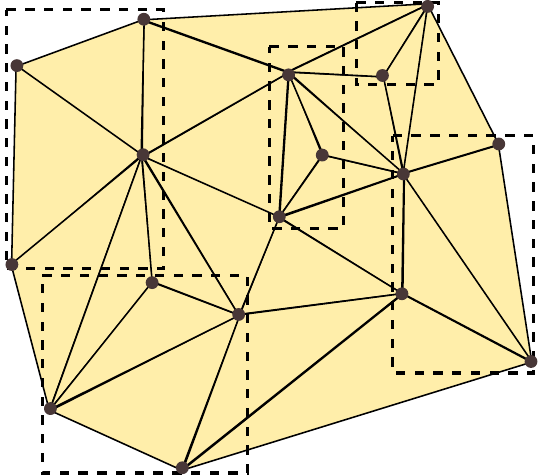}
		}
	}
	\hfil
	\subfloat[]{
		\resizebox{.25\textwidth}{!}{
			\includegraphics{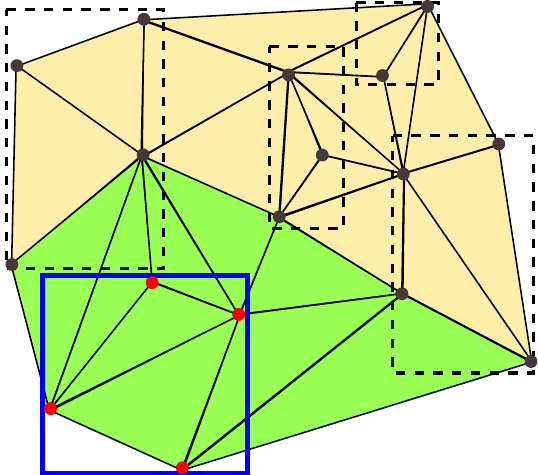}
		}
	}
	\hfil
	\subfloat[]{
		\resizebox{.25\textwidth}{!}{
			\includegraphics{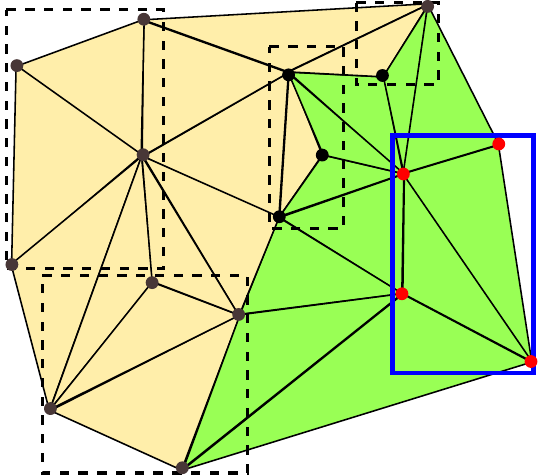}
		}
	}
	
	\caption{Mapping function \PhiMapTop\ for the decomposition \D\ from Figure~\ref{fig:mapping_verts_regions}. 
		Given a triangle mesh (a) and a vertex map \PhiMapVert\ on \D, \PhiMapTop\ associates the triangles in the star of the vertices in \PhiMapVert(\R) to \PhiMapTop(\R).
		(b) and (c) highlight the triangles (green) associated with two different regions (blue) of \D.
	} 
	\label{fig:mapping_tri_regions}
\end{figure*}

The Stellar decomposition gets its name from the properties of its top cell map \PhiMapTop. 
For each region \R\ of \D, \PhiMapTop(\R) is the set of all \topcpcells\ of \sCT\ incident in one or more vertices of \PhiMapVert(\R).
In other words, \PhiMapTop(\R) is defined by the union of cells in the \emph{star} of the vertices in \PhiMapVert(\R).
Formally, $\PhiMapTop: \D \rightarrow \mathcal{P}(\sCT)$  is a function from 
\D\ to the powerset of \sCT, where
\begin{equation} \label{eq:phitop_blocks}
	\forall \R \in \D, \PhiMapTop \left( \R \right) = \{\simplex \in \sCT | \exists \vertex \in \relation{k,0} \left( \simplex \right) : \vertex \in \PhiMapVert \left( \R \right)\}.
\end{equation}

Figure~\ref{fig:mapping_tri_regions} illustrates mapping \PhiMapTop\ for two regions of the decomposition of Figure~\ref{fig:mapping_verts_regions}(b) on a triangle mesh defined over its vertices.
Note that \PhiMapTop\ is based on a topological rather than a spatial property:  
A \topcp\ \simplex\ is only associated with a region \R\ when one (or  more) of its vertices is associated with \R\ under \PhiMapVert.

To characterize this representation, we define the \emph{spanning number} \ChiSimplex\ of a top CP cell in a Stellar decomposition
as the number of regions to which it is associated.

\begin{defn}  \label{def:chisimplex}
  Given Stellar decomposition $\sDec = (\sC,\D,\PhiMap)$,
  the \emph{spanning number} \ChiSimplex\ of a top CP cell $\simplex \in \sCT$
	is the number of regions in \D\ that map to \simplex. Formally,
	\begin{equation} \label{eq:chisimplex}
		\forall \simplex \in \sCT,\ \ChiSimplex = | \{ \R \in \D | \simplex \in \PhiMapTop \left( \R \right) \} |.
	\end{equation}
\end{defn}
A consequence of the unique mapping of each vertex in \PhiMapVert\ 
is that it provides an upper bound on the spanning number of a top CP cell in a Stellar decomposition. 
Specifically, the spanning number \ChiSimplex\ of a top CP cell \simplex\ is bounded by the cardinality of its vertex incidence relation \relation{k,0}:
$1 \leq \ChiSimplex \leq |\relation{k,0}(\simplex)|$.

It is also interesting to consider the \emph{average spanning number} \Chi\
as a global characteristic of the efficiency of a Stellar decomposition
over a complex, measuring the average number of times each \topcp\ is represented.
\begin{defn} \label{def:chi}
  The \emph{average spanning number} \Chi\ of a Stellar decomposition \sDec\ is the average number of regions indexing its top cells.
  Formally, 
	\begin{equation} \label{eq:chi}
		\Chi = \left( \sum_{\simplex \in \sCT} \ChiSimplex \right) / |\sCT| = \left( \sum_{\R \in \D} |\PhiMapTop \left( \R \right)| \right) / |\sCT|.
	\end{equation}
\end{defn}


\subsection{Encoding}
\label{sec:stellar_dec_enc}

In this section, we describe how we represent the two components of a Stellar decomposition, providing a detailed description of the data structures for representing a CP complex (subsection \ref{sec:mesh_structure}),
and a compressed encoding for the regions of the partitioning (subsection \ref{sec:leaf_encodings}).
We do not describe how the decomposition \D\ is represented, as this is specific to each concrete realization of the Stellar decomposition model.

\subsubsection{Indexed representation of the CP complex}
\label{sec:mesh_structure}

We represent the underlying CP complex as an indexed complex~\cite{Lawson1977Software}, 
which encodes the vertices, \ktopcpcells\ and the boundary relation \relation{\tDim,0} of each \ktopcell\ in \sC. 
\NOTA{ 
  which encodes the attributes associated with the vertices and the boundary relation \relation{\tDim,0} of each \ktop\ in \sC. 
  If the complex is embedded in the Euclidean, as we consider in this paper, the spatial position is associated with each vertex. 
}
In the following, we assume a \cDim-dimensional CP complex \sC\ embedded in \eSpace.


We use an array-based representation for the vertices and for the top cells of \sC.
Since the arrays are stored contiguously, each vertex \vertex\ has a unique position index \vIndex\ in the vertex array, that we denote as \sCV.
Similarly, each \topcp\ \simplex\ has a unique position index \tIndex.
%
%
The \topcpcells\ of \sC\ are encoded using separate arrays \sCTk\ for each dimension $\tDim \le \cDim$ that has \topcpcells\ in \sC. 
\sCTk\ encodes the boundary connectivity from the \ktopcpcells\ of \sC\ to their vertices, i.e., relation \relation{\tDim,0}
in terms of the indices \vIndex\ of the vertices of its cells within \sCV.
This requires $|\relation{\tDim,0}(\simplex)| $ references for a top \tDim-cell \simplex,
e.g., \tDim+1 vertex indices for a \tDim-simplex and $2^{\tDim}$ references for a \tDim-cube.
%
Thus, the total storage cost of \sCT\ is: 
\begin{equation}
\sum\limits_{\tDim=1}^d \sum\limits_{\simplex \in \sCTk} |\relation{\tDim,0}(\simplex)|.
\label{eq:storage_indexed_simplex}
\end{equation}
%
We note that when \sC\ is pure (i.e., its \topcpcells\ all have the same dimension \cDim), the encoding of \sC\ requires only two arrays:
one for the vertices and one for the top cells.
For simplicity, we refer to the top cell arrays collectively as \sCT.

\subsubsection{A compressed region representation}
\label{sec:leaf_encodings}

In this subsection, we discuss two encoding strategies for the data maps in each region of the partition \D. 
We begin with a simple strategy that explicitly encodes the arrays of vertices and \topcpcells\ associated with each region and work our way to a compressed representation of these arrays. 
Coupling this compressed representation with a reorganization of the vertices and cells of the CP complex 
(as we will describe in Section~\ref{sec:stellar_dec_generation})
yields a significant reduction in storage requirements. 
We will demonstrate this claim in Section~\ref{sec:storage_encodings} on a data structure instantiating the Stellar decomposition.

\begin{figure}[t]
	\centering
	\subfloat[]{
		\def\svgwidth{.45\columnwidth}
		{\scriptsize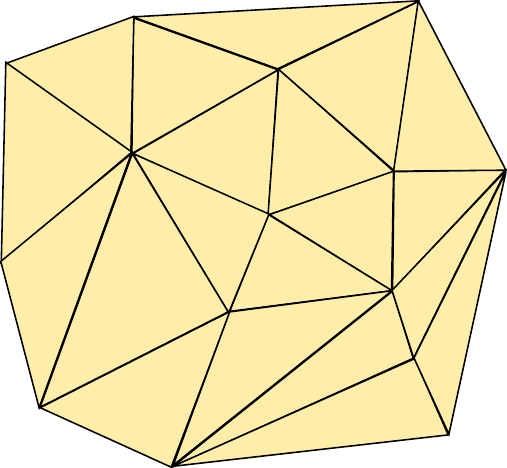}
	}
	\hfil
	\subfloat[]{
		\def\svgwidth{.45\columnwidth}
		{\scriptsize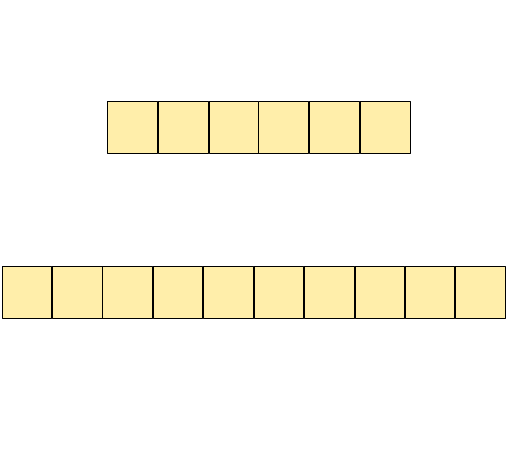}
	}
	\caption{\datasetName{explicit} encoding for triangles within a region (dotted square). 
		The arrays explicitly encode the 6 vertices and 20 triangles in the region.} 
	\label{fig:encoding_explicit}
\end{figure}

\NOTA{\kennyComment{ The \vR\ and \tR\ notation had not yet been introduced for readers to recall.
  I did it here by adding a ``\ldots which we denote as \vR\ \ldots''
  Alternatively, we could do it earlier (see my previous comment)
}}
Recall that under \PhiMap, each region \R\ in \D\ maps to an array of vertices and an array of \topcpcells\ from the complex \sC\ which we denote as \vR\ and \tR, respectively.
A straightforward strategy would be to encode arrays of vertices and \topcpcells\ that explicitly enumerate the associated elements for each region \R. We refer to this as the \datasetName{explicit} Stellar decomposition encoding.
An example of this encoding for a single region 
with six vertices in \vR\ and twenty triangles in \tR\ is shown in Figure \ref{fig:encoding_explicit}. 


It is apparent that the above encoding can be very expensive due to the redundant encoding of \topcpcells\ with vertices in multiple regions.
A less obvious redundancy is that it does not account for the ordering of the elements.

\begin{figure}[t]
	\centering
	\subfloat[RLE]{
		\def\svgwidth{.4\columnwidth}
		{\footnotesize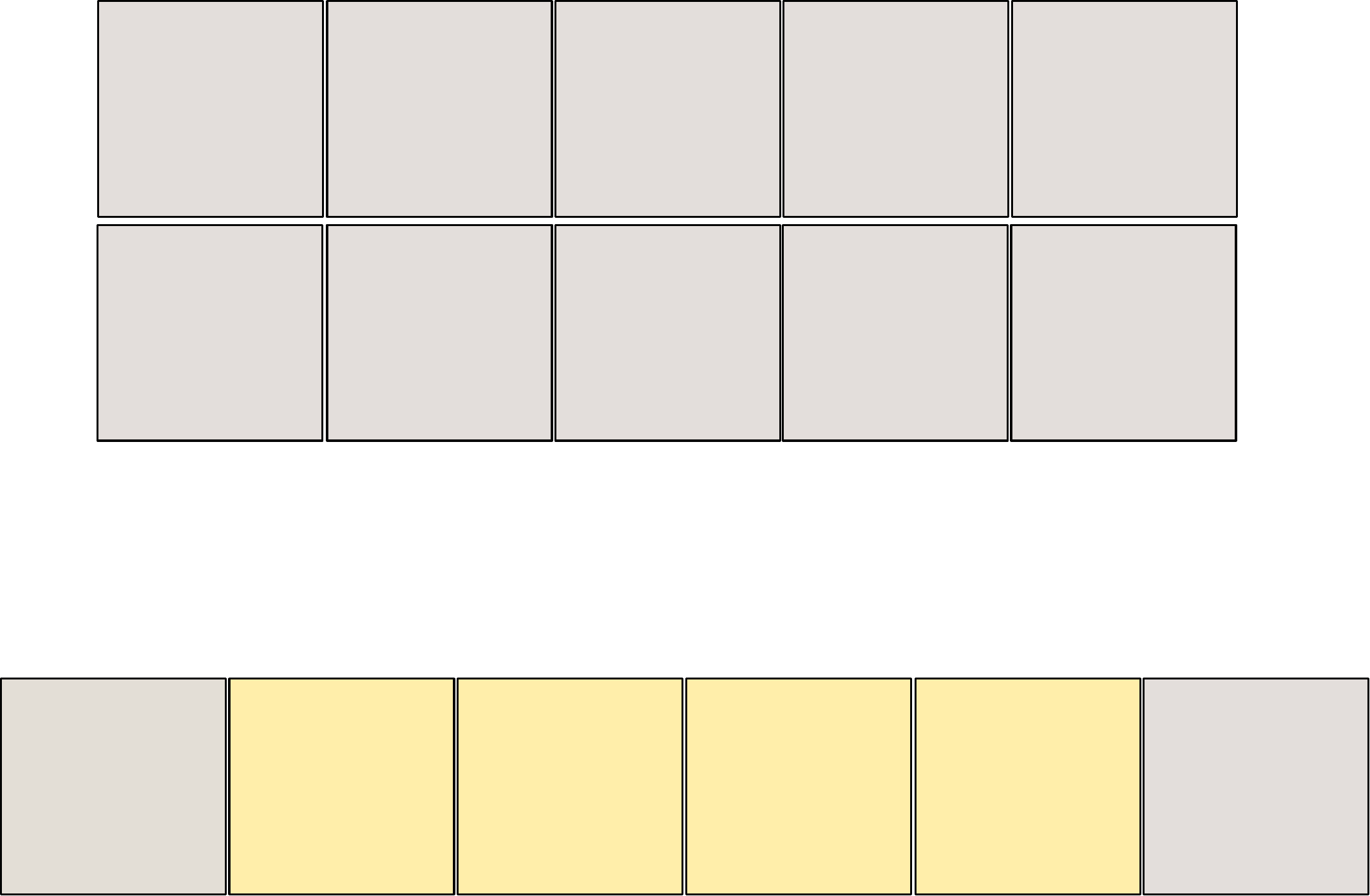}
		\label{fig:rle_example}
	}
	\hfil
	\subfloat[SRE]{
		\def\svgwidth{.4\columnwidth}
		{\footnotesize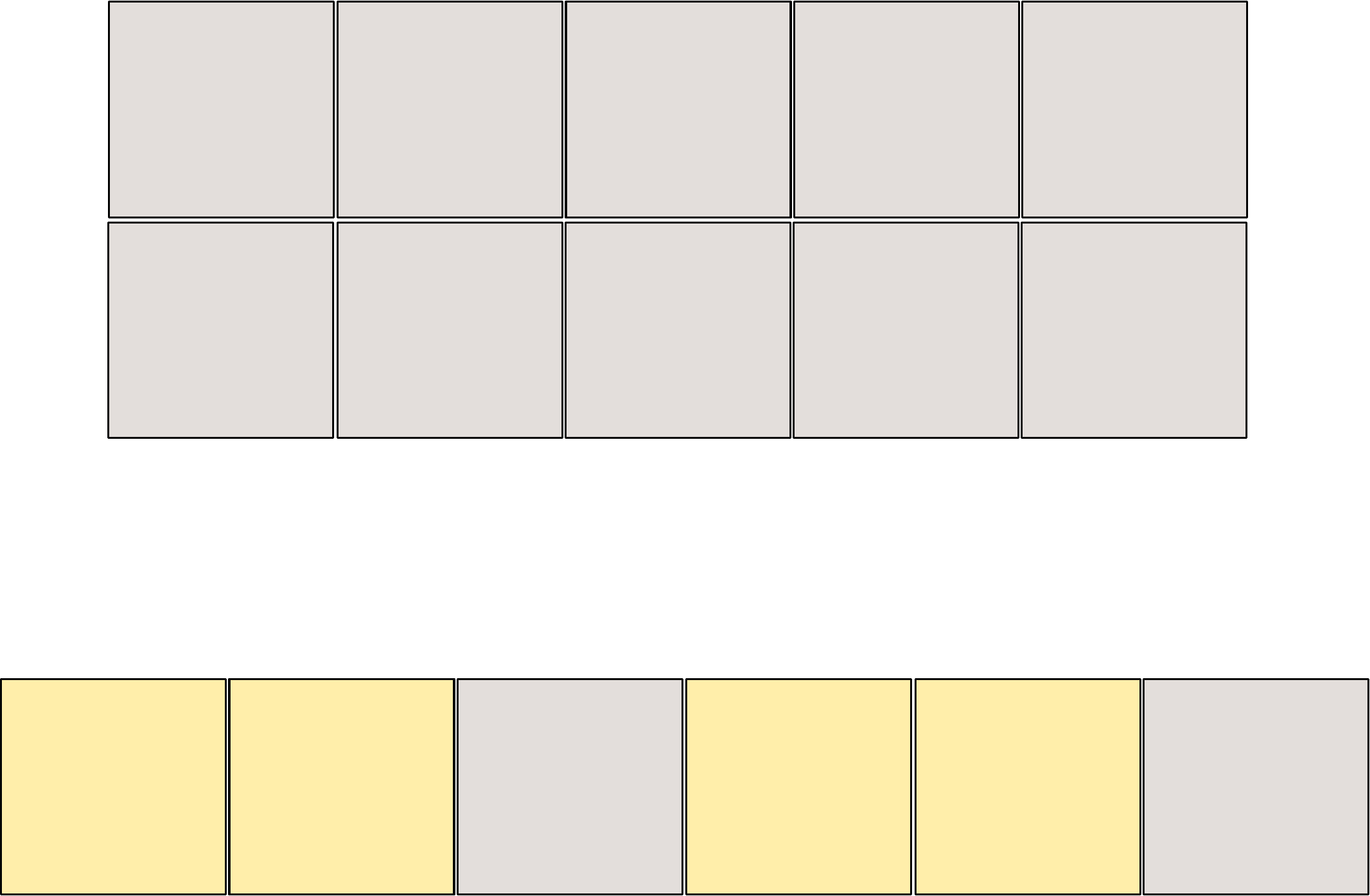}
		\label{fig:sre_example}
	}
	\caption{Compressed arrays of non-negative integers using (a) \emph{Run Length Encoding} (RLE) and (b) \emph{Sequential Range Encoding} (SRE).
	} 
	\label{fig:rle}
\end{figure}

We now consider a \datasetName{compressed} Stellar decomposition encoding that compacts the vertex and \topcpcells\ arrays 
in each region \R\ by exploiting the \emph{locality} of the elements within \R. 
%
The \datasetName{compressed} encoding reduces the storage requirements within region arrays by replacing \emph{runs} of incrementing consecutive sequences of indices using a generalization of \emph{run-length encoding (RLE)}~\cite{Held1991Data}. 
%
RLE is a form of data compression in which \emph{runs} of consecutive identical values
are encoded as pairs of integers representing the value and repetition count, rather than as multiple copies of the original value.
For example, in Figure~\ref{fig:rle_example}, the four entries with value `$2$' 
are compacted into a pair of entries $[\text{-}2,3]$, where a negative first number indicates the start of a run and its value,
while the second number indicates the remaining elements of the run in the range.

While we do not have such duplicated runs in our indexed representation, we often have increasing sequences of indexes, 
such as \{40,41,42,43,44\}, within a local vertex array \vR\ or \topcpcells\ array \tR.
We therefore use a generalized RLE scheme to compress such sequences, which we refer to as \emph{Sequential Range Encoding (SRE)}.
SRE encodes a run of \emph{consecutive} non-negative indexes using a pair of integers, 
representing the starting index, and the number of remaining elements in the range.
As with RLE, we can intersperse runs (sequences) with non-runs in the same array 
by negating the starting index of a run (e.g.\ $[\text{-}40,4]$ for the above example).
Thus, it is easy to determine whether or not we are in a run while we iterate through a sequential range encoded array.
A nice feature of this scheme is that it allows us to dynamically append individual elements or runs to an SRE array without any storage overhead (other than occasional array reallocations).
Furthermore, we can easily \emph{expand} a compacted range by replacing its entries with the first two values of the range and appending the remaining values to the end of the array.
After the updates are finished, we can sort the array and reapply SRE compaction to recover space.
Figure~\ref{fig:sre_example} shows an example SRE array over an array, where, e.g., sequence \{1,2,3,4\} is represented  as $[\text{-}1,3]$.

\begin{figure}[t]
	\centering
	\subfloat[]{
		\def\svgwidth{.45\columnwidth}
		{\scriptsize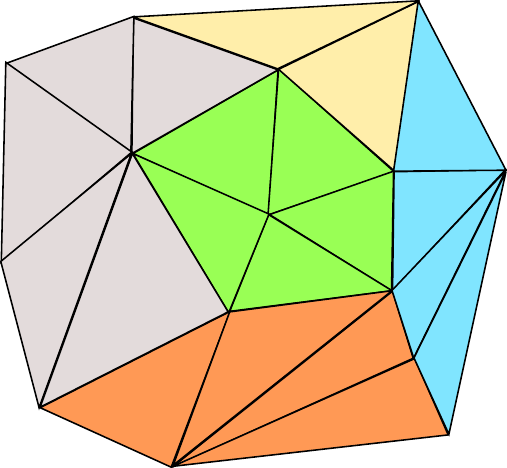}
	}
	\hfil
	\subfloat[]{
		\def\svgwidth{.45\columnwidth}
		{\scriptsize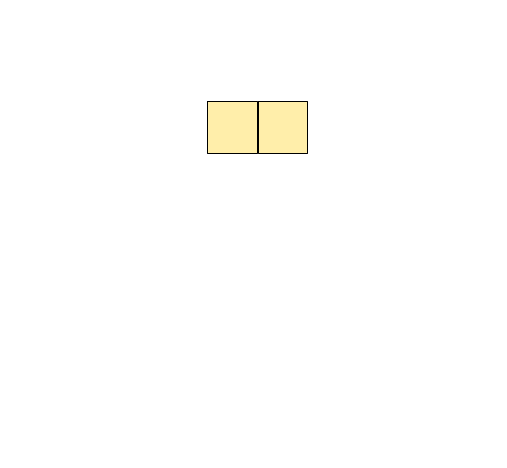}
	}
	\caption{\datasetName{compressed} encoding within a region (dotted square) 
           after reindexing the vertices and triangles of the mesh from Figure~\ref{fig:encoding_explicit}.} 
	\label{fig:encoding_compressed}
\end{figure}

To facilitate comparisons between the \datasetName{explicit} and \datasetName{compressed} representations of a Stellar decomposition, 
we introduce a global characteristic that measures the average storage requirements to represent a \topcp. 
\begin{defn} \label{def:mu}
	The \emph{average reference number} \Mu\ of a Stellar decomposition is the average number of references 
	required to encode a \topcp\ in the \tR\ arrays of the regions in \D.
	Formally:
	\begin{equation} \label{eq:mu}
	\Mu = \left( \sum_{\R \in \D} |\tR| \right) / |\sCT|
	\end{equation}
	where $|\tR|$ is the size of the \topcpcells\ array in a region \R.
\end{defn}
In contrast to the average spanning number \Chi, which is a property of the decomposition, the average reference number \Mu\ is a property of how the decomposition is encoded.
An \datasetName{explicit} representation is equivalent to a \datasetName{compressed} representation without any compressed runs,
and, thus, it is always the case that $\Mu \leq \Chi$.  In the \datasetName{explicit} representation (i.e., without any sequence-based compression), $\Mu = \Chi$,
while in the \datasetName{compressed} representation, \Mu\ decreases as the compression of the \vR\ and \tR\ arrays becomes more effective.
Figure~\ref{fig:encoding_compressed} illustrates a \datasetName{compressed} representation
of the mesh from Figure~\ref{fig:encoding_explicit}
after its vertex and triangle arrays have been reordered (in an external process) and highlights its sequential ranges, 
where \vR\ is encoded by a single run 
and \tR\ is encoded by four sequential runs as well as several non-run indices.

\subsection{Generating a Stellar decomposition}
\label{sec:stellar_dec_generation}

We now describe how to generate a \datasetName{compressed} Stellar decomposition from an indexed CP complex \sC\
%
and a given partition \D\ on its vertices \sCV. This process consists of three phases:
\begin{compactenum}
	\item reindex the vertices of \sC\ following a traversal of the regions of \D\ and SRE-compress the \vR\ arrays;
	\item insert the \topcpcells\ of \sC\ into \D;  
	\item reindex the \topcpcells\ of \sC\ based on locality within common regions of \D\ and SRE-compress the regions \tR\ arrays. 
\end{compactenum}

As it can be noted, the generation process ignores how the partitioning on the vertices is obtained, since this step is defined by the data structure instantiating a Stellar decomposition.
The reindexing of the vertices follows a traversal of the regions of \D\ in such a way that 
all vertices associated with a region have a contiguous range of indices in the reindexed global vertex array \sCV\ (as detailed in the Appendix~\ref{sec:verts_reordering}).
%

The second phase inserts each \ktopcp\ \simplex, with index \tIndex\ in \sCTk, into all the regions of \D\ that index its vertices. 
This is done by iterating through the vertices of \simplex\  and inserting \tIndex\ into the \tR\ array
of each region \R\ whose vertex map \PhiMapVert(\R) contains at least one of these vertices.
As such, each \ktopcp\ \simplex\ appears in at least one and at most $|\relation{\tDim,0}(\simplex)|$ regions of \D.
Due to the vertex reindexing of step 1, this operation is extremely efficient. 
Determining if a vertex of a given cell lies in a block requires only a range comparison on its index \vIndex.

Finally, we reindex the \topcp\ arrays \sCT\ to better exploit the locality induced by the vertex-based partitioning and compress the local \tR\ arrays using a sequential range encoding over this new index.
The reindexing and the compression of the \topcpcells\ is obtained following a traversal of the regions of \D\ in such a way that all \topcpcells\ associated with the same set of regions have a contiguous range of indices in the reindexed arrays \sCT.
This last step is detailed in the Appendix~\ref{sec:tops_reordering} and \ref{sec_app:tops_reordering_long}.
As we demonstrate in Section~\ref{sec:storage}, this compression yields significant storage savings in a wide range of mesh datasets.

\section{Stellar trees}
\label{sec:stellar_tree}


The Stellar decomposition is a general model that is agnostic about how the partitioning is attained and about its relationship with the underlying CP complex.
Thus, for example, we can define a Stellar decomposition using Voronoi diagrams, or based on a nearest neighbor clustering of the vertices of a given CP complex.
In this section, we introduce \emph{Stellar trees} as a class of Stellar decompositions defined over nested spatial decompositions of the CP complex 
and discuss some of our design decisions.
Before defining a Stellar tree (Section~\ref{sec:stellar_tree_def}), its encoding (Section~\ref{sec:stellar_tree_enc}) and its generation procedure (Section~\ref{sec:stellar_tree_generation}), we review some underlying notions.
%

\begin{figure}[t]
	\centering
	\subfloat[]{
		\resizebox{.4\columnwidth}{!}{
			\includegraphics{imgs/mapping_function_v_start}
		}
	}
	\hfil
	\subfloat[]{
		\resizebox{.4\columnwidth}{!}{
			\includegraphics{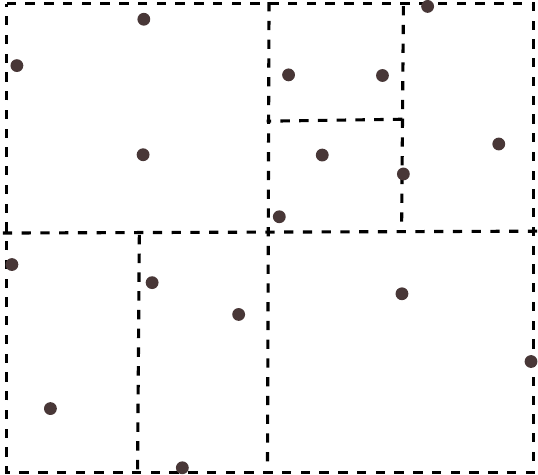}
		}
	}
	
	\caption{
		A mapping function \PhiMapVert\ over a nested spatial decomposition \D. 
		The vertices (a) are partitioned into regions by \D's leaf blocks (b)
    using a bucketing threshold, $\kv = 4$, i.e.\ at most 4 vertices can be in a region.
	} 
	\label{fig:mapping_verts_quad}
\end{figure}


The \emph{ambient space} \aSpace\ is the subset of $\eSpace$ in which the data is embedded. 
We consider the region bounding the ambient space to be a hyper-rectangular \emph{axis-aligned bounding block}, 
which we refer to simply as a \emph{block}.
%
A \tDim-dimensional \emph{closed} block \R\ in $\eSpace$, with $\tDim \leq \sDim$, is the Cartesian product of closed intervals $[l_i,u_i]$ 
where exactly \tDim\ of them are non-degenerate, i.e., 
\begin{math}
\R = \{ (x_1,\ldots,x_n)\in \eSpace  \,\, | \,\, x_i \in [l_i,u_i]\}
\end{math}
and $|\{ i\,\, |\, l_i<u_i\}| = \tDim$. 

Given two blocks $\R:=[l_i,u_i]$ and $\R':=[l'_i,u'_i]$, 
$\R'$ is a \emph{face} of $\R$ if, for each dimension $i$, either their intervals overlap (i.e., $l'_i=l_i$ and $u'_i=u_i$)
or the $i$-th interval of $\R'$ is degenerate (i.e., $l'_i=u'_i=l_i$, or $l'_i=u'_i=u_i$). 
%
Given a block \R, we refer to its 0-dimensional face of degenerate intervals $x_i = l_i$ as its \emph{lower corner}
and to its 0-dimensional face where $x_i = u_i$ as its \emph{upper corner}.
The above block definition describes \emph{closed} blocks.
It can be useful to allow some faces of \R\ to be \emph{open}, 
especially on faces of neighboring blocks that overlap only on their boundaries.
A \tDim-dimensional \emph{half-open} block $\R$ in $\eSpace$ is defined as
\begin{math}
\R = \{ (x_1,\ldots,x_n)\in \eSpace  \,\, | \,\, x_i \in [l_i,u_i)\}
\end{math}
and $|\{ i\,\, |\, l_i<u_i\}| = \tDim$. 
%
Note that the faces of a half-open block \R\ incident in its lower corner are \emph{closed},
while all other faces of \R\ are \emph{open}. 

We now focus on \emph{nested decompositions}, hierarchical space-based decompositions
whose overlapping blocks are nested and whose leaf blocks \DL\ (i.e., those without any nested blocks) 
form a non-overlapping cover of the ambient space \aSpace.
The nesting relationship 
defines a \emph{containment hierarchy} \h, which can be described using a rooted \emph{tree}.
The tree's root \hR\ covers the ambient space \aSpace; 
the tree's leaves \hL\ encode the regions of the decomposition \D;
and its internal nodes \hI\ provide access to the regions of the decomposition.

Nested decompositions can adopt different hierarchical refinement strategies.
Among the most popular are those based on \emph{regular} refinement 
and \emph{bisection} refinement of simple primitives (e.g., simplices and cubes).
An $\sDim$-dimensional block \R\ is regularly refined by adding vertices at all edge and face midpoints of \R\ 
and replacing \R\ with $2^{\sDim}$ disjoint blocks covering \R. 
This generates \emph{quadtrees} in 2D, and \emph{octrees} in 3D~\cite{Samet2006Foundations}.
In bisection refinement, a block is bisected along an axis-aligned hyperplane into two blocks, generating \emph{kD-trees}~\cite{Bentley1975Multidimensional}. 

\subsection{Definition}
\label{sec:stellar_tree_def}

Since a Stellar tree \sTree\ is a type of Stellar decomposition, it consists of three components:
\begin{inparaenum}[(1)]
	\item a \emph{CP complex} \sC\ embedded in an \emph{ambient space} \aSpace; 
	\item a \emph{nested decomposition} \D\ covering the domain of \sC; and
	\item a \emph{map} \PhiMap\ from blocks of \D\ to entities of \sC.
\end{inparaenum}
The nested decomposition is described by a containment hierarchy \h, represented by a \emph{tree} 
whose blocks use the \emph{half-open} boundary convention 
to ensure that every point in the domain is covered by exactly one leaf block.

Since Stellar trees are defined over nested spatial decompositions that cover the ambient space, 
we customize the vertex mapping function \PhiMapVert\ to partition the vertices of \sC\ according to spatial containment: 
each vertex is associated with its single containing leaf block. 
Formally, 
\begin{equation} \label{eq:phivert_blocks}
	\forall \R \in \DL, \PhiMapVert(\R) = \{\vertex \in \sCV : \vertex \cap \R \neq \emptyset \}.
\end{equation}

A two-dimensional example is shown in Figure~\ref{fig:mapping_verts_quad}, where a set of points
are associated with the leaf blocks of \D\ through \PhiMapVert.

\begin{figure}[t]
	\centering
	\subfloat[]{
		\resizebox{.45\columnwidth}{!}{
			\includegraphics{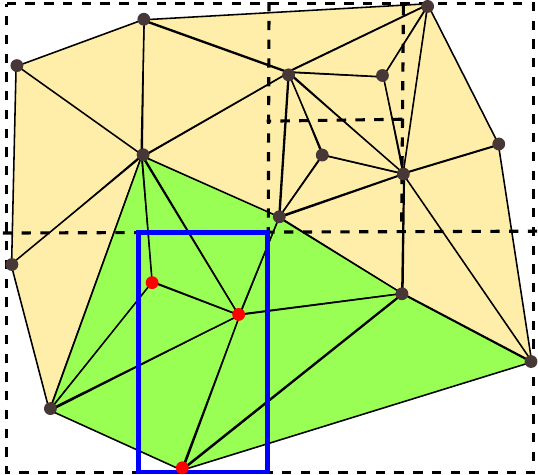}
		}
	}
	\hfil
	\subfloat[]{
		\resizebox{.45\columnwidth}{!}{
			\includegraphics{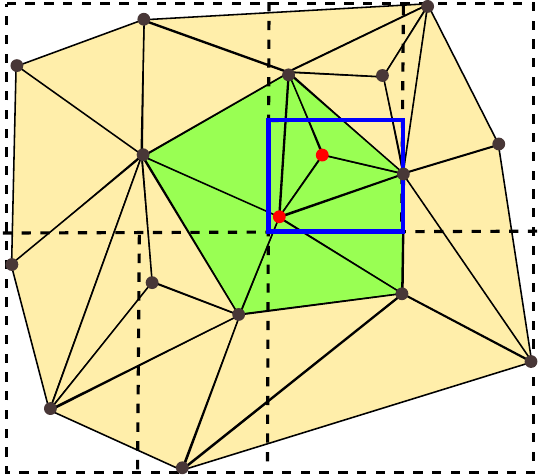}
		}
	}
	\caption{Top cell mapping function \PhiMapTop\ for two blocks (blue) 
    of the nested decomposition from Figure~\ref{fig:mapping_verts_quad}
	  on the triangle mesh from Figure~\ref{fig:mapping_tri_regions}. 
		\PhiMapTop(\R) maps the triangles in the star of the vertices in \PhiMapVert(\R).
	} 
	\label{fig:mapping_tri_quad}
\end{figure}

The \topcpcells\ mapping function \PhiMapTop\ for a Stellar tree has the same definition as for the Stellar decomposition (see Equation~\ref{eq:phitop_blocks}).
Figure \ref{fig:mapping_tri_quad} shows the mapping \PhiMapTop\ for two blocks of the nested kD-tree decomposition of Figure~\ref{fig:mapping_verts_quad}(b)
over the triangle mesh from Figure~\ref{fig:mapping_tri_regions}.

%
Since the nested decomposition \D, and, consequently, the tree \h\ describing it, 
are determined by the number of vertices indexed by a block, we utilize a \emph{bucket PR tree}~\cite{Samet2006Foundations} to drive our decomposition.
This provides a single tuning parameter, the \emph{bucketing threshold} \kv, 
that uniquely determines the decomposition for a given complex \sC.

Recall that a (leaf) block \R\ in a bucket PR-tree is considered \emph{full} when it indexes more than \kv\ vertices (in our case, when $|\PhiMapVert(\R)| > \kv$).
Insertion of a vertex into a full block causes the block to refine
and to redistribute its indexed vertices among its children.
As such, the domain decomposition of a Stellar tree depends only on the bucketing threshold \kv.
%
Smaller values of \kv\ yield deeper hierarchies whose leaf blocks index relatively few vertices and \topcpcells,
while larger values of \kv\ yield shallower hierarchies with leaf blocks that index more vertices and \topcpcells. 
Thus, \kv\ and the average spanning number \Chi\ of a Stellar tree are inversely correlated. 
%

In practice, we use different spatial indexes to represent \h\ based on the dimension \sDim\ of the ambient space \aSpace.
  In lower dimensions, we use a quadtree-like subdivision, i.e., a quadtree in 2D, and an octree in 3D, while in higher-dimensions, we switch to a kD-tree subdivision. 
As discussed in~\cite{Samet2006Foundations}, while quadtree-like subdivisions are quite efficient in low dimensions, 
the data becomes sparser in higher dimensions (due to the \emph{curse of dimensionality}~\cite{Bellman1966Dynamic}), and tends to be better encoded by kD-trees. 

\subsection{Encoding}
\label{sec:stellar_tree_enc}

We represent the containment hierarchy \h\ using an explicit pointer-based data structure, 
in which the blocks of \h\ use a type of \texttt{Node} structure that changes state from leaf to internal block
during the generation process of a Stellar tree. 

We use a \emph{brood-based} encoding~\cite{Hunter1991Classification}, where each block in \h\ encodes a pointer to its parent block and a single pointer to its brood of children. This reduces the overall storage since leaves do not need to encode pointers to their children, and also allows us to use the same representation for n-dimensional quadtrees and kD-trees.
We explicitly encode all internal blocks, but only represent leaf blocks \R\ in \h\ with non-empty maps \PhiMap(\R).

\NOTA{
	\kennyComment{Do either of you have access to \cite{Hunter1991Classification}? 
		I found the reference in \cite{Lacoste2007Appearance}, but couldn't access the paper. \\
		\textbf{R}: no chance also from UMD library.. we do not have the Wiley subscription}
}

The mapped entities of the CP complex \sC\ are encoded in the leaf blocks \hL\ using the mapping arrays \PhiMap.
%
Note that each leaf block \R\ encodes the arrays of vertices \vR\ and of \topcpcells\ \tR\ in terms of the indices \vIndex\ and \tIndex, 
respectively, that identify \vertex\ and \simplex\ in the \sCV\ and \sCT\ arrays. 
%
%
For each block \R, we have:
\begin{inparaenum}[(1)]
	\item three pointers for the hierarchy: one to its parent, another to its list of children and it is pointed to by one parent;
	\item a pointer to an array of vertices \vR\ and the size of this array;
	\item a pointer to an array of \topcpcells\ \tR\ and the size of this array.
\end{inparaenum}
Thus, the hierarchy \h\ of a Stellar tree requires $7 |\h|$ storage.  
%
\NOTA{
  \felleComment{removed the figure illustrating the hierarchy representation. I do agree with Leila that it does not add anything to the text}
}

By considering the encodings, defined in Section \ref{sec:leaf_encodings}, for the CP complex \sC, and for the vertices and top cp-cells associated with the regions of \h, we can estimate the storage requirements for the \datasetName{explicit} and \datasetName{compressed} Stellar trees.
An \datasetName{explicit} Stellar tree requires a total of $|\sCV|$ references for its vertex arrays, since each vertex is indexed by a single leaf block, and a total of $\Chi|\sCT|$ references for all \topcpcells\ arrays.
Thus, the total cost of the \datasetName{explicit} Stellar tree, including the hierarchy (but excluding the cost of the indexed mesh) is:
$
	7 |\h| + |\sCV| + \Chi|\sCT|.
$

Conversely, in a \datasetName{compressed} Stellar tree, 
we can reindex the vertex array \sCV\ in such a way that all vertices associated with the same leaf block are indexed consecutively (see Section \ref{sec:verts_reordering} in the Appendix for additional details).
Thus, we can encode the \vR\ arrays using only two integers per leaf block for a total cost of $2|\hL|$ rather than $|\sCV|$.
Moreover, since leaf blocks no longer need to reference an arbitrary array, these two references can be folded into the block's hierarchical representation for \vR: 
instead of a pointer to a array and a size of the array, we simply encode the range of vertices in the same space.
As the cost of representing the \tR\ arrays is $\Mu|\sCT|$,
the total cost for encoding a \datasetName{compressed} Stellar tree 
(excluding the cost of the indexed mesh representation) is:
$
	7 |\h| + \Mu |\sCT|.
$

\begin{figure}[t]
	\centering
	\subfloat[Vertices inserted]{
		\resizebox{.45\columnwidth}{!}{
			\includegraphics{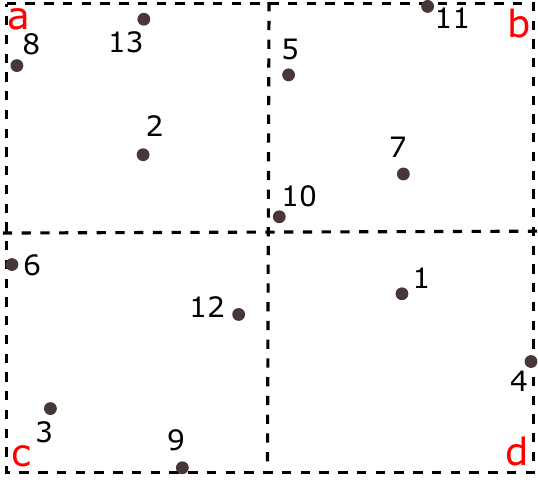}
		} 
	}
	\hfil
	\subfloat[Vertices reindexed]{
		\resizebox{.45\columnwidth}{!}{
			\includegraphics{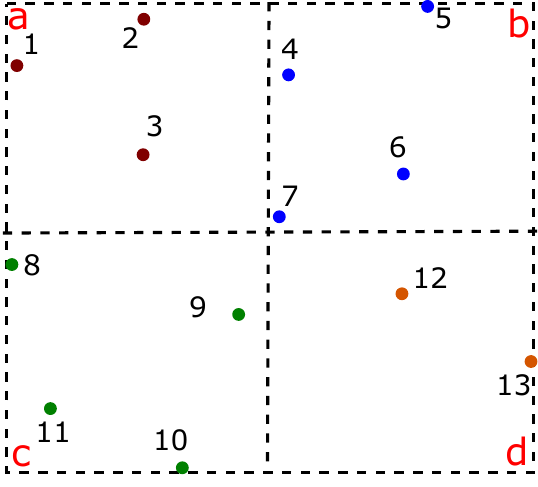}
		} 
	}
	\caption{Generating a nested hierarchy \h\ with $\kv=4$ over vertices. 
		After inserting the vertices (a), we reindex \sCV\ according to \hL\ (b).} 
	\label{fig:stellar_tree_vertices_reindexing}
\end{figure}

\subsection{Generating a Stellar tree}
\label{sec:stellar_tree_generation}


In this section, we describe how to generate a \datasetName{compressed} Stellar tree from an indexed CP complex \sC\ and a given bucketing threshold \kv.
%
We can also deal with input complexes that are not already indexed. For example, if our input is a ``soup'' of CP cells in which each CP cell is specified by a list of coordinates, we can generate an indexed representation of the complex as we insert the vertices and generate the decomposition. 

%
First, given a user-defined bucketing threshold \kv, we generate a bucket PR-tree 
over the vertices of $\Sigma$. 
%
\NOTA{
  Note that this is the only phase of the generation process that depends on the geometry of \sC.
  Although we do not maintain the spatial extent of each tree block, we can reconstruct it by tracking the split planes as we descend the tree
  (based on a bounding box enclosing \sC\ defined as the root \hR\ of the hierarchy).
}
The procedure for inserting a vertex \vertex\ with index \vIndex\ in \sCV\ into \h\ is recursive.  
We use the geometric position of \vertex\ to traverse the internal blocks to reach the unique leaf block \R\ containing \vertex.
After adding \vertex\ to \R\ (i.e., appending \vIndex\ into the \vR\ array of \R),
we check if this causes an overflow in \R. 
If it does, we refine \R\ and reinsert its indexed vertices into its children.
Once all the vertices in \sC\ have been inserted, the decomposition is fixed.

The rest of the Stellar tree generation process follows the strategy described in Section~\ref{sec:stellar_dec_generation} and detailed in the Appendix \ref{sec_app:stellar_dec_generation}. One key optimization between a generic partitioning on the vertices and a nested hierarchical decomposition relates to  extracting the vertex index ranges. 
In a Stellar tree, this step is performed through a depth-first traversal of the tree, which, for each leaf block \R, generates a contiguous range of indices for the vertices in \R, and, for each internal block, provides a single contiguous index range for the vertices in all descendant blocks. 
For example, in Figure~\ref{fig:stellar_tree_vertices_reindexing}, after executing this step on leaf block \textit{b}, we have $\vstart = 4$ and $\vend= 7$. Similarly, at the end of this step the root \hR\ has $\vstart=1$ and $\vend=13$.

We provide an experimental evaluation of the timings for generating a Stellar tree in Section~\ref{sec:generation_timings}.

\NOTA{
	Recall that, we assume that all the blocks of \h\ are \emph{half-open} blocks unless a block \R\ is incident 
	in the block representing the ambient space \aSpace, as in this case we consider that face of \R\ as \emph{closed}.
}

\NOTA{
	We insert the vertices sequentially, over which we start a visit of \h\ to insert a vertex \vertex\ in a leaf block \R\ containing it.
	We recall that we do not explicitly encode within each block its domain, but we compute it at runtime, keeping track of the split planes. 
	We assume that the root block \hR\ completely covers the complex domain and
	represents it as a \emph{closed block}. 
}

\NOTA{
	We insert each \top\ \simplex\ at once and for each boundary vertex \vertex\ of a \ktop\ \simplex, we find the leaf block \R\ of \hL\ that indexes \vertex\ 
	and then, we add \simplex\ to \top\ array of \R\ (i.e., we insert \tIndex\ index in \tB\ array).
}

\NOTA{
	\begin{enumerate}
		\item first, we navigate the tree to compress it and to get the spatially coherent ordering of the vertices (see Algorithm \ref{alg:get_vertices_ordering});
		\item then, for each \simplex\ in \sCT, we update its boundary relation \relation{k,0} (see rows 4 to 6 of Algorithm \ref{alg:main_vertices_reordering});
		\item finally, we update the \sCV\ array according to the new position ordering on the vertices.
	\end{enumerate}
	As auxiliary data structure for this procedure, we need an array of integer references, that we call $v\_permutation$. 
	This array contains exactly $|\sCV|$ entries, and at each entry $i$, it is associated the $i$-th vertex \vertex\ in \sCV. 
	This entry in $v\_permutation$ contains the new \emph{spatially coherent} position of \vertex\ in \sCV. 
	Thus, the extra storage overhead is exactly $|\sCV|$.
	$v\_permutation$ is the output of Algorithm \ref{alg:get_vertices_ordering}, and it is required as input during the update 
	of the \topcp\ boundary relations (rows 4 to 6 of Algorithm \ref{alg:main_vertices_reordering}) and during the update of \sCV\ array.
	%
	Algorithm \ref{alg:get_vertices_ordering} is a recursive procedure in which we visit all the blocks in \h. 
	If we are in an internal block (see rows 1 to 5 of the Algorithm), we recursively visit all the children, while if we are in a leaf block \R\ 
	(see rows 7 to 12 of the Algorithm), we visit the vertices array \PhiMapVert(b), we set up a consecutive indexes run in \R\ 
	and we update the corresponding entries in $v\_permutation$.
}

\NOTA{
	After this stage, each leaf block contains a contiguous range of vertices and each internal block contains a continuous range 
	of vertices equal to the union of the ranges contained in its descendant.
	Moreover, we do not require extra structures to encode this range, thanks to the \emph{sequential run-length} compression, and, thus, 
	in each vertex array, we require just two entries that have been initialized in order to represent the unique vertices run into the leaf block. 
	We denote the extreme vertices of this run as \vstart\ and \vend.
	Then, we proceed with the updating of boundary relation \relation{k,0} on all \ktops\ in \sCT\ (see rows 4 to 6 in Algorithm \ref{alg:main_vertices_reordering}). 
	For each vertex \vertex\ (with index \vIndex\ in \sCV) in \relation{k,0} of a \ktop\ \simplex, we get its spatial coherent position 
	in \sCV\ from $v\_permutation$, by accessing the \vIndex\ entry. Once we get this position we update the entry, associated to \vertex, 
	in \relation{k,0}(\simplex) with the value in $v\_permutation[\vIndex]$.
	Finally, we update the \sCV\ array accordingly with the spatially coherent ordering (in Algorithm \datasetName{update\_array}). 
	During the algorithm, we iteratively swap the index positions, updating at each swap operation  a vertex \vertex, 
	for which we gather its new spatially coherent position from $v\_permutation[\vIndex]$. This algorithm does exactly $|\sCV|$ swaps without requiring any extra storage.
}

\NOTA{
	In order to update $M$ and $t\_position$, we need to extract the tuple of leaf blocks indexing \simplex.
	If \simplex\ is completely indexed into \R, then we have its tuple already available. Otherwise, we have to visit the tree to find the leaf blocks that index the other vertices of \simplex.
	Once we get the tuple of leaf blocks indexing \simplex, we have to check if the tuple is already into $M$, and if it is not present we insert it. Conversely, if it is already into $M$, we simply increment the counter that keeps track of the \topcpcells\ indexed into that tuple.
}

\section{Processing paradigm for Stellar trees}
\label{sec:general_strategy}

	\begin{table*}[tb]
		\centering
		\caption{Overview of experimental datasets.
				 For each CP complex \sC, we list the number of vertices $|\sCV|$ and of top CP-cells $|\sCT|$.
		}
		{
		\resizebox{\textwidth}{!}{
			\begin{tabular}{c|c|c|c|c|c|c|c|c|c|c|c|c|c|c|c|c|c|c}
				\toprule
				
				 & \begin{sideways}\datasetName{neptune}\end{sideways} & \begin{sideways}\datasetName{statuette}\end{sideways} & \begin{sideways}\datasetName{lucy}\end{sideways} &
				 \begin{sideways}\datasetName{neptune}\end{sideways} & \begin{sideways}\datasetName{statuette}\end{sideways} & \begin{sideways}\datasetName{lucy}\end{sideways} &
				 \begin{sideways}\datasetName{bonsai}\end{sideways} & \begin{sideways}\datasetName{vismale}\end{sideways} & \begin{sideways}\datasetName{foot}\end{sideways} &
				 \begin{sideways}\datasetName{f16}\end{sideways} & \begin{sideways}\datasetName{san fern}\end{sideways} & \begin{sideways}\datasetName{vismale}\end{sideways} &
				 \begin{sideways}\datasetName{5D}\end{sideways} & \begin{sideways}\datasetName{7D}\end{sideways} & \begin{sideways}\datasetName{40D}\end{sideways} &
				 \begin{sideways}\datasetName{vismale 7D}\end{sideways} & \begin{sideways}\datasetName{foot 10D}\end{sideways} & \begin{sideways}\datasetName{lucy 34D}\end{sideways} \\
				 \midrule
				 & \multicolumn{3}{c|}{\datasetName{triangular}} & \multicolumn{3}{c|}{\datasetName{quadrilateral}} & \multicolumn{3}{c|}{\datasetName{tetrahedral}} & \multicolumn{3}{c|}{\datasetName{hexahedral}} & \multicolumn{3}{c|}{\datasetName{probabilistic}} & \multicolumn{3}{c}{\datasetName{v-rips}} \\
				 \midrule
				 \textbf{$|\sCV|$} & 2.00M & 5.00M & 14.0M & 12.0M & 30.0M & 84.1M & 4.25M & 4.65M & 5.02M & 27.9M & 61.3M & 136M & 385K & 239K & 204K & 4.65M & 5.02M & 14.0M \\
				 \midrule
				 \textbf{$|\sCT|$} & 4.01M & 10.0M & 28.1M & 12.0M & 30.0M & 84.2M & 24.4M & 26.5M & 29.5M & 25.4M & 55.9M & 125M & 26.5M & 258M & 16.5M & 6.39M & 63.9M & 41.1M \\
				 \bottomrule
			\end{tabular}
		}
		}
		\label{tab:datasets}
	\end{table*}

\NOTA{
  \leilaComment{In Section \ref{sec:stellar_ecosystem} we should say when we use this paradigm in the context of the various applications}
}

\NOTA{ 
An advantage of the Stellar tree is that it enables one to defer decisions 
about the details of the topological data structure.
Thus, one can easily customize the structure and layout of the representation to better suit the needs of a given application.
We note that processing individual mesh elements in a Stellar tree can be expensive due to the compressed leaf block format.
For example, to reconstruct the star of a given vertex \vertex, we must first identify the leaf block \R\ indexing  \vertex\ and visit the top cells in \R\ to identify those that are incident in \vertex. 
To amortize the reconstruction costs, we adopt a \emph{batched} processing strategy for Stellar tree applications,
in which we reconstruct and process local subsets of the complex. 
This tends to work well in practice since mesh processing applications are often applied to the entire complex
or within spatial regions of interest.
}


Mesh processing applications rarely process individual mesh elements. 
Rather, they typically operate on the entire complex, or on large regions of interest within the complex.
%
The structure of a Stellar tree naturally supports a \emph{batched} processing strategy, i.e., a strategy in which portions of the complex are reconstructed and processed within each block of the tree. 
As these local blocks are processed, their representation and extracted topological relations can be customized
to suit the needs of the application. 
This helps in amortizing the reconstruction costs and, thus, processing the entire complex efficiently.
%

%
The general paradigm for executing application algorithms on a Stellar tree is to iterate through the 
leaf blocks of hierarchy \h, locally processing the encoded complex in a streaming manner.
For each leaf block \R\ in \h, a local topological data structure catered to the application's needs is constructed
and used to process the indexed subcomplex. 
We refer to this local data structure in a block \R\ as an \emph{expanded leaf-block representation}, and we denote it as \eR. 
Once we finish processing leaf block \R, we discard \eR\ and begin processing the next block.

For efficiency and at relatively low storage overhead, we can cache the expanded leaf block representation \eR, using a \emph{Least-Recent-Used (LRU)} cache.
This is especially advantageous in applications that require processing portions of the complex in neighboring leaf blocks.
Adopting a fixed-size cache allows us to amortize the extraction costs of the local data structures, with a controllable storage overhead.

Algorithm~\ref{alg:stellar_generic_application} outlines the general strategy for processing
a Stellar tree. 
The algorithm recursively visits all the blocks of the hierarchy \h. 
For each leaf block \R, we either recover \eR\ from the \emph{LRU} cache (rows 5--8),
or construct the desired application-dependent local topological data structure \eR.
After using this local data structure to process the local geometry in \R\ (row 9),
we either cache or discard \eR\ (rows 10--13).

\NOTA{ \kennyComment{
		Do we use caching in any of the applications in this paper?
    If not, can (should?) we point out a specific application benefits from caching?
} }

\NOTA{
\kennyComment{CHECK:
	\\ Do we have an experiment that measures the effects of caching?}
\felleComment{not really.. we have used the cache in Morse and during the geometric simplification.. but never test the behavior extensively.. i.e., we only check for some sizes which was the best one.. and simply use it}
\kennyComment{ Do we have an experiment that measures the effects on memory and compute times for a broad range of different $k_v$ values? E.g. reconstruction of a relation, and its application to a larger problem? }
\felleComment{also here.. not really..  can we do this after the initial thesis submission for \relation{0,k} extraction?}
}

\begin{algorithm}[tb]			
	\caption{ $\AlgoName{stellar\_tree\_processing\_paradigm}(\R,\emph{c})$ }
	\label{alg:stellar_generic_application}				
	\begin{algorithmic}[1]			
		\Input {$\R$ is a block in \h}
		\Input {\emph{c} is a fixed-size \emph{LRU}-cache} 
		\If { $\R$ is an internal block in \h}
			\ForAll {blocks $\R_C$ in \AlgoName{children}(\R)}
					\State {\AlgoName{stellar\_tree\_processing\_paradigm}($\R_C$,\emph{c})}
			\EndFor
		\Else \SideComment{$\R$ is a leaf block in \h}
			\If {\R\ is in $c$}
				\State {\eR\ $\leftarrow$ \AlgoName{get}($c$,\R)}
			\Else
				\State {\eR\ $\leftarrow$ \AlgoName{expand}(\R)} \SideComment{expand \R\ into \eR}
			\EndIf
			
			\State {execute application algorithm using \eR}
			\If {\AlgoName{max\_size}($c$) $> 0$} \SideComment{we are using a cache}
				\State {save \eR\ in \emph{c}}
			\Else
				\State {discard \eR}
			\EndIf
		\EndIf
	\end{algorithmic}
\end{algorithm}

%

%
Within this general processing paradigm, we can have two different approaches, that we call \textit{local} and \textit{global},  depending on how auxiliary data structures are encoded and maintained.  In a \textit{local} approach, the scope of these auxiliary data structures is limited to that of a single leaf block \R, or to a restricted subset of its neighbors.
In general, a local approach is preferred for applications that extract, or analyze local features,
such as those that depend only on the link or star of cells. These includes, for instance, the extraction of geometric features, like the curvature at a vertex, or the extraction of morphological features, like critical points, when the complex is a discretization of the domain of a scalar field. In these examples, the auxiliary data structures are just needed within the scope of a leaf block \R, and thus, immediately discarded after extracting the corresponding feature in \R.
%
Conversely, in a \textit{global} approach, data structures are maintained over the entire complex and updated during the visit of the tree.
A global approach can be preferable for applications that require the analysis or the processing of the entire complex, like
geometric simplification, morphological segmentation, or validation of geometric and topological properties. 
In these examples, 
auxiliary data structures are used to represent partial results over the complex.
The decision between using a local and global approach can be driven by the needs of the application or as a tradeoff balancing memory usage and execution times.
Due to the limited scope of auxiliary data structures in the local approach, the storage overhead is typically proportional 
to the complexity of the local complex but requires an increased number of memory allocations compared to a global approach 
since each leaf block expansion requires memory allocations.
Conversely, while auxiliary data structures in the global approach are allocated only once, 
these structures can require significantly more storage space compared to the local approach.


\NOTA{
\felleComment{updated this last paragraph, since: (1) we are no longer presenting how to generate a topological d.s. using the Stellar tree, while (2) we give an overview of all the applications defined in a stellar tree}
}
%
In Section~\ref{sec:stellar_ecosystem}, we present applications on mesh processing and analysis, based on Stellar trees, on which these two paradigms have been extensively applied.

%
\NOTA{
\felleComment{I commented a sentence that can be used in Section \ref{sec:stellar_ecosystem}. \emph{Thus, Stellar trees can be also used as an intermediary representation by applications that expect a specific topological data structure, or on very large meshes, when there are insufficient resources to generate the original data structure}}
}



\section{Experimental setup}
\label{sec:threshold_calibration}

\NOTA{ Kenny moved the include dataset table to the previous section }

\NOTA{
\felleComment{give the motivation of these experiments and expand the evaluation}
\felleComment{here I am not sure if we should introduce the experimental datasets, since we are just using 3 of them. Also, moving Table 1 here does make sense.. since it gives \kv\ thresholds and trees statistics have not been introduced at this point. \\ 
		Possibilities: (1) just introduce here the datasets we are using, delegating the description later in section \ref{sec:storage} \\
		(2) describe the datasets in a separate section and create a smaller table showing just dataset statistics (the update Table 1 to just show kv and trees statistics)}
}
	
In this section, we describe our experimental setup, 
including the datasets used in our evaluation (Section~\ref{sec:stellar_experimental_plan}). 
We also evaluate how the bucketing threshold \kv\ affects the quality of a Stellar tree's decomposition and its performance in extracting topological queries (Section \ref{sec:calibration_results}).

\begin{figure*}[t]
	\centering
	\subfloat[\datasetName{neptune} triangle complex]{
		\resizebox{.32\textwidth}{!}{
			\begin{tikzpicture}
	\pgfplotstableread[col sep=comma]{charts/chart_neptune_tri.csv}\tableND
	
	\begin{axis}[
	separate axis lines,
	y axis line style={darkerBlue},
	y label style={darkerBlue},
	y tick style={darkerBlue},
	ymode=log,
	log ticks with fixed point,
	axis y line*=left,
	ylabel = time (seconds),
	xlabel = block threshold ($k_v$),
	font=\small,
  ymin=.1,
  ymax=110,
%
	]
	\draw [kvs,thick] ({axis cs:100,0}|-{rel axis cs:0,1}) -- ({axis cs:100,0}|-{rel axis cs:0,0});
	\draw [kvl,thick] ({axis cs:500,0}|-{rel axis cs:0,1}) -- ({axis cs:500,0}|-{rel axis cs:0,0});
	\addplot[kvs,thick] coordinates{(100,0.5)} node {\ks};
	\addplot[kvl,thick] coordinates{(500,0.5)} node {\kl};
	\addplot[solid,darkerBlue,very thick] table [x expr={\thisrow{kv}}, y expr={\thisrow{gen}}, col sep=comma, meta=kv] {charts/chart_neptune_tri.csv}; \addlegendentry[color=darkerBlue]{generation time} \label{gen}
	\addplot[solid,lightBlue,very thick] table [x expr={\thisrow{kv}}, y expr={\thisrow{vt}}, col sep=comma, meta=kv] {charts/chart_neptune_tri.csv}; \addlegendentry[color=lightBlue]{top co-boundary time} \label{vt}
	\end{axis}
	
	\begin{axis}[
	ymode=log,
	separate axis lines,
	y axis line style={red},
	y label style={red},
	y tick style={red},
	axis y line*=right,
	axis x line=none,
	ylabel = number of blocks,
	font=\small,
  ymin=4000,
  ymax=10000000,
	]
	\addlegendimage{/pgfplots/refstyle=gen,darkerBlue}\addlegendentry{generation time}
	\addlegendimage{/pgfplots/refstyle=vt,lightBlue}\addlegendentry{top co-boundary time}
	\addplot[solid,red,very thick] table [x expr={\thisrow{kv}}, y expr={\thisrow{nodes}}, col sep=comma, meta=kv] {charts/chart_neptune_tri.csv}; \addlegendentry{number of blocks}
	\end{axis}
	
\end{tikzpicture}
		}
		\label{chart:neptune}
	}
	\hfil
	\subfloat[\datasetName{bonsai} tetrahedral complex]{
		\resizebox{.32\textwidth}{!}{
			\begin{tikzpicture}
	\pgfplotstableread[col sep=comma]{charts/chart_bonsai_tet.csv}\tableND
	\begin{axis}[
	separate axis lines,
	y axis line style={darkerBlue},
	y label style={darkerBlue},
	y tick style={darkerBlue},
	ymode=log,
	log ticks with fixed point,
	axis y line*=left,
	ylabel = time (seconds),
	xlabel = block threshold ($k_v$),
	font=\small,
  ymin=.1,
  ymax=110,  
	]
	\draw [kvs,thick] ({axis cs:400,0}|-{rel axis cs:0,1}) -- ({axis cs:400,0}|-{rel axis cs:0,0});
	\draw [kvl,thick] ({axis cs:800,0}|-{rel axis cs:0,1}) -- ({axis cs:800,0}|-{rel axis cs:0,0});
	\addplot[kvs,thick] coordinates{(400,2.5)} node {\ks};
	\addplot[kvl,thick] coordinates{(800,2.5)} node {\kl};
	\addplot[solid,darkerBlue,very thick] table [x expr={\thisrow{kv}}, y expr={\thisrow{gen}}, col sep=comma, meta=kv] {charts/chart_bonsai_tet.csv}; 
	\addplot[solid,lightBlue,very thick] table [x expr={\thisrow{kv}}, y expr={\thisrow{vt}}, col sep=comma, meta=kv] {charts/chart_bonsai_tet.csv}; 
	\end{axis}
	
	\begin{axis}[
	ymode=log,
	separate axis lines,
	y axis line style={red},
	y label style={red},
	y tick style={red},
	axis y line*=right,
	axis x line=none,
	ylabel = number of blocks,
	font=\small,
  ymin=   4000,
  ymax=10000000,   
	]
	\addplot[solid,red,very thick] table [x expr={\thisrow{kv}}, y expr={\thisrow{nodes}}, col sep=comma, meta=kv] {charts/chart_bonsai_tet.csv}; 
	\end{axis}
\end{tikzpicture}
		}
		\label{chart:bonsai}
	}
	\hfil
	\subfloat[\datasetName{vismale 7D} V-Rips complex]{
		\resizebox{.32\textwidth}{!}{
			\begin{tikzpicture}
	\pgfplotstableread[col sep=comma]{charts/chart_vismale_vr.csv}\tableND
	\begin{axis}[
	separate axis lines,
	y axis line style={darkerBlue},
	y label style={darkerBlue},
	y tick style={darkerBlue},
	ymode=log,
	log ticks with fixed point,
	axis y line*=left,
	ylabel = time (seconds),
	xlabel = block threshold ($k_v$),
	font=\small,
  ymin=.1,
  ymax=110,
	]
	\draw [kvs,thick] ({axis cs:400,0}|-{rel axis cs:0,1}) -- ({axis cs:400,0}|-{rel axis cs:0,0});
	\draw [kvl,thick] ({axis cs:800,0}|-{rel axis cs:0,1}) -- ({axis cs:800,0}|-{rel axis cs:0,0});
	\addplot[kvs,thick] coordinates{(400,1.9)} node {\ks};
	\addplot[kvl,thick] coordinates{(800,1.9)} node {\kl};
	\addplot[solid,darkerBlue,very thick] table [x expr={\thisrow{kv}}, y expr={\thisrow{gen}}, col sep=comma, meta=kv] {charts/chart_vismale_vr.csv}; \addlegendentry[color=darkerBlue]{generation time} \label{gen}
	\addplot[solid,lightBlue,very thick] table [x expr={\thisrow{kv}}, y expr={\thisrow{vt}}, col sep=comma, meta=kv] {charts/chart_vismale_vr.csv}; \addlegendentry[color=lightBlue]{top co-boundary time} \label{vt}
	\end{axis}
	
	\begin{axis}[
	ymode=log,
	separate axis lines,
	y axis line style={red},
	y label style={red},
	y tick style={red},
	axis y line*=right,
	axis x line=none,
	ylabel = number of blocks,
	font=\small,
  ymin=   4000,
  ymax=10000000, 
	]
	\addlegendimage{/pgfplots/refstyle=gen,darkerBlue}\addlegendentry{generation time}
	\addlegendimage{/pgfplots/refstyle=vt,lightBlue}\addlegendentry{top co-boundary time}
	\addplot[solid,red,very thick] table [x expr={\thisrow{kv}}, y expr={\thisrow{nodes}}, col sep=comma, meta=kv] {charts/chart_vismale_vr.csv}; \addlegendentry{number of blocks}
	\end{axis}
\end{tikzpicture}
		}
		\label{chart:vismale_vr}
	} 
	\caption{Bucketing threshold calibration experiments comparing the number of Stellar tree blocks (red, right y-axis) 
          and generation and top-coboundary extraction times (blue, left y-axis) against bucket threshold values (\kv). 
          The vertical bars (gray) represent the \kv\ values selected for our experiments.} 
	\label{chart:bucketing_calibration}
\end{figure*}
\begin{figure*}[t]
	\centering
	\subfloat[\datasetName{neptune} triangle complex]{
		\resizebox{.3\textwidth}{!}{
			\begin{tikzpicture}
	\pgfplotstableread[col sep=comma]{charts/chart_neptune_tri.csv}\tableND
	
	\begin{axis}[
	%
	xlabel = block threshold ($k_v$),
	font=\small,
  ymin=0.0078125,
  ymax=9,
  ymode=log,
  log basis y={2},
  ytick={8,4,2,1,.5,.25,.125,.0625,0.03125,0.015625,0.0078125},
	]	
	\addplot[solid,brewer_div_gold,very thick] table [x expr={\thisrow{kv}}, y expr={\thisrow{chi}}, col sep=comma, meta=kv] {charts/chart_neptune_tri.csv}; \addlegendentry{\Chi}
	\addplot[solid,brewer_div_green,very thick] table [x expr={\thisrow{kv}}, y expr={\thisrow{mu}}, col sep=comma, meta=kv] {charts/chart_neptune_tri.csv}; \addlegendentry{\Mu} 
	\draw [kvs,thick] ({axis cs:100,0}|-{rel axis cs:0,1}) -- ({axis cs:100,0}|-{rel axis cs:0,0});
	\draw [kvl,thick] ({axis cs:500,0}|-{rel axis cs:0,1}) -- ({axis cs:500,0}|-{rel axis cs:0,0});
	\addplot[kvs,thick] coordinates{(100,0.5)} node {\ks};
	\addplot[kvl,thick] coordinates{(500,0.5)} node {\kl};
	\end{axis}
	

\end{tikzpicture}
		}
		\label{chart:neptune_chi}
	}
	\hfil
	\subfloat[\datasetName{bonsai} tetrahedral complex]{
		\resizebox{.3\textwidth}{!}{
			\begin{tikzpicture}
	\pgfplotstableread[col sep=comma]{charts/chart_bonsai_tet.csv}\tableND
	
	\begin{axis}[
	%
	xlabel = block threshold ($k_v$),
	font=\small,
  ymin=0.0078125,
  ymax=9,
  ymode=log,
  log basis y={2},
  ytick={8,4,2,1,.5,.25,.125,.0625,0.03125,0.015625,0.0078125},
	]	
	\addplot[solid,brewer_div_gold,very thick] table [x expr={\thisrow{kv}}, y expr={\thisrow{chi}}, col sep=comma, meta=kv] {charts/chart_bonsai_tet.csv}; \addlegendentry{\Chi} 
	\addplot[solid,brewer_div_green,very thick] table [x expr={\thisrow{kv}}, y expr={\thisrow{mu}}, col sep=comma, meta=kv] {charts/chart_bonsai_tet.csv}; \addlegendentry{\Mu} 
	\draw [kvs,thick] ({axis cs:400,0}|-{rel axis cs:0,1}) -- ({axis cs:400,0}|-{rel axis cs:0,0});
	\draw [kvl,thick] ({axis cs:800,0}|-{rel axis cs:0,1}) -- ({axis cs:800,0}|-{rel axis cs:0,0});
	\addplot[kvs,thick] coordinates{(400,0.5)} node {\ks};
	\addplot[kvl,thick] coordinates{(800,0.5)} node {\kl};
	\end{axis}
	
\end{tikzpicture}
		}
		\label{chart:bonsai_chi}
	}
	\hfil
	\subfloat[\datasetName{vismale 7D} V-Rips complex]{
		\resizebox{.3\textwidth}{!}{
			\begin{tikzpicture}
	\pgfplotstableread[col sep=comma]{charts/chart_vismale_vr.csv}\tableND
	
	\begin{axis}[
	%
	xlabel = block threshold ($k_v$),
	font=\small,
  ymin=0.0078125,
  ymax=9,
  ymode=log,
  log basis y={2},
  ytick={8,4,2,1,.5,.25,.125,.0625,0.03125,0.015625,0.0078125},
	]
	\addplot[solid,brewer_div_gold,very thick] table [x expr={\thisrow{kv}}, y expr={\thisrow{chi6}}, col sep=comma, meta=kv] {charts/chart_vismale_vr.csv}; \addlegendentry{$\Chi$}
	\addplot[solid,brewer_div_green,very thick] table [x expr={\thisrow{kv}}, y expr={\thisrow{mu6}}, col sep=comma, meta=kv] {charts/chart_vismale_vr.csv}; \addlegendentry{$\Mu$}
	\draw [kvs,thick] ({axis cs:400,0}|-{rel axis cs:0,1}) -- ({axis cs:400,0}|-{rel axis cs:0,0});
	\draw [kvl,thick] ({axis cs:800,0}|-{rel axis cs:0,1}) -- ({axis cs:800,0}|-{rel axis cs:0,0});
	\addplot[kvs,thick] coordinates{(400,0.5)} node {\ks};
	\addplot[kvl,thick] coordinates{(800,0.5)} node {\kl};
	\end{axis}
	
\end{tikzpicture}
		}
		\label{chart:vismale_vr_chi}
	}
	\caption{Bucketing threshold calibration experiments comparing the evolution of the average spanning number \Chi\ 
        and of the average reference number \Mu\ against bucket threshold values (\kv) for three datasets. 
        The vertical bars (gray) represent the \kv\ values we selected for our experiments on these datasets.} 
	\label{chart:bucketing_calibration_chi}
\end{figure*}

\subsection{Experimental datasets}
\label{sec:stellar_experimental_plan}

We have performed experiments on a range of CP complexes consisting of
triangle, quadrilateral, tetrahedral and hexahedral meshes in $\eucl^3$
as well as pure non-manifold simplicial complexes in higher dimensions 
and higher dimensional non-manifold simplicial complexes (embedded in $\eucl^3$).
Table~\ref{tab:datasets} summarizes the datasets used in our experiments and their numbers of vertices and top cells.

Our triangle and tetrahedral meshes are \emph{native} models ranging from 4 to 28 million triangles and from 24 to 29 million tetrahedra,
where we use the term native to refer to models from public domain repositories discretizing objects in space.
Since we only had access to relatively small native quadrilateral and hexahedral meshes (with tens to hundreds of thousand elements), 
we have generated some larger models ranging from 12 to 125 million elements from our triangle and tetrahedral models. 
The generation procedure refines each triangle into three quadrilaterals and each tetrahedron into four hexahedra 
by adding vertices at the face centroids.

To experiment with \emph{pure} non-manifold models in higher dimensions, we have generated some models based on a process that we call 
\emph{probabilistic Sierpinski filtering}, where we regularly refine all simplices in the complex 
and randomly remove a fixed proportion of the generated simplices in each iteration. 
For our experiments, we have created 5-, 7- and 40-dimensional models using different levels of refinement
and a filtering threshold of 65\%, yielding pure simplicial complexes with 16.5 million to 258 million \tops.

Finally, to experiment with general simplicial complexes in higher dimensions, we have generated several 
(non-pure)  \emph{Vietoris-Rips} complexes, which we embed in a lower dimensional space. 
A Vietoris-Rips (V-Rips) complex is the \emph{flag} complex defined by a neighborhood graph over a point cloud
whose arcs connect pairs of points with distance less than a user-provided parameter $\epsilon$.
Given the neighborhood graph, the simplices of the V-Rips complexes are defined by its \emph{cliques}, 
subsets of the graph vertices that form a complete subgraph.
We refer to \cite{Zomorodi2010Fast} for further details.
For our experiments, we have generated V-Rips complexes 
over the vertices of a triangle model (\datasetName{lucy}) 
and of two tetrahedral models (\datasetName{vismale} and \datasetName{foot}) from our manifold datasets
and set our distance threshold $\epsilon$ to $\{0.1\%,0.5\%,0.4\%\}$ of the bounding box diagonal, respectively. 
%
The range of top simplices in the generated complexes goes from 6.4 million to 64 million and their dimension from 7 to 34.
%
Although the generated complexes are synthetic, they provide a good starting point to demonstrate the efficiency of the Stellar tree in higher dimensions.

\NOTA{\felleComment{this is said later now}}

\NOTA{\felleComment{saying when we use a different spatial decomposition does not matter a lot at this point (in case add it later). Also, a reviewer criticized this, since we do not have an heuristic (or experimental evaluation) justifying this choice}}

%
All tests have been performed on a PC equipped with a 3.2 gigahertz Intel i7-3930K CPU with 64 gigabytes of RAM.
The source code will be made available at~\cite{Fellegara_StellarTree_Github}.

\subsection{Calibrating Stellar tree bucket thresholds}
\label{sec:calibration_results}

Spatial indexes typically involve a careful balance among index generation times, storage costs and query performances.
Stellar trees provide users with a single tuning parameter \kv\ to control the maximum number of vertices indexed by each block of the tree.
%
%
In the following, we calibrate \kv\ on a characteristic subset of three of our experimental datasets:
  \datasetName{neptune} triangle mesh, 
  \datasetName{bonsai} tetrahedral mesh, and 
  \datasetName{vismale} Vietoris-Rips complex. 
For each dataset, we generated 195 Stellar trees using \kv\ values ranging from 1 to 1500 and compared Stellar tree generation and query times as well as the number of blocks as a proxy for the complexity of the generated tree.
Within this range, we increment \kv\ by 1 for values between 1 and 50, and by 10 for values between 60 and 1500. 
This allows us to evaluate the decomposition quality and the extraction performance for a fundamental topological query at different scales. 
For the latter, we use the vertex co-boundary extraction, i.e., the top cells incident in each vertex (which we describe in Section~\ref{sec:coboundary_extraction}).

\NOTA{
  \felleComment{the vertical bars in the charts represent the $k_v$ values used in the experiments.
  I added this since I think it can help justify why we have chosen such values 
  (i.e., we excluded $k_v$ values generating either too deep indices or indices in which the $k_v$ value is not influencing significantly the spatial decomposition).
  The y-axis are two: the one on the left is for \emph{generation} and \emph{vt} functions, while the one on the right is for the node number.}
  \leilaComment{I think that we need to have a different graph for the number of nodes. I do not understand these charts}
  \kennyComment{I think we should plot the y-axes on a log scale.
      I looked back at your original email and it is much easier to see what is going on with a log scale.
      In particular, the coboundary times does not appear to depend too strongly on \kv.
      }  
  \kennyComment{These results suggest to me that there are three separate performance regimes and that our k\_small is too large.  
    It should be closer to the first cusp.  I'd imagine that this would be around the avg./max. cardinality of the vertex coboundary relation.
    I am not suggesting that we should rerun the experiments (due to the time it would take, and having to reanalyze everything), but we might get questions about it.}    
}
\NOTA{
  \kennyComment{I updated the axes on the calibration figures to be consistent.  
    The legend was covering the lines for the Bonsai figure, so I commented out the legend for that subplot}
}
The results are summarized in the charts of Figures~\ref{chart:bucketing_calibration} and~\ref{chart:bucketing_calibration_chi}, which compare the complexity of the generated Stellar tree (in terms of number of blocks), its generation and query times, and the average spanning ($\Chi$) and reference ($\Mu$) numbers as a function of the threshold value \kv.

To better compare different units (i.e., number of blocks and timings), each chart in Figure~\ref{chart:bucketing_calibration}
has two logarithmic y-axes, showing the time scale (blue curves using the left y-axes) and the number of blocks (red curves using the right y-axes), respectively. In this way, we can directly compare, for each dataset, how the \kv\ value influences the decomposition and the timing performances.
After an initial rapid decrease in the generation time and block number, the curves begin to level off for increasingly large \kv\ values. While there are more than a million blocks when \kv\ is less than 10, the number of blocks rapidly decreases to hundreds of thousands for \kv's between 50 and 200, and grows even smaller for large \kv\ values (e.g., above 500), where the number of blocks remains steadily in the thousands to tens of thousands. 
%
This trend appears to be related to the point distribution within each dataset, which induces finer decompositions for \kv\ values between 1 and 50, and coarser decompositions for larger \kv\ values. 
%
This trend can also be observed for the generation times, which reduce by a factor of two for \kv\ values between 1 and 100, 
and by another factor of two for large bucketing thresholds. 
%
While the topological extraction query is largely unaffected by \kv\ size, it gets slightly faster for larger \kv\ values.
\NOTA{
  \felleComment{I do not think that the max VT cardinality influences here.. since the decomposition is vertex based. 
    Here the point density and distribution play a key role.. I think the "jumps" that we see in Figures 12(b) and 12(c) are related to "grouping" points that are spatially close}
  \kennyComment{ Interesting -- maybe it is related to the ratio between bounding box and average edge size ?}
  \kennyComment{ Note: Both BONSAI and VISMALE were derived from regular grids of (roughly) resolution $256^3$ }
}
When comparing the influence of \kv\ on \Chi\ and \Mu\ (shown in Figure \ref{chart:bucketing_calibration_chi}), we observe that the behavior of these two variables is very similar to that of the number of blocks.
This is expected, since the top cells distribution is directly linked to the number of blocks in the tree. As mentioned in Section~\ref{sec:stellar_dec_def}, the number of leaf blocks indexing a top cell is bounded from above by the number of its vertices,
and this defines a topological upper bound that reduces
the overall storage requirements. We note that the SRE compression is able to reduce the number of references per top cell (\Mu), even for very small \kv\ values.
\NOTA
{
  Similarly to the charts in Figure~\ref{chart:bucketing_calibration}, after an initial rapid decrease, both \Chi\ and \Mu\ values remains overall constant, around 1.5 and 0.1, respectively.
}

\begin{figure}[t]
	\centering
		\resizebox{\columnwidth}{!}{
			\includegraphics{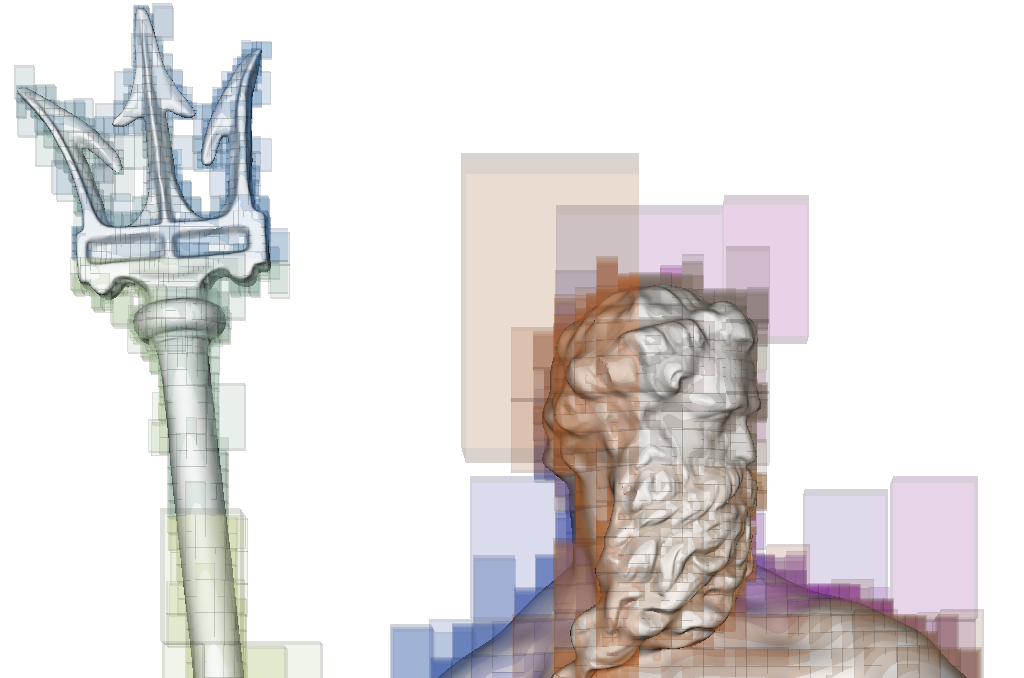}
		}
	\caption{Leaf blocks for a Stellar tree decompositions over \datasetName{neptune} triangle mesh. 
    Each leaf block indexes up to $\kv=100$ mesh vertices.}
	\label{fig:neptune_octree}
\end{figure}

Our calibration experiments indicate that, while there are slight differences in timing and storage costs, Stellar tree performance is relatively stable over a wide range of \kv\ values. However, threshold values that are either too small or too large should be avoided, since in the first case the storage requirements and time performances are heavily affected, while in the latter case the benefit of having a hierarchical decomposition is limited, as both storage requirements and time performance are not clearly influenced by it.
In the rest of this paper, for every model, we build two Stellar trees to compare how their performances depend on parameter \kv. 
These two \kv\ values are chosen in order to: (i) have a hierarchical decomposition that still plays a critical role in the storage requirements and time performances; and (ii) obtain trees with different characteristics: one deeper and another relatively shallower. 
In the following, we use \ks\ to refer to the smaller \kv\ value and \kl\ to the larger one.
Since there is a direct correlation between the decomposition quality and \Chi,
these calibration choices are also reflected in the \Chi\ values across our experimental datasets.
Table~\ref{tab:stellar_trees_summary} summarizes statistics on the Stellar trees obtained from each input data set by considering two values of the vertex threshold \kv, namely \ks\ and  \kl. Figure~\ref{fig:neptune_octree} illustrates the \ks\ octree decomposition for the 4M triangle \datasetName{Neptune} dataset.

	\begin{table}[t]
		\centering
		\caption{Overview of our generated Stellar trees for each dataset.
				 For each Stellar tree, we list the thresholds \kv, the number of blocks in the index (total $|\h|$ and leaf $|\hL|$) 
						and the average spanning number $\Chi$.}
		{
		\resizebox{0.7\columnwidth}{!}{
			\begin{tabular}{cccrrrr}
				\toprule
				\textbf{Data} & &  &  \textbf{\kv} & \textbf{$|\h|$} & \textbf{$|\hL|$} &
				\textbf{\Chi} \\
 				\midrule

\multirow{2}{*}{\datasetName{neptune}} & \multirow{7}{*}{\begin{sideways}\datasetName{tri.}\end{sideways}} 
 & \ks & 100 & 73.7K & 58.8K & 1.37 \\
 &  &  \kl & 500 & 15.0K & 12.2K & 1.17 \\
\cmidrule(lr){1-1} \cmidrule(lr){3-7}
 
\multirow{2}{*}{\datasetName{statuette}} &  
 & \ks & 100 & 182K & 147K & 1.36 \\ 
 &  &  \kl & 500 & 39.8K & 32.7K & 1.17 \\
\cmidrule(lr){1-1} \cmidrule(lr){3-7}
 
\multirow{2}{*}{\datasetName{lucy}} &  
 & \ks & 100 & 464K & 374K & 1.35 \\ 
 &  &  \kl & 500 & 88.8K & 70.3K & 1.16 \\
\midrule
 
\multirow{2}{*}{\datasetName{neptune}} & \multirow{7}{*}{\begin{sideways}\datasetName{quad.}\end{sideways}}
 & \ks & 100 & 407K & 322K & 1.47 \\
 &  &  \kl & 800 & 55.0K & 44.3K & 1.17 \\ 
\cmidrule(lr){1-1} \cmidrule(lr){3-7}
 \multirow{2}{*}{\datasetName{statuette}} &  
 & \ks & 100 & 1.10M & 883K & 1.47 \\
 &  &  \kl & 800 & 146K & 120K & 1.17 \\
\cmidrule(lr){1-1} \cmidrule(lr){3-7}
\multirow{2}{*}{\datasetName{lucy}} &  
 & \ks & 100 & 3.53M & 2.85M & 1.54 \\ 
 &  &  \kl & 800 & 329K & 265K & 1.17 \\ 
\midrule

\multirow{2}{*}{\datasetName{bonsai}} & \multirow{7}{*}{\begin{sideways}\datasetName{tetra.}\end{sideways}} 
 & \ks & 400 & 45.2K & 39.5K & 1.58 \\ 
 &  &  \kl & 800 & 17.9K & 15.7K & 1.44 \\ 
 \cmidrule(lr){1-1} \cmidrule(lr){3-7}
\multirow{2}{*}{\datasetName{vismale}} &  
 & \ks & 400 & 32.8K & 28.7K & 1.52 \\ 
 &  &  \kl & 800 & 17.7K & 15.5K & 1.45 \\ 
 \cmidrule(lr){1-1} \cmidrule(lr){3-7}
\multirow{2}{*}{\datasetName{foot}} &  
 & \ks & 400 & 88.8K & 77.7K & 1.75 \\
 &  &  \kl & 800 & 17.1K & 15.0K & 1.43 \\
 \midrule
 
\multirow{2}{*}{\datasetName{f16}} & \multirow{7}{*}{\begin{sideways}\datasetName{hexa.}\end{sideways}}   
 & \ks & 100 & 1.11M & 972K & 3.08 \\ 
 &  &  \kl & 1000 & 113K & 99.0K & 1.90 \\
\cmidrule(lr){1-1} \cmidrule(lr){3-7} 
\multirow{2}{*}{\datasetName{san fern}} &  
 & \ks & 100 & 2.02M & 1.77M & 3.15 \\
 &  &  \kl & 1000 & 247K & 216K & 1.88 \\
\cmidrule(lr){1-1} \cmidrule(lr){3-7}
 \multirow{2}{*}{\datasetName{vismale}} &  
 & \ks & 100 & 7.39M & 6.46M & 2.80 \\ 
 &  &  \kl & 1000 & 800K & 700K & 1.72 \\ 
\midrule
\midrule
 
\multirow{2}{*}{\datasetName{5D}} & \multirow{7}{*}{\begin{sideways}\datasetName{prob.}\end{sideways}}
 & \ks & 100 & 37.4K & 36.1K & 4.39 \\
 &  & \kl & 500 & 2.79K & 2.68K & 2.55 \\
\cmidrule(lr){1-1} \cmidrule(lr){3-7} 
\multirow{2}{*}{\datasetName{7D}} &  
 & \ks & 100 & 10.8K & 4.87K & 4.98 \\ 
 &  &  \kl & 500 & 2.02K & 1.00K & 3.78 \\ 
\cmidrule(lr){1-1} \cmidrule(lr){3-7}
\multirow{2}{*}{\datasetName{40D}} &  
 & \ks & 100 & 15.2K & 4.32K & 36.2 \\
 &  & \kl & 1000 & 1.56K & 550 & 34.0 \\
\midrule

\multirow{2}{*}{\datasetName{vismale 7D}} & \multirow{7}{*}{\begin{sideways}\datasetName{v-rips}\end{sideways}} 
 & \ks & 400 & 32.8K & 28.7K & 1.44 \\
 &  &  \kl & 800 & 17.7K & 15.5K & 1.37 \\
 \cmidrule(lr){1-1} \cmidrule(lr){3-7}
\multirow{2}{*}{\datasetName{foot 10D}} &  
 & \ks & 400 & 88.8K & 77.7K & 2.02 \\
 &  & \kl & 800 & 17.1K & 15.0K & 1.56 \\
 \cmidrule(lr){1-1} \cmidrule(lr){3-7}
\multirow{2}{*}{\datasetName{lucy 34D}} &  
 & \ks & 100 & 464K & 374K & 2.47 \\
 &  & \kl & 500 & 88.8K & 70.3K & 1.73 \\
\bottomrule

			\end{tabular}
		}
		}
		\label{tab:stellar_trees_summary}
	\end{table}

\section{Evaluation of storage costs and generation times}
\label{sec:storage}

\NOTA{ 
Our implementation is based on the observation that the \iastar\ representation encodes its topological connectivity information in a \emph{stratified} manner.
Specifically, the \iastar\ representation can be decomposed into a union of data structures on the pure \tDim-dimensional subcomplexes of \sC\ defined by its \ktopcells.
For each \ktopcell\ \simplex\ in \sC, all topological information pertaining to \simplex\ is contained in the following relations, all restricted to the \ktopcells\ of \sC:
	boundary relation \relation{\tDim,0}(\simplex), 
	adjacency relation $\relation{\tDim,\tDim}(\simplex)$,
	co-boundary relation $\relation{\tDimMinusOne,\tDim}(\tau)$ 
	and restricted vertex co-boundary relation $\relation{0,\tDim}(\vertex)$.
}

In this section, we evaluate the storage costs and generation times of Stellar trees.
%
First, we compare the cost of different Stellar tree encodings (Section~\ref{sec:storage_encodings}), then
we compare the Stellar tree against several state-of-the-art topological mesh data structures (Section~\ref{sec:storage_other_structures}), and, finally, we evaluate the generation times of the Stellar tree (Section~\ref{sec:generation_timings}).

\NOTA{
\kennyComment{ Riccardo, it's been a while since I last looked at this... are we using Octrees or Kd-trees for our experiments? 
    I think it is octrees, but we should be explicit since earlier in the paper, e.g. in the encoding (Section~\ref{sec:stellar_tree_enc}), 
    we mention that Stellar trees can be either kd-trees or octrees. }
\kennyComment{ Also, do we want to comment on the difference b/w octree and kd-tree, if any? }
\kennyComment{ ...and do we use kd-trees in any of our applications in Section~\ref{sec:stellar_ecosystem}?}
}

\subsection{Storage comparison among Stellar tree encodings}
\label{sec:storage_encodings}

We begin by comparing the \datasetName{explicit} and \datasetName{compressed} Stellar tree encodings
as well as a \datasetName{vertex-compressed} encoding, similar to the PR-star encoding for tetrahedral meshes~\cite{Weiss2011PR},
that compresses the vertex array but not the top cells arrays.
Table~\ref{tab:storage_encodings} lists the storage costs for the indexed representation of the complex (`Base Complex')
as well as the additional costs required for the three Stellar tree encodings, in terms of megabytes ($MBs$). 
In the following, 
we assume that pointers require 64 bits and indices 32 bits, 
the de-facto standard in modern computing hardware.
Stellar trees based on the \datasetName{compressed} encoding are always the most compact.

\begin{table}[t]
			\centering
			\caption{
			Storage costs (in $MB$s) and average spanning (\Chi) and reference (\Mu) numbers
			for different Stellar tree encodings.
			}
			{
			\resizebox{\columnwidth}{!}{
				\begin{tabular}{cccc|cc|cc|cc}
					\toprule
					\multicolumn{1}{c}{\multirow{4}{*}{\textbf{Data}}}  &
					\multicolumn{1}{c}{\multirow{4}{*}{\textbf{}}}  &
					\multicolumn{1}{c}{\multirow{4}{*}{}} &
					\multicolumn{1}{c}{\multirow{3}{*}{\textbf{Base}}}  &					
					\multicolumn{6}{c}{\textbf{Stellar tree}} \\
					\cmidrule(lr){5-10}
					 & & & \multicolumn{1}{c}{\multirow{2}{*}{\textbf{Complex}}} & \multicolumn{2}{c}{\textbf{\datasetName{explicit}}} & \multicolumn{2}{c}{\textbf{\datasetName{v\_compr.}}}  & \multicolumn{2}{c}{\textbf{\datasetName{compr.}}} \\
					\cmidrule(lr){5-6} \cmidrule(lr){7-8} \cmidrule(lr){9-10}
					 & & & & 
					 \textbf{cost} & \textbf{\Chi} & \textbf{cost} & \textbf{\Chi} & 
					 \textbf{cost} & \textbf{\Mu} \\
					\midrule

\multirow{2}{*}{\datasetName{neptune}} & \multirow{7}{*}{\begin{sideways}\datasetName{tri.}\end{sideways}} &  \ks & \multirow{2}{*}{45.9} & 32.0 & 1.37 & 24.3 & 1.37 & 5.76 & 0.16 \\
&  &  \kl & & 26.2 & 1.17 &	18.6 & 1.17 & 1.24 & 0.04 \\
\cmidrule(lr){1-1} \cmidrule(lr){3-10}

\multirow{2}{*}{\datasetName{statuette}} & &  \ks & \multirow{2}{*}{114} & 79.2 & 1.36 & 60.2 & 1.36 & 14.6 & 0.17 \\
&  &  \kl & & 65.6 & 1.17 & 46.6 & 1.17 & 3.41 & 0.04 \\
\cmidrule(lr){1-1} \cmidrule(lr){3-10}

\multirow{2}{*}{\datasetName{lucy}} & &  \ks & \multirow{2}{*}{321} & 220 & 1.35 & 166 & 1.35 & 34.5 & 0.12 \\
&  &  \kl & & 181 & 1.16 & 128 & 1.16 & 6.18 & 0.02 \\
\midrule

\multirow{2}{*}{\datasetName{neptune}} & \multirow{7}{*}{\begin{sideways}\datasetName{quad.}\end{sideways}} &  \ks & \multirow{2}{*}{183} & 132 & 1.47 & 86.0 & 1.47 & 28.0 & 0.20 \\
&  &    \kl & & 102 & 1.17 & 56.3 & 1.17 & 3.86 & 0.03 \\
\cmidrule(lr){1-1} \cmidrule(lr){3-10}

\multirow{2}{*}{\datasetName{statuette}} & &  \ks & \multirow{2}{*}{458} & 333 & 1.47 & 219 & 1.47 & 76.0 & 0.22 \\
&  &  \kl & &  255 & 1.17 & 141 & 1.17 & 10.4 & 0.03 \\
\cmidrule(lr){1-1} \cmidrule(lr){3-10}

\multirow{2}{*}{\datasetName{lucy}} & &  \ks & \multirow{2}{*}{1.3K} & 976 & 1.54 & 656 & 1.54 & 245 & 0.26 \\
&  &    \kl & & 710 & 1.17 & 389 & 1.17 & 23.1 & 0.03 \\
\midrule
\multirow{2}{*}{\datasetName{bonsai}} & \multirow{7}{*}{\begin{sideways}\datasetName{tetra.}\end{sideways}} &  \ks & \multirow{2}{*}{373} & 166 & 1.58 & 150 & 1.58 & 6.55 & 0.05 \\
&  &    \kl & & 151 & 1.44 & 135 & 1.44 & 2.65 & 0.02 \\
\cmidrule(lr){1-1} \cmidrule(lr){3-10}

\multirow{2}{*}{\datasetName{vismale}} & &  \ks & \multirow{2}{*}{405} & 173 & 1.52 & 156 & 1.52 & 4.87 & 0.03 \\
&  &    \kl & & 165 & 1.45 & 147 & 1.45 & 2.69 & 0.02 \\
\cmidrule(lr){1-1} \cmidrule(lr){3-10}

\multirow{2}{*}{\datasetName{foot}} & &  \ks & \multirow{2}{*}{450} & 220 & 1.75 & 201 & 1.75 & 13.0 & 0.08 \\
&  &    \kl & & 181 & 1.43 & 161 & 1.43 & 2.60 & 0.02 \\
\midrule
\multirow{2}{*}{\datasetName{f16}} & \multirow{7}{*}{\begin{sideways}\datasetName{hexa.}\end{sideways}} &  \ks & \multirow{2}{*}{775} & 456 & 3.08 & 349 & 3.08 & 151 & 1.03 \\
&  &    \kl & & 296 & 1.90 & 189 & 1.90 & 18.0 & 0.13 \\
\cmidrule(lr){1-1} \cmidrule(lr){3-10}

\multirow{2}{*}{\datasetName{san fern}} & &  \ks & \multirow{2}{*}{1.7K} & 999 & 3.15 & 765 & 3.15 & 275 & 0.86 \\
&  &    \kl & & 646 & 1.88 & 412 & 1.88 & 33.1 & 0.10 \\
\cmidrule(lr){1-1} \cmidrule(lr){3-10}
	
\multirow{2}{*}{\datasetName{vismale}} & &  \ks & \multirow{2}{*}{3.8K} & 2.2K & 2.89 & 1.7K & 2.89 & 887 & 1.15 \\
&  &    \kl & & 1.4K & 1.72 & 858 & 1.72 & 106 & 0.15 \\
\midrule
\midrule
 
\multirow{2}{*}{\datasetName{5D}} & \multirow{7}{*}{\begin{sideways}\datasetName{prob.}\end{sideways}} &  \ks & \multirow{2}{*}{607} & 448 & 4.39 & 446 & 4.39 & 63.7 & 0.61 \\
  &  &  \kl & & 259 & 2.55 & 258 & 2.55 & 3.57 & 0.03 \\
\cmidrule(lr){1-1} \cmidrule(lr){3-10}
 
\multirow{2}{*}{\datasetName{7D}} &  & \ks & \multirow{2}{*}{7.9K} & 4.9K & 4.98 & 4.9K & 4.98 & 101 & 0.10 \\
 &  &  \kl & & 3.7K & 3.78 & 3.7K & 3.78 & 12.2 & 0.01 \\
\cmidrule(lr){1-1} \cmidrule(lr){3-10}
 
\multirow{2}{*}{\datasetName{40D}} &  & \ks & \multirow{2}{*}{2.6K} & 2.3K & 36.2 & 2.3K & 36.2 & 55.7 & 0.87 \\
  &  &  \kl & & 2.1K & 34.0 & 2.1K & 34.0 & 0.45 & 0.01 \\
\midrule
\multirow{2}{*}{\datasetName{vismale 7D}} & \multirow{7}{*}{\begin{sideways}\datasetName{v-rips}\end{sideways}} &  \ks & \multirow{2}{*}{134} & 56.2 & 1.44 & 37.0 & 1.44  & 7.38 & 0.26 \\
&  &    \kl & & 53.7 & 1.37 & 34.6 & 1.37 & 4.54 & 0.18 \\
\cmidrule(lr){1-1} \cmidrule(lr){3-10}

\multirow{2}{*}{\datasetName{foot 10D}} & &  \ks & \multirow{2}{*}{2.1K} & 604 & 2.02 & 586 & 2.02 & 65.1 & 0.33 \\
&  &    \kl & & 431 & 1.56 & 413 & 1.56 & 11.5 & 0.12 \\
\cmidrule(lr){1-1} \cmidrule(lr){3-10}
	
\multirow{2}{*}{\datasetName{lucy 34D}} & &  \ks & \multirow{2}{*}{2.0K} & 416 & 2.47 & 363 & 2.47 & 86.2 & 0.92 \\
&  &    \kl & & 292 & 1.73 & 238 & 1.73 & 19.0 & 0.53 \\
%
%
%
\bottomrule

				\end{tabular}
			}
			}
			\label{tab:storage_encodings}
\end{table}


%
We first consider the storage requirements of the hierarchical structures with respect to our tuning parameter \kv\ 
and observe that higher values of \kv\ always yield reductions in memory requirements. As expected,
this effect is more pronounced for the \datasetName{compressed} encoding than for the other two encodings.
Specifically, the \datasetName{explicit} and \datasetName{vertex-compressed} \kl\ trees achieve a 20-50\% reduction in storage requirements
compared to their \ks\ counterparts, while the \datasetName{compressed} \kl\ trees are a factor of 3-10 smaller than their \ks\ counterparts.
For example, on the triangular \datasetName{neptune} dataset, storage requirements for the \datasetName{explicit} Stellar tree 
reduces from 32.0 MB (\ks) to 26.2 MB (\kl), while the \datasetName{compressed} Stellar trees
reduces by more than a factor of 4 from 5.76 MB (\ks) to 1.24 MB (\kl).

When comparing the three encodings, we see that compressing the vertices alone, as in the \datasetName{vertex-compressed} representation,
achieves only 10-20\% reduction in storage requirements compared to the \datasetName{explicit} representation, in most cases.  
In contrast, compressing the vertices and top cells, as in our \datasetName{compressed} representation,
yields an order of magnitude improvement, requiring a factor of 10-20 less storage than their \datasetName{explicit} counterparts.
This trend is nicely tracked for each dataset by the 
differences between its average references number \Mu\ and its average spanning number \Chi.
This is particularly evident on our \textit{probabilistic} datasets, for which it is difficult to calibrate \kv\ in order to reduce \Chi\ values. 
However, after SRE compression, \Mu\ values are always very small, leading to significant storage reductions in the \datasetName{compressed} representation.

Considering the hierarchical storage requirements against those of the original indexed base complex,
we observe that \datasetName{explicit} Stellar trees require about 50\% to 80\% the storage of the base complex,
while \datasetName{compressed} Stellar trees require only around 10\% (\ks) and 1\% (\kl) the storage of the base complex. 
Thus, for reasonable \kv\ values, \datasetName{compressed} Stellar trees impose only a negligible storage overhead 
with respect to the underlying indexed complex, which the Stellar tree representation does not modify.
In the remainder of this paper, we restrict our attention to the \datasetName{compressed} Stellar Tree, 
which we refer to as the Stellar tree, for simplicity.

\subsection{Storage comparison with respect to other data structures}
\label{sec:storage_other_structures}

We compare the Stellar tree with several dimension-independent topological data structures
as well as dimension-dependent topological data structures for 2D and 3D simplicial complexes.
Figures~\ref{hist:storage_nD_norm}, \ref{hist:storage_quad_hexa_norm} and \ref{hist:storage_tri_tetra_norm} %
compare the storage requirements for the different data structures
normalized against the storage costs of the indexed base complex.
%
%
The analysis compares the topological overhead of the data structures, 
and thus, we omit the cost of the geometry of the underlying complex, which is common to all the data structures. 
%

\begin{figure}[t]
\centering
\resizebox{\columnwidth}{!}{
   \begin{tikzpicture}
      \pgfplotstableread[col sep=comma]{charts/histogram_storage_nD.csv}\tableND
      \centering
      \begin{axis}[
			  reverse legend,
		width=7cm,
		height=8.5cm,
        enlarge y limits=0.1,
        legend cell align = left,
        legend style={
					at={(1.25,0.05)},
        	anchor=north east,
        },
        xmin=0, xmax=3,
        xbar=2pt,
        ytick=data,
        bar width=3.5pt, 
        font=\scriptsize,
        yticklabels from table= \tableND{overlap},
        y tick label style= {text width=1.2cm, align=right}, 
        nodes near coords,
        every node near coord/.append style={anchor=mid west, yshift=-0.12ex,
        	black!95, font=\scriptsize},
        xlabel = $|\Sigma|$,
        y=1.1cm,
        restrict x to domain*=0:3.25, 
        visualization depends on=rawx\as\rawx, 
        nodes near coords={%
        	\pgfmathprintnumber{\rawx}
        },
        after end axis/.code={ 
        	\draw [ultra thick, white, decoration={snake, amplitude=1pt}, decorate] (rel axis cs:1.05,0) -- (rel axis cs:1.05,1);
        },
        axis lines*=left,
        clip=false
      ]
      \addplot [ksStellarStyle]   table [y expr=-\coordindex, x expr={\thisrow{hit5}/\thisrow{hit7}}, meta=str5] {\tableND};\addlegendentry{\ks\ Stellar tree}
      \addplot [klStellarStyle]   table [y expr=-\coordindex, x expr={\thisrow{hit6}/\thisrow{hit7}}, meta=str6] {\tableND};\addlegendentry{\kl\ Stellar tree}
      \addplot [simplexTreeStyle] table [y expr=-\coordindex, x expr={\thisrow{hit4}/\thisrow{hit7}}, meta=str4] {\tableND};\addlegendentry{Simplex tree}
      \addplot [iaStarStyle]      table [y expr=-\coordindex, x expr={\thisrow{hit1}/\thisrow{hit7}}, meta=str1] {\tableND};\addlegendentry{\iastar}
    \end{axis}      
  \end{tikzpicture}
}
\caption{Storage costs for high dimensional \emph{probabilistic-refinement} simplicial complexes (\datasetName{prob.5D}, \datasetName{prob.7D} and \datasetName{prob.40D})
         and \emph{V-Rips} simplicial complexes (\datasetName{vismale7D}, \datasetName{foot10D} and \datasetName{lucy34D}). 
				 Costs (labels to right of each bar) are normalized to the indexed mesh representation (listed along y-axis). 
				 Note that: (1) the x-axis is truncated to a factor of 3; 
                    (2) datasets marked with $\odot$ or $\otimes$ could not be directly generated on our test machine 
                        for the Simplex tree or \iastar\ (respectively); 
                    and (3) the Simplex tree results for the \datasetName{Prob.40D} and \datasetName{Lucy34D} dataset are partial ($>$). 
		}
\label{hist:storage_nD_norm}
\end{figure}

\begin{figure}[t]
\centering
\resizebox{\columnwidth}{!}{
	\begin{tikzpicture}
        \pgfplotstableread[col sep=comma]{charts/histogram_storage_quad_hexa.csv}\histCube
        \centering
        \begin{axis}[
			  reverse legend,
		width=7cm,
		height=8.5cm,
        enlarge y limits=0.1,
        legend cell align = left,
        legend style={
			at={(1.1,0.05)},
            anchor=north east,
        },
        xmin=0, 
        xbar=2pt,
        ytick=data,
        bar width=3.5pt,
        font=\scriptsize,
        yticklabels from table= \histCube{overlap},
        y tick label style= {text width=1.5cm, align=right},
        nodes near coords,
       every node near coord/.append style={anchor=mid west, yshift=-0.12ex,
       black!95, font=\scriptsize},
       xlabel = $|\Sigma|$,
       axis lines*=left,
       y=0.8cm
        ]
        \addplot [ksStellarStyle] table [y expr=-\coordindex, x expr={\thisrow{hit5}/\thisrow{hit7}}, meta=str5] {\histCube};\addlegendentry{\ks\ Stellar tree}
        \addplot [klStellarStyle] table [y expr=-\coordindex, x expr={\thisrow{hit6}/\thisrow{hit7}}, meta=str6] {\histCube};\addlegendentry{\kl\ Stellar tree}
        \addplot [iaStarStyle]    table [y expr=-\coordindex, x expr={\thisrow{hit1}/\thisrow{hit7}}, meta=str1] {\histCube};\addlegendentry{\iastar}
      \end{axis}
      
  \end{tikzpicture}
}
\caption{Storage costs for manifold quadrilateral (\datasetName{neptune}, \datasetName{statuette} and \datasetName{lucy})  
        and hexahedral (\datasetName{bonsai}, \datasetName{vismale} and \datasetName{foot}) complexes. 
				Costs (labels to right of each bar) are normalized to the indexed mesh representation (listed along y-axis). 
				Datasets marked with $\otimes$ could not be directly generated on our test machine using the standalone \iastar. }
\label{hist:storage_quad_hexa_norm}
\end{figure}      
\begin{figure}[t]
\centering
\resizebox{\columnwidth}{!}{
	\begin{tikzpicture}
        \pgfplotstableread[col sep=comma]{charts/histogram_storage_tri_tetra.csv}\tableSimplex
        \centering
        \begin{axis}[
          reverse legend,
          width=7cm,
          height=8.5cm,
          enlarge y limits=0.1,
          legend cell align = left,
          legend style={
            at={(0.65,0.37)},
            anchor=north,
          },
          xmin=0, 
          xbar=2pt,
          ytick=data,
          bar width=3.5pt,
          font=\scriptsize,
          yticklabels from table= \tableSimplex{overlap},
          y tick label style= {text width=1.5cm, align=right},   
          nodes near coords,
          every node near coord/.append style={anchor=mid west, yshift=-0.12ex,
                                               black!95, font=\scriptsize},
          xlabel = $|\Sigma|$,
          axis lines*=left,
          y=1.4cm,
        ]
        \addplot [ksStellarStyle]   table [y expr=-\coordindex, x expr={\thisrow{hit5}/\thisrow{hit7}}, meta=str5] {\tableSimplex};\addlegendentry{\ks\ Stellar tree}
        \addplot [klStellarStyle]   table [y expr=-\coordindex, x expr={\thisrow{hit6}/\thisrow{hit7}}, meta=str6] {\tableSimplex};\addlegendentry{\kl\ Stellar tree}
        \addplot [simplexTreeStyle] table [y expr=-\coordindex, x expr={\thisrow{hit4}/\thisrow{hit7}}, meta=str4] {\tableSimplex};\addlegendentry{Simplex tree}
        \addplot [sotStyle]         table [y expr=-\coordindex, x expr={\thisrow{hit3}/\thisrow{hit7}}, meta=str3] {\tableSimplex};\addlegendentry{SOT}
        \addplot [cotStyle]         table [y expr=-\coordindex, x expr={\thisrow{hit2}/\thisrow{hit7}}, meta=str2] {\tableSimplex};\addlegendentry{CoT} 
        \addplot [iaStarStyle]      table [y expr=-\coordindex, x expr={\thisrow{hit1}/\thisrow{hit7}}, meta=str1] {\tableSimplex};\addlegendentry{\iastar}
      \end{axis}
	\end{tikzpicture}
}
  \caption{Storage costs for manifold triangle (\datasetName{neptune}, \datasetName{statuette} and \datasetName{lucy}) 
        and tetrahedral (\datasetName{bonsai}, \datasetName{vismale} and \datasetName{foot}) complexes. 
				Costs (labels to right of each bar) are normalized to the indexed mesh representation (listed along y-axis). }
  \label{hist:storage_tri_tetra_norm}
\end{figure}   


Based on our analysis of the literature (see Section~\ref{sec:related_topological_ds}), 
the most relevant dimension-independent topological data structures that scale to our experimental datasets are:
	the Incidence Graph (IG)~\cite{Edelsbrunner1987Algorithms}, 
	the Incidence Simplicial (IS)~\cite{DeFloriani2010dimension}, 
	the Simplex tree~\cite{Boissonnat2014simplex}, 
	and the Generalized Indexed data structure with Adjacencies (\iastar)~\cite{Canino2011IA}.
Since Canino\etal~\cite{Canino2014Representing} demonstrated that the \iastar\ data structure is more compact 
than the IG and IS data structures for low and high-dimensional datasets, 
we restrict our comparisons to the \iastar\ and Simplex tree data structures.

\NOTA{
  \leilaComment{for \textbf{Kenny}: please say why we do not compare with the other data structures by Boissonnat (like $MST$ and $SAL$)}
  \kennyComment{for \textbf{Leila}: I addressed this when we discuss MST and SAL in Section 3 and in the conclusions.  
    I also added the phrase `that scale to our experimental datasets'.
    Please let me know if that is sufficient and/or update it. }
}

The \emph{\iastar\ data structure} has been defined for dimension-independent simplicial complexes, 
and for our experiments, we have extended it to dimension-independent CP complexes.
It explicitly encodes all vertices and \ktopcpcells\ in $\Sigma$, with $0 < \tDim \leq \cDim$, as well as the following topological relations:
%
\begin{compactenum}[(i)]
	\item boundary relation $\relation{\tDim,0}(\simplex)$, for each \ktopcp\ $\simplex$;
	\item adjacency relation $\relation{\tDim,\tDim}(\sigma)$, for each \ktopcp\ $\simplex$; 
	\item co-boundary relation $\relation{\tDimMinusOne,\tDim}(\tau)$, for each non-manifold (\tDimMinusOne)-cell $\tau$ bounding a \ktopcp; 
	\item partial co-boundary relation $\partialrelation{0,\tDim}(\vertex)$, 
        for each vertex $\vertex$, consisting of an arbitrary \ktopcp\  $\simplex$
				from each \emph{\tDim-cluster} in the star of \vertex. 
        A \tDim-cluster is a ($\tDim{-}1$)-connected component of the star of $v$ restricted to its top CP $k$-cells.
\end{compactenum}
Note that for pure CP complexes, the non-manifold co-boundary relation $\relation{\tDimMinusOne,\tDim}$ is empty.
Further, for pseudo-manifold complexes, the partial vertex co-boundary relation $\partialrelation{0,\tDim}$ has cardinality 1,
and the \iastar\ is identical to the IA data structure~\cite{Paoluzzi1993Dimension}.

The \emph{Simplex tree} encodes all $j$-simplices in $\Sigma$, with $0 \leq j \leq \cDim$, like the IG, while storing a subset of the incidence relations encoded by the IG. 
The Simplex tree is defined over a total order on the vertices of \sC, and thus, each simplex \simplex\ is uniquely represented as an ordered path in a trie whose nodes correspond to the boundary vertices of \simplex. 
Thus, the nodes are in bijection with the simplices of the complex, and a Simplex tree over a simplicial complex with $|\Sigma|$ simplices (of any dimension) contains exactly $|\Sigma|$ nodes.
This provides an efficient representation for extracting all boundary relations of simplices in \sC.
We compare the Stellar tree to the implementation of the Simplex tree provided in~\cite{GUDHI}, 
where each node of a Simplex tree requires 
  a reference to the label of the vertex
  and three references to the tree structure 
  (pointers to the parent node, to the first child and to the next sibling node) 
for a total of $4|\Sigma|$ references. 
\NOTA{
  \felleComment{removed the sentence in which it was claimed that the Simplex tree cannot compute co-boundary relations.}
  \NOTA{ 
    We note that this implementation only supports the efficient extraction of boundary relations.
    An implementation that supports extraction of co-boundary and adjacency relations, as proposed in~\cite{Boissonnat2014simplex}, would require additional storage.
  }
}

Note that the Stellar tree and our extended \iastar\ data structure can both represent CP complexes in arbitrary dimension and, 
thus, have the same expressive power, while the Simplex tree can represent only simplicial complexes.
%
%
Another difference is that Stellar trees require the complex to be embedded in an ambient space \aSpace, 
while the other data structures are purely topological and do not require a spatial embedding.
We note, however, that while this is a requirement for Stellar trees, it is not a requirement for the more general Stellar decomposition.

In terms of storage requirements, the Stellar tree is always more compact than the \iastar\ data structure, requiring approximately half of the storage,
nearly all of which is used for encoding boundary relation \relation{\tDim, 0} for the top cells (i.e., the indexed representation that they share in common).
%
%
It is worth noting that we were unable to directly generate the \iastar\ data structure for several of our larger datasets 
on our 64 GB test machine. We generated the \iastar\ data structure on these datasets indirectly using our Stellar tree representation (see Section \ref{sec_app:stellar_build_structures} in the Appendix) 
and we have marked these datasets with an $\otimes$ in Figures~\ref{hist:storage_nD_norm} and~\ref{hist:storage_quad_hexa_norm}.



When comparing the Stellar tree to the Simplex tree, we observe that the Stellar tree is significantly more compact:
by an order of magnitude on manifold and pure models, 
and by two orders of magnitude or more on non-manifold models.
Here too, we were unable to generate Simplex trees for several of the higher dimensional models on our test machine.
For these datasets (marked with $\odot$ in Figure~\ref{hist:storage_nD_norm}), we estimated the storage requirements based on the number of simplices of each dimension in the model. On two of these datasets, \datasetName{prob 40D} and \datasetName{lucy 34D}, 
we were unable to extract all simplices in all dimensions (even indirectly, see Section~\ref{sec:implicit_cell_extraction}), 
and thus, the storage shown in Figure~\ref{hist:storage_nD_norm} is a lower bound of the real storage requirements.
In contrast, we had no difficulty generating Stellar Trees for any of our test datasets.

%
\NOTA{
  \felleComment{(for \textbf{Kenny}): added a sentence here for enforcing that on two dataset the Simplex tree storage requirements are partial, but the trend is still confirmed}
}


	
For our dimension-dependent comparisons on manifold simplicial complexes, we also considered the \emph{Corner Table} ($CoT$)~\cite{Rossignac20013D} 
and the \emph{Sorted Opposite Table} ($SOT$)~\cite{Gurung2009SOT} data structures, both defined only for manifold triangle and tetrahedral complexes.
The $CoT$ data structure is similar to the IA data structure and explicitly encodes
	boundary relation $\relation{\cDim,0}(\simplex)$ and adjacency relation $\relation{\cDim,\cDim}(\simplex)$ of each top \cDim-simplex \simplex.
The $SOT$ extends the $CoT$ by implicitly encoding boundary relation $\relation{\cDim,0}(\simplex)$.
	It only explicitly encodes adjacency relation $\relation{\cDim,\cDim}(\simplex)$.

\NOTA{=== TABLES ===}

When comparing the Stellar tree to corner-based data structures,
we observe that the $CoT$ data structure has similar storage requirements as the IA and is roughly twice as large as the Stellar tree,
while the $SOT$ has similar storage requirements as the Stellar tree, requiring about 1\% to 10\% less space.




\NOTA{ 
All of these topological data structures involve a tradeoff between storage (e.g.\ compact encodings) 
and compute time (e.g.\ to reconstruct the desired topological relations).
The $SOT$ achieves its compactness by dropping its boundary relations, which must be recomputed from its adjacency relations.
The \iastar\ explicitly encodes these relations and is able to more efficiency at reconstructing boundary and co-boundary relations at run-time. 
In the same way, as the Stellar tree encodes the \relation{k,0} boundary relation for each \top\ and a very compact indexing structure only,  
we have to consider an extra computational time in reconstructing the adjacency and co-boundary relations at run-time.
}

Finally, we consider the effect of different bucketing threshold on the size and efficiency of the Stellar tree representation.
For our experimental datasets, there was only about a 10\% difference in storage requirements between the large (\kl) and small (\ks) bucketing factors.
Clearly, this is not always true, especially in the limit cases, i.e., with $\kv = 1$ and $\kv = \infty$. 
%
Very low bucketing thresholds (with $\kv$ near 1) yield deeper trees whose leaf blocks index only a few entities, leading to a high topological overhead
but more efficient execution for individual mesh processing operations.
%
Conversely, really large bucketing threshold values lead to lower storage overhead
at the expense of increased query and execution times for individual operations.
At the limit, when $\kv=\infty$, the Stellar tree is effectively identical to the indexed representation. 

These results confirm that the Stellar tree can efficiently represent both low- and high-dimensional complexes
with only a slight storage overhead relative to that of the indexed base complex.
This is largely due to the Stellar tree's exploitation of the complex's spatial locality via SRE compression.

\NOTA{
\kennyComment{Should we identify the datasets that can't be directly generated?  Perhaps adding an asterisk (*) to the sizes in the figures.}
\felleComment{added $\odot$ or $\otimes$ next to datasets that cannot be generated directly using Simplex tree or \iastar, respectively. I didn't found a way to do that next to the corresponding bar..}
\kennyComment{We should emphasize the large overhead of the Simplex tree, and the low overhead of the Stellar trees, especially for the V-Rips complexes, which the Simplex tree is explicitly targeting.}
}


%

\subsection{Evaluation of Stellar tree generation times}
\label{sec:generation_timings}

In this section, we evaluate the generation times for the Stellar tree.
%
Table~\ref{tab:generation_timings} shows the timings of the four generation phases and the overall \emph{total} timings. 
The two \emph{insert} columns show the time for creating the base indexing structure \h\ over the vertices \sCV\ of the complex \sC, 
or the time for inserting the top cells \sCT\ into \h, 
while \emph{reindex} columns show the timings for reordering and SRE compressing the indexed lists and arrays in \h\ and \sC.

We first consider the relative cost of each of the generation phases.
In general, the vertex reindexing phase consumes less than 10\% of the overall timings. 
For the \emph{triangle}, \emph{quadrilateral}, \emph{hexahedral} complexes, and the lower dimensional \emph{Vietoris-Rips} complex,
generating \h\ is the most expensive phase, while for the \emph{tetrahedral}, \emph{probabilistic-refinement} and the two higher dimensional \emph{Vietoris-Rips} models,
reindexing the top cells is the most expensive phase.
These results can be understood by considering the relative sizes of \sCV\ and \sCT.
When the number of vertices is greater than or equal to the number of top cells, it is more expensive to generate the spatial hierarchy \h.
Otherwise, reindexing and compressing the top cells arrays dominates. 

Finally, considering the effect of the bucketing thresholds (\kv) on generation times, 
we find that Stellar trees with higher bucketing thresholds (\kl) can be generated in less time than those with lower bucketing thresholds (\ks).
This is expected since high values of \kv\ tend to produce coarser spatial subdivisions with lower average spanning numbers \Chi.

\begin{table}[t]
			\centering
			\caption{Generation timings (in seconds) for the Stellar tree.
			}
			{
			\resizebox{0.85\columnwidth}{!}{
				\begin{tabular}{ccc|cc|cc|c}
					\toprule
					\multicolumn{1}{c}{\multirow{4}{*}{\textbf{Data}}}  &
					\multicolumn{1}{c}{\multirow{4}{*}{\textbf{}}}  &
					\multicolumn{1}{c}{\multirow{4}{*}{\kv}} & 
					\multicolumn{5}{c}{\textbf{Timings}}  \\	
					\cmidrule(lr){4-8}
					 \cmidrule(lr){4-7}
					  & & & \multicolumn{2}{c}{\textbf{vertices}} & \multicolumn{2}{c}{\textbf{top CP cells}} &  \multirow{2}{*}{\textbf{total}}  \\
					 \cmidrule(lr){4-5} \cmidrule(lr){6-7}
					  & & & \textbf{insert} & \textbf{reindex} & \textbf{insert} & \textbf{reindex} & \\
					\midrule

\multirow{2}{*}{\datasetName{neptune}} & \multirow{7}{*}{\begin{sideways}\datasetName{tri.}\end{sideways}} 
& \ks & 4.52 & 0.68 & 1.64 & 3.23 & 10.1 \\
&  &  \kl & 3.83 & 0.67 & 1.24 & 2.77 & 8.51 \\
\cmidrule(lr){1-1} \cmidrule(lr){3-8}

\multirow{2}{*}{\datasetName{statuette}} & 
& \ks & 11.6 & 1.77 & 3.42 & 7.99 & 24.8 \\
&  &  \kl & 10.1 & 1.74 & 2.74 & 6.70 & 21.3 \\
\cmidrule(lr){1-1} \cmidrule(lr){3-8}

\multirow{2}{*}{\datasetName{lucy}} & 
& \ks & 34.6 & 1.32 & 8.85 & 21.9 & 66.7 \\
&  &   \kl & 30.3 & 0.48 & 7.45 & 18.1 & 56.3 \\
\midrule

\multirow{2}{*}{\datasetName{neptune}} & \multirow{7}{*}{\begin{sideways}\datasetName{quad.}\end{sideways}} 
& \ks & 32.2 & 4.39 & 6.64 & 11.3 & 54.5 \\
&  &  \kl & 27.5 & 4.36 & 4.63 & 8.58 & 45.1 \\
\cmidrule(lr){1-1} \cmidrule(lr){3-8}

\multirow{2}{*}{\datasetName{statuette}} & 
& \ks & 82.7 & 12.3 & 14.0 & 29.1 & 138  \\
&  &  \kl & 73.8 & 12.2 & 10.7 & 22.7 & 119 \\
\cmidrule(lr){1-1} \cmidrule(lr){3-8}

\multirow{2}{*}{\datasetName{lucy}} & 
& \ks & 263 & 2.17 & 37.0 & 61.8 & 364 \\
&  &  \kl & 223 & 2.02 & 29.5 & 35.5 & 290 \\
\midrule

\multirow{2}{*}{\datasetName{bonsai}} & \multirow{7}{*}{\begin{sideways}\datasetName{tetra.}\end{sideways}} 
& \ks & 6.69 & 1.66 & 7.99 & 20.8 & 37.2 \\
&  &  \kl & 6.25 & 1.65 & 7.12 & 19.3 & 34.3 \\
\cmidrule(lr){1-1} \cmidrule(lr){3-8}

\multirow{2}{*}{\datasetName{vismale}} & 
& \ks & 7.25 & 1.82 & 8.35 & 22.1 & 39.6 \\
&  &  \kl & 6.96 & 1.81 & 7.88 & 21.2 & 37.8 \\
\cmidrule(lr){1-1} \cmidrule(lr){3-8}

\multirow{2}{*}{\datasetName{foot}} & 
& \ks & 8.55 & 2.00 & 10.8 & 27.9 & 49.2 \\
&  &  \kl & 7.34 & 1.97 & 8.52 & 23.4 & 41.2 \\
\midrule

\multirow{2}{*}{\datasetName{f16}} & \multirow{7}{*}{\begin{sideways}\datasetName{hexa.}\end{sideways}} 
& \ks & 103 & 14.2 & 77.7 & 53.9 & 249 \\
&  &  \kl & 94.1 & 13.9 & 46.7 & 35.1 & 190 \\
\cmidrule(lr){1-1} \cmidrule(lr){3-8}

\multirow{2}{*}{\datasetName{san fern}} & 
& \ks & 154 & 27.6 & 52.1 & 102 & 336 \\
&  &  \kl & 140 & 27.5 & 37.1 & 67.8 & 273 \\
\cmidrule(lr){1-1} \cmidrule(lr){3-8}
	
\multirow{2}{*}{\datasetName{vismale}} & 
& \ks & 337 & 72.8 & 118 & 222 & 751 \\
&  &  \kl & 324 & 71.8 & 85.3 & 147 & 628 \\
\midrule
\midrule

\multirow{2}{*}{\datasetName{5D}} & \multirow{7}{*}{\begin{sideways}\datasetName{prob.}\end{sideways}} 
& \ks & 0.50 & 0.58 & 40.9 & 53.0 & 95.0 \\
 &  &  \kl & 0.37 & 0.58 & 20.9 & 32.7 & 54.5  \\
\cmidrule(lr){1-1} \cmidrule(lr){3-8}
 
\multirow{2}{*}{\datasetName{7D}} & 
& \ks &  0.55 & 5.98 & 332 & 612 & 950 \\
 &  &  \kl & 0.45 & 5.97 & 203 & 471 & 681  \\
\cmidrule(lr){1-1} \cmidrule(lr){3-8}
 
\multirow{2}{*}{\datasetName{40D}} &  
& \ks & 1.32 & 1.73 & 972 & 769 & 1.7K \\
 &  &  \kl & 1.02 & 1.73 & 529 & 448 & 980  \\
\midrule
%
%
%
%

\multirow{2}{*}{\datasetName{vismale 7D}} & \multirow{7}{*}{\begin{sideways}\datasetName{v-rips}\end{sideways}} 
& \ks & 7.20 & 1.80
& 2.65 & 3.04 & 13.0 \\
 &  & \kl & 6.94 & 1.81
& 2.50 & 2.75 & 12.3  \\
\cmidrule(lr){1-1} \cmidrule(lr){3-8} 
\multirow{2}{*}{\datasetName{foot 10D}} & & \ks & 9.01 & 1.99
& 41.0 & 57.1 & 108 \\
 &  & \kl & 7.91 & 1.98
 & 30.6 & 35.9 & 75.2  \\
\cmidrule(lr){1-1} \cmidrule(lr){3-8} 
\multirow{2}{*}{\datasetName{lucy 34D}} & & \ks & 35.9 & 1.63 & 36.8 & 42.2 & 117 \\
 &  & \kl & 30.7 & 0.81 & 28.7 & 24.3 & 84.5  \\

\bottomrule
				\end{tabular}
			}
			}
			\label{tab:generation_timings}
\end{table}

	

\section{Topological queries on a Stellar tree}
\label{sec:local_topo_rels}

In this section, we describe how to perform 
topological queries on a CP complex \sC\ in the Stellar tree representation.
These queries 
are the fundamental building blocks for locally traversing and processing the underlying complex. 
%
\NOTA{
  \kennyComment{I removed this b/c it repeated the contents of Section 6}
  In contrast to traditional topological data structures 
  which operate on individual elements of the mesh, the Stellar tree adopts a \emph{batched} strategy, 
  in which we concurrently reconstruct the local topological relation or auxiliary data structures on the portion of the complex indexed by each block of the tree.
  This strategy is particularly effective when processing large datasets where we need to apply our analysis to the entire complex 
  or to large portions of it.
  In this common scenario, the cost of constructing these local data structures can be amortized 
  over the processing costs of the geometry within the block.
}
%
%
\NOTA{
	In the remainder of this section, we focus on 
	batched algorithms that reconstruct topological relations on the cells within a leaf block \R\ of a Stellar tree \sTree.
}
\NOTA{
		It is a bit awkward to refer to the non-top cells in the complex.
		Is there a good name for these, other than 'all $p$-cells (top or non-top)'?
}

Since these queries often depend on \emph{all} cells in the complex, rather than just the explicitly represented top cells, 
we first describe how we obtain and represent the implicitly encoded boundary cells of the complex 
from the Stellar tree representation (Section~\ref{sec:implicit_cell_extraction}). 
We then present algorithms for extracting the co-boundary (Section~\ref{sec:coboundary_extraction}).
For brevity, we omit a description of how to extract adjacency relations, but in the Appendix \ref{sec_app:stellar_build_structures} 
we describe how to extract the \relation{\cDim,\cDim} adjacency relations to generate the \iastar\ data structure from a Stellar tree.

\subsection{Extracting boundary relations}
\label{sec:implicit_cell_extraction}

The Stellar tree's underlying indexed representation of a CP complex \sC\ explicitly encodes 
only the vertices and top CP \tDim-cells of \sC\ for $\tDim \leq \cDim$ (see Section~\ref{sec:mesh_structure}).
However, many applications require access to non-top cells within the complex.
Since they are implicitly encoded within the Stellar tree representation,
we must create a local (explicit) representation 
to support algorithms for processing and attaching data to such cells.
%
\begin{algorithm}[t]			
  \renewcommand{\algorithmicensure}{\textbf{Optional:}}
	\renewcommand{\algorithmiccomment}[1]{\bgroup\hfill//\textit{~#1}\egroup}
	\caption{ \AlgoName{extract\_p\_cells}($p$,\R,\sC) }
	\label{alg:extract-p-cells}				
	\begin{algorithmic}[1]			
		\medskip
		\Input {$p$ is the cell dimension to extract}
		\Input {$\R$ is a leaf block in \h}
		\Input {$\sC$ is the CP complex indexed by $\h$}
		\Variable { $m\_p$ maps a $p$-cell vertex tuple to its local index }
		\Require {Extract boundary $p$-cells of top \tDim-cells, $0\!<\!p\!\leq\!k\!\leq\!d$ }

		\ForAll {\ktopcpcells\ $\simplex$ in $\PhiMapTop(\R)$ (with index \tIndex\ in \sCT)}
			\ForAll { $p$-faces $\tau$ in \relation{\tDim,p}($\simplex$) (with face index \emph{i$_{\tau}$ in \simplex})}
				
				\LineComment{Rearrange $\tau$'s vertices into a canonical order}
				\State { $v\_tuple$ $\leftarrow$ \AlgoName{canonical\_tuple}($\relation{p,0}(\tau)$) }
				
				\LineComment{If $\tau$ is indexed by \R, add it to the local p-faces map}
				\If {there exists $\vertex\ \in \relation{p,0}(\tau)$ such that $\vertex\ \in\ \PhiMapVert(\R)$}
					\LineComment{Insert $\tau$ as a new $p$-cell, if not already present}
					\If { $v\_tuple$ is not in $m\_p$ }
						\State{ $id_\tau$ $\leftarrow$ \AlgoName{size}($m\_p$) } \SideComment{$id_\tau$ is $\tau$'s local index in \R}
						\State { $m\_p[v\_tuple]$ $\leftarrow$ $id_\tau$ } 
					\EndIf
				\EndIf
			\EndFor
		\EndFor 
	\end{algorithmic}
\end{algorithm}

Our strategy for extracting all $p$-cells is to iterate through the top \tDim-cells of a leaf block
for each dimension $k$, $0 < p \leq k \leq d$ 
and to extract an ordered set of $p$-cells 
(see Algorithm~\ref{alg:extract-p-cells}).
We use an associative array $m\_p$ to track the unique set of encountered $p$-cells with at least one vertex indexed by \R\ (row 4).
Array $m\_p$ maps the tuple of vertices for a $p$-cell $\tau$ to an integer index $id_\tau$ in the set,
accounting for changes in ordering and orientation through the \AlgoName{canonical\_tuple} routine (row 3).
In some applications, it is useful to also explicitly maintain the boundary relation \relation{p,0} for the $p$-cells 
and/or the incidence relations \relation{\tDim,p} or \relation{p,\tDim} for the top \tDim-cells.
These are encoded using the local indices within the ordered set of extracted $p$-cells.

We note that, for truly high-dimensional datasets, it is not feasible to extract $p$-cells in all cases.
For example, there are ${41}\choose{21}$ 20-simplices within a single 40-simplex. 
Encoding these 269 billion simplices would require more than 40TB of storage.
However, even on these datasets, we can still extract the lowest and highest dimensional $p$-cells.
This highlights an advantage of only encoding the top cells of the complex (as in the Stellar tree and \iastar\ data structure)
compared to representations that encode all cells of the complex (as in the IG or Simplex tree).
Stellar trees have no difficulty encoding and processing such high-dimensional complexes, 
despite the combinatorial explosion in the number of overall cells.

\begin{table}[t]
	\centering
	\caption{
		Summed timings (seconds) and additional storage requirements (number of references) to extract boundary $p$-cells from Stellar tree, \iastar\ and Simplex tree data structures.
		Datasets marked with an $\otimes$ could not be directly generated on our test machine by the \iastar.
		\NOTA{ \felleComment{we extract/encode a p-face type at a time, but this table shows the summation of the timings and storage requirements} }
	}	
	{
	\resizebox{0.9\columnwidth}{!}{
		\begin{tabular}{cccccc|cc}
			\toprule
			\multicolumn{1}{c}{\multirow{3}{*}{\textbf{Data}}} & \multicolumn{1}{c}{\multirow{2}{*}{\textbf{}}}  & 
			\multicolumn{1}{c}{\multirow{3}{*}{\textbf{\kv}}} &
			\multicolumn{3}{c}{\multirow{1}{*}{\textbf{Time}}} & 
			\multicolumn{2}{c}{\multirow{1}{*}{\textbf{Storage}}} \\
			\cmidrule(lr){4-6} \cmidrule(lr){7-8}
			& & & \multirow{2}{*}{\textbf{IA$^*$}} & \textbf{Simplex} & \textbf{Stellar} & \textbf{IA$^*$ $/$} & \textbf{Stellar} \\
			& & & & \textbf{tree} & \textbf{tree} & \textbf{Simplex} & \textbf{tree} \\
			\midrule
			
\multirow{2}{*}{\datasetName{neptune}} & \multirow{7}{*}{\begin{sideways}\datasetName{tri.}\end{sideways}} & k$_S$ & \multirow{2}{*}{4.93} & \multirow{2}{*}{1.82} & 1.90 & \multirow{2}{*}{12.0M} & 0.70K \\
 & & k$_L$ &  &  & 2.20 &  & 3.24K \\
\cmidrule(lr){1-1} \cmidrule(lr){3-8}
\multirow{2}{*}{\datasetName{statuette}} &  & k$_S$ & \multirow{2}{*}{9.21} & \multirow{2}{*}{3.73} & 4.90 & \multirow{2}{*}{30.0M} & 0.72K \\
 & & k$_L$ &  &  & 5.55 &  & 3.22K \\
\cmidrule(lr){1-1} \cmidrule(lr){3-8}
\multirow{2}{*}{\datasetName{lucy}} &  & k$_S$ & \multirow{2}{*}{25.3} & \multirow{2}{*}{9.94} & 13.8 & \multirow{2}{*}{84.1M} & 0.82K \\
 & & k$_L$ &  &  & 16.2 &  & 3.28K \\
\midrule

\multirow{2}{*}{\datasetName{neptune}} & \multirow{7}{*}{\begin{sideways}\datasetName{quad.}\end{sideways}} & k$_S$ & \multirow{2}{*}{40.8} & \multirow{2}{*}{n/a} & 6.61 & \multirow{2}{*}{96.2M} & 0.52K \\
 & & k$_L$ &  &  & 7.43 &  & 3.37K \\
\cmidrule(lr){1-1} \cmidrule(lr){3-8}
\multirow{2}{*}{\datasetName{statuette}} &  & k$_S$ & \multirow{2}{*}{91.3} & \multirow{2}{*}{n/a} & 15.9 & \multirow{2}{*}{240M} & 0.50K \\
 & & k$_L$ &  &  & 19.0 &  & 3.38K \\
\cmidrule(lr){1-1} \cmidrule(lr){3-8}
\multirow{2}{*}{\datasetName{lucy}} &  & k$_S$ & \multirow{2}{*}{251} & \multirow{2}{*}{n/a} & 43.2 & \multirow{2}{*}{673M} & 0.53K \\
 & & k$_L$ &  &  & 53.4 &  & 3.41K \\
\midrule

\multirow{2}{*}{\datasetName{bonsai}} & \multirow{7}{*}{\begin{sideways}\datasetName{tetra.}\end{sideways}} & k$_S$ & \multirow{2}{*}{49.6} & \multirow{2}{*}{22.7} & 45.6 & \multirow{2}{*}{204M} & 20.9K \\
 & & k$_L$ &  &  & 47.8 &  & 42.5K \\
\cmidrule(lr){1-1} \cmidrule(lr){3-8}
\multirow{2}{*}{\datasetName{vismale}} &  & k$_S$ & \multirow{2}{*}{54.5} & \multirow{2}{*}{25.1} & 52.2 & \multirow{2}{*}{222M} & 21.4K \\
 & & k$_L$ &  &  & 53.7 &  & 36.5K \\
\cmidrule(lr){1-1} \cmidrule(lr){3-8}
\multirow{2}{*}{\datasetName{foot}} &  & k$_S$ & \multirow{2}{*}{59.5} & \multirow{2}{*}{29.7} & 50.9 & \multirow{2}{*}{246M} & 21.2K \\
 & & k$_L$ &  &  & 57.5 &  & 43.3K \\
\midrule

\multirow{2}{*}{\datasetName{f16}} & \multirow{7}{*}{\begin{sideways}\datasetName{hexa.}\end{sideways}} & k$_S$ & \multirow{2}{*}{OOM} & \multirow{2}{*}{n/a} & 49.6 & \multirow{2}{*}{OOM} & 2.64K \\
 & & k$_L$ &  &  & 71.1 &  & 18.9K \\
\cmidrule(lr){1-1} \cmidrule(lr){3-8}
\multirow{2}{*}{\datasetName{san fern}} &  & k$_S$ & \multirow{2}{*}{OOM} & \multirow{2}{*}{n/a} & 109 & \multirow{2}{*}{OOM} & 2.89K \\
 & & k$_L$ &  &  & 143 &  & 21.1K \\
\cmidrule(lr){1-1} \cmidrule(lr){3-8}
\multirow{2}{*}{\datasetName{vismale}$^\otimes$} &  & k$_S$ & \multirow{2}{*}{OOM} & \multirow{2}{*}{n/a} & 263 & \multirow{2}{*}{OOM} & 1.77K \\
 & & k$_L$ &  &  & 340 &  & 17.4K \\
\midrule
\midrule

\multirow{2}{*}{\datasetName{5D}} & \multirow{5}{*}{\begin{sideways}\datasetName{prob.}\end{sideways}} & k$_S$ & \multirow{2}{*}{456} & \multirow{2}{*}{123} & 316 & \multirow{2}{*}{970M} & 152K \\
 & & k$_L$ &  &  & 425 &  & 1.94M \\
\cmidrule(lr){1-1} \cmidrule(lr){3-8}
\multirow{2}{*}{\datasetName{7D}$^\otimes$} &  & k$_S$ & \multirow{2}{*}{OOM} & \multirow{2}{*}{OOM} & 21.2K & \multirow{2}{*}{OOM} & 51.3M \\
 & & k$_L$ &  &  & 24.6K &  & 167M \\
\midrule
 
\multirow{2}{*}{\datasetName{vismale 7D}} & \multirow{5}{*}{\begin{sideways}\datasetName{v-rips}\end{sideways}} & k$_S$ & \multirow{2}{*}{179} & \multirow{2}{*}{149} & 156 & \multirow{2}{*}{1.43B} & 267K \\
 & & k$_L$ &  &  & 162 &  & 318K \\
\cmidrule(lr){1-1} \cmidrule(lr){3-8}
\multirow{2}{*}{\datasetName{foot 10D}} &  & k$_S$ & \multirow{2}{*}{OOM} & \multirow{2}{*}{OOM} & 16.6K & \multirow{2}{*}{OOM} & 12.0M \\
 & & k$_L$ &  &  & 21.4K &  & 15.9M \\
\bottomrule

\end{tabular}
	}
	}
	\label{tab:pcells_results}
\end{table}


\paragraph*{Experimental results}

We now analyze the effectiveness of the Stellar tree representation for (batched) $p$-cell extractions against our implementation of the \iastar\ data structure and the Simplex tree (as implemented in the \emph{GUDHI} framework~\cite{GUDHI}).
\NOTA{
  \color{red}
  We do not compare against other topological data structures, like the Corner Table and the Sorted Opposite Table, since these data structures can only represent simplicial complexes in low dimensions, and being adjacency-based, they have similar performances to the manifold version of the \iastar\ data structure, the IA data structure.
  \color{black}
}
Table~\ref{tab:pcells_results} lists the aggregate times and storage requirements for extracting all non-top $p$-cells from our experimental datasets. Notice that we do not consider the higher dimensional \emph{probabilistic} dataset and the \datasetName{lucy 34D} V-Rips complex, as extracting all $p$-cells on these datasets is unfeasible due to its computational and storage requirements.


First, we analyze the influence of the bucketing threshold \kv\ for Stellar trees. Smaller \kv\ values lead to faster extractions on all our experimental datasets.
This speedup increases with the dimension of the complex since the auxiliary data structure encoding a $p$-face type becomes smaller, 
and thus, checking for the presence of duplicates has a lower computational cost.

The \iastar\ data structure follows a similar strategy as the Stellar trees for extracting its implicit $p$-cells
since both data structures use an indexed representation for encoding the boundary relations of a CP complex. 
Table~\ref{tab:pcells_results} demonstrates the computational and storage advantages of the Stellar tree over the \iastar\ for this task.
Namely, Stellar trees require from 20\% to 55\% less time for the two-dimensional datasets and approximately 10\% less time on the higher dimensional ones.
Notice, however, that the \iastar\ data structure is a global data structure over the entire complex and 
runs out of memory (OOM) on our \emph{hexahedral} datasets and on the 7D \emph{probabilistic} and \datasetName{foot 10D} V-Rips datasets.
In addition, the Stellar tree's auxiliary storage requirements are negligible compared to those of the \iastar\ data structure. 

The Simplex tree explicitly encodes all simplices of a simplicial complex, thus, its $p$-cells can be enumerated by traversing all simplices at the $p$-th level of the tree. Explicitly encoding boundary relation \relation{p,0} would require the same auxiliary storage as the \iastar\ data structure, since both data structures require global structures. Table~\ref{tab:pcells_results} demonstrates that Stellar trees are slower than Simplex trees at boundary cell extraction, but, still, competitive with respect to a representation that explicitly encodes all cells. This is possible thanks to the smaller local auxiliary data structures used by Stellar trees. 
Note that the Simplex tree runs out of memory (OOM) on our workstation for the 7D \emph{probabilistic} dataset and the \datasetName{foot 10D} V-Rips complex.
Since a Simplex tree can only represent simplicial complexes, it does not support $p$-cell extraction on \emph{quad} and \emph{hexahedral} datasets.

\subsection{Extracting co-boundary relations}
\label{sec:coboundary_extraction}


Co-boundary queries arise in a variety of mesh processing applications, 
including those requiring mesh simplification and refinement~\cite{Garland1997Surface,Natarajan2004Simplification,Zorin00},
or the \emph{dual} of a complex~\cite{Hirani2003,Mullen11,Weiss2013primaldual}.

Co-boundary queries are naturally supported by the Stellar decomposition model.
By definition, all regions of the decomposition that contain at least one vertex of a CP cell $\tau$ 
must index all CP cells in the star of $\tau$ (see Equation~\ref{eq:phitop_blocks}).
Since the top cells are explicitly represented in \sC, 
we first describe how to extract the vertex co-boundary relation \relation{0,\tDim} restricted to the top \tDim-cells of \sC,
which we will refer to as the \emph{restricted co-boundary relation \relation{0,\tDim}}.
We  will then discuss how to extend this to extract vertex co-boundary relation \relation{0,p} over \emph{all} $p$-cells in \sC,
and the general co-boundary relation \relation{p,q} with $0 \leq p < q \leq d$. 

The \emph{restricted vertex co-boundary relation} \relation{0,\tDim} in a leaf block \R\ is generated by inverting boundary relation \relation{\tDim,0} on the \ktopcpcells\ in \PhiMapTop(\R).
Since the indexed vertices in the leaf blocks of a \datasetName{compressed} Stellar tree are contiguous, 
with indices in the range $[\vstart,\vend)$, 
we encode our local data structure using an array of size $|\PhiMapVert(\R)| = \vend - \vstart$. 
Each position in the array corresponds to a vertex indexed by $\R$ and points to an (initially empty) list of indexes from \sCT. 
As shown in Algorithm \ref{alg:vertex-tops}, we populate these arrays by iterating through relation \relation{\tDim,0} of the \ktopcpcells\ in \PhiMapTop(\R).
For each cell \simplex\ such that relation $\relation{\tDim,0}(\simplex)$ contains a vertex \vertex\ with index $\vIndex \in [\vstart,\vend)$, the index of \simplex\ is added to vertex $v$'s list.

\begin{algorithm}[t]			
	\caption{ \AlgoName{extract\_restricted\_vertex\_cbdry}(\R,\sC) }
	\label{alg:vertex-tops}				
	\begin{algorithmic}[1]			
		\medskip
		\Input {$\R$ is a leaf block in \h}
		\Input {$\sC$ is the mesh indexed by $\h$}
		\Variable {$r\_0\_k$ encodes \relation{0,k} relation for the vertices in \R}
		\Ensure	{Relation \relation{0,\tDim} is \emph{locally} reconstructed $\forall \simplex \in \PhiMapVert(\R)$} 
		\ForAll {\ktop\ $\simplex$ in $\PhiMapTop(\R)$ (with index \tIndex\ in \sCT)}
			\ForAll {vertices $\vertex$ in $\simplex$ (with index \vIndex\ in \sCV)}
				\If {$\vertex\ \in\ \PhiMapVert(\R)$}
					\State{add \tIndex\ to $r\_0\_k[\vIndex]$}
				\EndIf
			\EndFor
		\EndFor
	\end{algorithmic}
\end{algorithm}

\NOTA{
  \felleComment{I added back the \relation{0,k} algorithm.. I think that it can ease the description in the text}
  \felleComment{\emph{implementation detail}: actually we encode the restricted vertex coboundary in an array or arrays, 
    in which each array encodes only a type of top simplex. I think that this detail is not meaningful..}
}

Extending the vertex co-boundary relation to all $p$-cells in \R\
is complicated by the fact that we only have an explicit representation for the top cells in \sC.
%
%
A simple strategy we have developed for extracting $\relation{0,p}$ on all $p$-cells in \R\ is to first extract the explicit set of all $p$-cells in \R, as in Algorithm~\ref{alg:extract-p-cells} (see Section~\ref{sec:implicit_cell_extraction}).
We then invert $\relation{p.0}$ to obtain the complete relation $\relation{0,p}$ for the vertices in \R.

In some applications, we prefer to express \relation{0,p} entirely in terms of top cells from \sC.
Thus, another strategy we have developed is to extract the restricted co-boundary relation \relation{0,\tDim}
for all top \tDim-cells in \R, with $p \le \tDim \le \cDim$.
This redundant representation is thus used as an intermediate representation for $\relation{0,p}(v)$ since each \tDim-cell in $\relation{0,\tDim}(v)$ contains one (or more) $p$-face in the co-boundary of $v$.
For example, this provides a convenient representation for the star of a vertex $v$ as a union of restricted co-boundary relations $\relation{0,\tDim}(v)$, where $1 \le \tDim \le \cDim$.

Similarly, we have defined and implemented a strategy for generating the general co-boundary relation $\relation{p,q}$, where $p < q$.
First, the sets of all $q$-cells, which is expressed as $\relation{q,0}$, is extracted.
This implicitly provides also boundary relation $\relation{q,p}$.
Then, co-boundary relation $\relation{p,q}$ is extracted by inverting $\relation{q,p}$.
%
%

\begin{table}[t]
	\centering
	\caption{
		Times (seconds) and additional storage requirements (number of references) for restricted co-boundary relations \relation{0,k} extractions
		from Stellar tree and \iastar\ representations.
		Datasets marked with an $\otimes$ could not be directly generated on our test machine by the \iastar.
	}	
	{
	\resizebox{0.8\columnwidth}{!}{
		\begin{tabular}{ccccc|r}
			\toprule
			\multicolumn{1}{c}{\multirow{2}{*}{\textbf{Data}}} & \multicolumn{1}{c}{\multirow{2}{*}{\textbf{}}}  & 
			\multicolumn{1}{c}{\multirow{2}{*}{\textbf{\kv}}} &
			\multicolumn{2}{c}{\multirow{1}{*}{\textbf{Time}}} & \textbf{Storage} \\
			\cmidrule(lr){4-5} \cmidrule(lr){6-6}
			& & & \textbf{IA$^*$} & \textbf{Stellar} & \textbf{Stellar} \\
			\midrule

\multirow{2}{*}{\datasetName{neptune}} & \multirow{7}{*}{\begin{sideways}\datasetName{tri.}\end{sideways}} &
 k$_S$ & \multirow{2}{*}{5.02} & 0.66 & 0.61K \\
 &  &  k$_L$ & & 0.64 & 3.00K \\
  \cmidrule(lr){1-1} \cmidrule(lr){3-6} 
 
\multirow{2}{*}{\datasetName{statuette}} &  &  k$_S$ & \multirow{2}{*}{10.2} & 1.66 & 0.61K \\
 &  &  k$_L$ & & 1.58 & 3.01K \\
  \cmidrule(lr){1-1} \cmidrule(lr){3-6} 
 
\multirow{2}{*}{\datasetName{lucy}} &  &  k$_S$ & \multirow{2}{*}{24.8} & 4.20 & 0.61K \\ 
 &  &  k$_L$ & & 4.17 & 3.01K \\
  \midrule
  
\multirow{2}{*}{\datasetName{neptune}} & \multirow{7}{*}{\begin{sideways}\datasetName{quad.}\end{sideways}} &  k$_S$ & \multirow{2}{*}{27.5} & 2.86 & 0.41K \\
 &  &  k$_L$ & & 2.65 & 3.21K \\
  \cmidrule(lr){1-1} \cmidrule(lr){3-6} 

\multirow{2}{*}{\datasetName{statuette}} &  &  k$_S$ & \multirow{2}{*}{63.6} & 7.04 & 0.41K \\
 &  &  k$_L$ & & 7.22 & 3.22K\\
 \cmidrule(lr){1-1} \cmidrule(lr){3-6} 
 
\multirow{2}{*}{\datasetName{lucy}} &  &  k$_S$ & \multirow{2}{*}{156} & 20.4 & 0.42K \\
 &  &  k$_L$ & & 19.3 & 3.22K \\
  \midrule
  
  \multirow{2}{*}{\datasetName{bonsai}} & \multirow{7}{*}{\begin{sideways}\datasetName{tetra.}\end{sideways}} &  k$_S$ & \multirow{2}{*}{14.5} & 3.10 & 9.58K \\
   &  &  k$_L$ & & 2.81 & 18.5K \\
  \cmidrule(lr){1-1} \cmidrule(lr){3-6} 

  \multirow{2}{*}{\datasetName{vismale}} &  &  k$_S$ & \multirow{2}{*}{16.1} & 3.38 & 9.57K \\
   &  &  k$_L$ & & 3.07 & 18.2K\\
  \cmidrule(lr){1-1} \cmidrule(lr){3-6} 
  
  \multirow{2}{*}{\datasetName{foot}} &  &  k$_S$ & \multirow{2}{*}{17.3} & 3.83 & 9.62K \\
   &  &  k$_L$ & & 3.32 & 18.6K \\
 \midrule
 
\multirow{2}{*}{\datasetName{f16}} & \multirow{7}{*}{\begin{sideways}\datasetName{hexa.}\end{sideways}} &  k$_S$ & \multirow{2}{*}{145} & 11.8 & 0.83K \\
 &  &  k$_L$ & & 10.8 & 7.51K \\
 \cmidrule(lr){1-1} \cmidrule(lr){3-6} 

\multirow{2}{*}{\datasetName{san fern}} &  &  k$_S$ & \multirow{2}{*}{157} & 26.9 & 0.93K \\
 &  &  k$_L$ & & 22.0 & 8.51K \\
  \cmidrule(lr){1-1} \cmidrule(lr){3-6} 
 
\multirow{2}{*}{\datasetName{vismale}$^\otimes$} &  &  k$_S$ & \multirow{2}{*}{254} & 44.5 & 0.75K \\
 &  &  k$_L$ & & 47.7 & 7.54K \\
  \midrule
  \midrule
 
\multirow{2}{*}{\datasetName{5D}} & \multirow{7}{*}{\begin{sideways}\datasetName{prob.}\end{sideways}} & 
k$_S$ & \multirow{2}{*}{17.9} & 4.88 & 33.0K \\
 &  &  k$_L$ & & 2.73 & 243K \\
 \cmidrule(lr){1-1}  \cmidrule(lr){3-6}
 
\multirow{2}{*}{\datasetName{7D}$^\otimes$} &  & 
k$_S$ & \multirow{2}{*}{415} & 46.1 & 1.62M \\
 &  &  k$_L$ & & 35.7 & 9.01M \\
 \cmidrule(lr){1-1}  \cmidrule(lr){3-6}
 
\multirow{2}{*}{\datasetName{40D}$^\otimes$} &  & 
k$_S$ & \multirow{2}{*}{206} & 56.1 & 2.64M  \\
 &  &  k$_L$ & & 51.4 & 14.3M \\
 \midrule
 
\multirow{2}{*}{\datasetName{vismale 7D}} & \multirow{7}{*}{\begin{sideways}\datasetName{v-rips}\end{sideways}} & 
k$_S$ & \multirow{2}{*}{25.8} & 2.22 & 3.20K \\
 &  &  k$_L$ & & 2.16 & 5.04K \\
 \cmidrule(lr){1-1}  \cmidrule(lr){3-6}
 
\multirow{2}{*}{\datasetName{foot 10D}} &  & 
k$_S$ & \multirow{2}{*}{376} & 19.0 & 55.7K \\
 &  &  k$_L$ & & 16.0 & 72.6K \\
 \cmidrule(lr){1-1}  \cmidrule(lr){3-6}
 
\multirow{2}{*}{\datasetName{lucy 34D}$^\otimes$} &  & 
k$_S$ & \multirow{2}{*}{334} & 22.9 & 13.0K \\
 &  &  k$_L$ & & 23.2 & 43.8K \\
%
%
%
 \bottomrule
		\end{tabular}
	}
	}
	\label{tab:timings_vt}
\end{table}


\paragraph*{Experimental results} 

We now analyze the effectiveness of the Stellar tree representation for co-boundary extractions.
Specifically, since the main co-boundary extraction in our applications (see Section~\ref{sec:stellar_ecosystem}) 
is the restricted vertex co-boundary relation
and most of the other co-boundary extractions can be posed in terms of this primitive extraction,
we compare the performance of the Stellar tree against our implementation of the \iastar\ data structure for this query
and against the Simplex tree.
Table~\ref{tab:timings_vt} lists the extraction times and storage requirements for the vertex co-boundary relation \relation{0,\cDim} 
on our manifold (\emph{triangular}, \emph{quad}, \emph{tetrahedral} and \emph{hex}) and pure (\emph{probabilistic}) complexes
and the sum of extraction times for the restricted vertex co-boundary relations \relation{0,\tDim} for each dimension \tDim\ 
with top cells on our non-manifold (\datasetName{V-rips}) complexes. 

We first consider the influence of the bucketing threshold \kv\ for Stellar trees. 
While there is not much difference in extraction times for the two-dimensional complexes, 
larger \kv\ values lead to faster extractions for three-dimensional and non-manifold datasets in most cases.
While this comes with a slight increase in storage requirements for encoding the relation (see right column in Table~\ref{tab:timings_vt}), 
the overall storage cost per block is pretty low, requiring at most a few megabytes for the probabilistic models, and a few kilobytes in all other cases.

The \iastar\ data structure extracts co-boundary relations through a traversal along the face adjacencies of its top cells (encoded in the \relation{\tDim,\tDim} adjacency relation).
The traversal for a given vertex $v$ is seeded by one top \tDim-cell per \tDim-cluster 
(encoded by partial relation $\partialrelation{0,k}(v)$, see Section~\ref{sec:storage_other_structures}; we refer to~\cite{Canino2011IA} for more details).
Since each such traversal is run on demand, there is a negligible memory impact for this query.
Table~\ref{tab:timings_vt} demonstrates that Stellar trees are significantly faster at extracting \relation{0,k} relations, 
which can be performed in about one tenth of the time in most cases.
However, it is important to note that the Stellar tree extraction is batch-based (by leaf blocks of \h), 
and individual co-boundary extractions would likely be faster on the \iastar\ data structure.

\begin{figure}[t]
\centering
\resizebox{\columnwidth}{!}{
   \begin{tikzpicture}
      \pgfplotstableread[col sep=comma]{charts/histogram_vt_smaller_datasets.csv}\tableND
      \centering
      \begin{axis}[
			  reverse legend,
		width=7cm,
		height=8.5cm,
        enlarge y limits=0.25,
        legend cell align = left,
        legend style={
					at={(1,0.5)},
        	anchor=north east,
        },
        xmin=0, xmax=6.75,
        xbar=2pt,
        ytick=data,
        bar width=3.5pt, 
        font=\scriptsize,
        yticklabels from table= \tableND{overlap},
        y tick label style= {text width=1cm, align=right},
        nodes near coords,
        every node near coord/.append style={anchor=mid west, yshift=-0.12ex,
        	black!95, font=\scriptsize},
        xlabel = $time$ $(seconds)$,
		y=1.2cm,
        restrict x to domain*=0:7.25, 
        visualization depends on=rawx\as\rawx, 
        nodes near coords={%
        	\pgfmathprintnumber{\rawx}
        },
        after end axis/.code={ 
        	\draw [ultra thick, white, decoration={snake, amplitude=1pt}, decorate] (rel axis cs:1.05,0) -- (rel axis cs:1.05,1);
        },
        axis lines*=left,
        clip=false
      ]
      \addplot [ksStellarStyle]   table [y expr=-\coordindex, x expr={\thisrow{hit5}}, meta=str5] {\tableND};\addlegendentry{\ks\ Stellar tree}
      \addplot [klStellarStyle]   table [y expr=-\coordindex, x expr={\thisrow{hit6}}, meta=str6] {\tableND};\addlegendentry{\kl\ Stellar tree}
      \addplot [simplexTreeStyle] table [y expr=-\coordindex, x expr={\thisrow{hit4}}, meta=str4] {\tableND};\addlegendentry{Simplex tree}
      \addplot [iaStarStyle]      table [y expr=-\coordindex, x expr={\thisrow{hit1}}, meta=str1] {\tableND};\addlegendentry{\iastar}
    \end{axis}      
  \end{tikzpicture}
}
\caption{Extraction times (in seconds) for the restricted vertex co-boundary relations. 
  The top dataset is the triangle mesh used in our main comparison, 
  the second is a tetrahedral mesh with 256 thousand vertices and 1.4 million tetrahedra, 
  and the last dataset is a probabilistic-refinement CP complex with 7-dimensional top simplices.}
\label{hist:vt_small}
\end{figure}
The Simplex tree extracts co-boundary relations through a traversal of the underlying trie. 
Given a vertex \vertex, the procedure for extracting its restricted co-boundary first identifies the simplices incident in \vertex\ (i.e., its star),
and then extracts just the top simplices from the star. 
The former requires a trie traversal, with a worst-case complexity linear in the number of nodes in the trie, 
since, as stated in the GUDHI documentation~\cite{GUDHI}, this corresponds to a depth-first search of the trie starting from the node with value \vertex. 
Identifying the top simplices in the star of a vertex has a negligible cost on low dimensional meshes, while it becomes a costly operation on higher-dimensional ones, where it accounts for nearly 50\% of the overall extraction time. 
As with the \iastar, since this traversal is done on demand, this query imposes negligible memory impact.
On our experimental datasets, the Simplex tree was able to complete the extraction of restricted vertex co-boundary relations 
only on the smaller triangle mesh \datasetName{neptune}, for which it requires nearly $72$ hours. 
%
To provide a comprehensive performance comparison against the Stellar tree, 
we consider two additional smaller datasets for this query:
  a tetrahedral mesh (\datasetName{fighter2}) with 256 thousand vertices and 1.4 million tetrahedra, 
  and a probabilistic-refinement CP complex with six thousand vertices and two million top 7-simplices. 
The results, shown in Figure~\ref{hist:vt_small}, highlight the Stellar tree's significant advantage 
over the Simplex tree for restricted vertex co-boundary extraction (i.e., less than a second vs hours). 

\NOTA{
\kennyComment{These are really nice results Riccardo, but reviewers might ask why we would want the restricted co-boundary for the Simplex tree, 
    when it already encodes all cells.  Do you have a breakdown of the additional costs, if any, to filter out the non-top cells? }
}

\section{A brief tour of applications in the Stellar universe}
\label{sec:stellar_ecosystem}

\NOTA{
    \felleComment{shall we add back a short description of how to generate topological data structures with Stellar tree?}
    \kennyComment{I don't think this is necessary.}
}

\begin{figure*}[ht]
	\centering
	\hfil
	\subfloat[60\% simplification]{    
		\resizebox{.49\columnwidth}{!}{
			\includegraphics{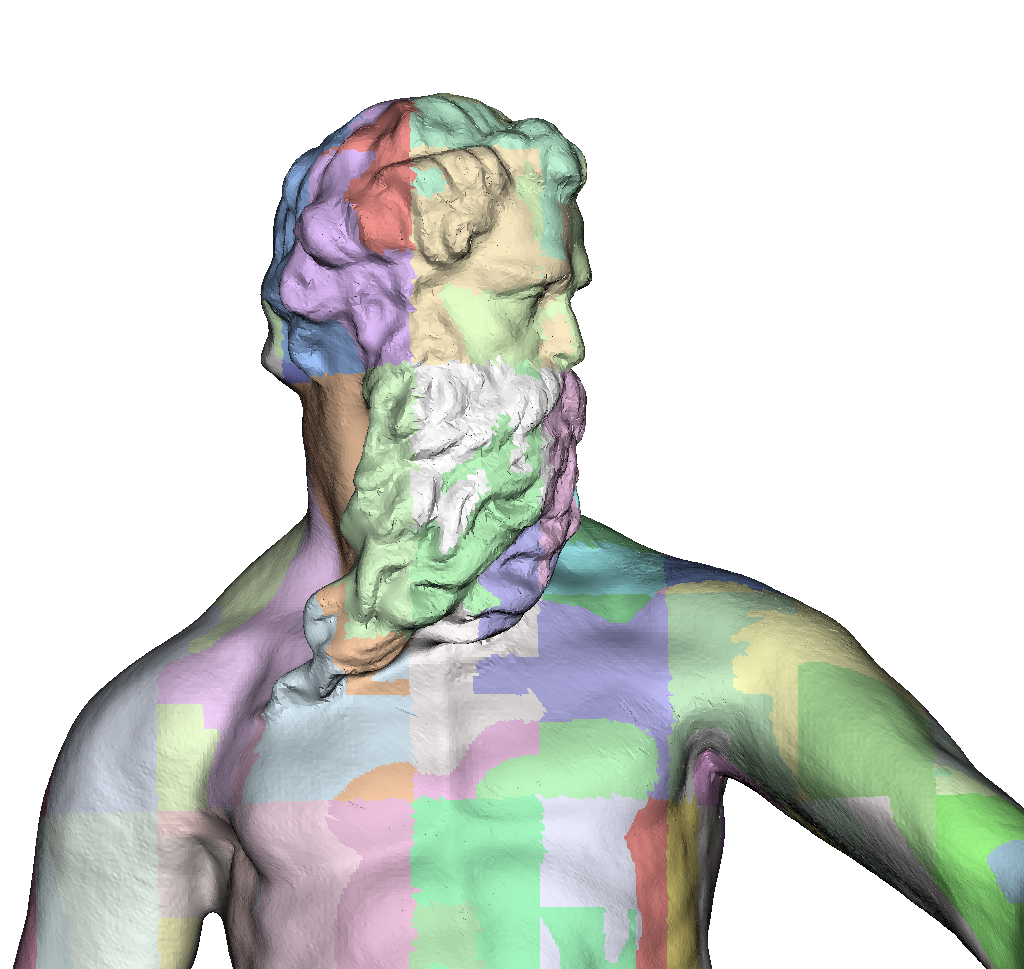}
		}
	}
	\hfil
	\subfloat[90\% simplification]{    
		\resizebox{.49\columnwidth}{!}{
			\includegraphics{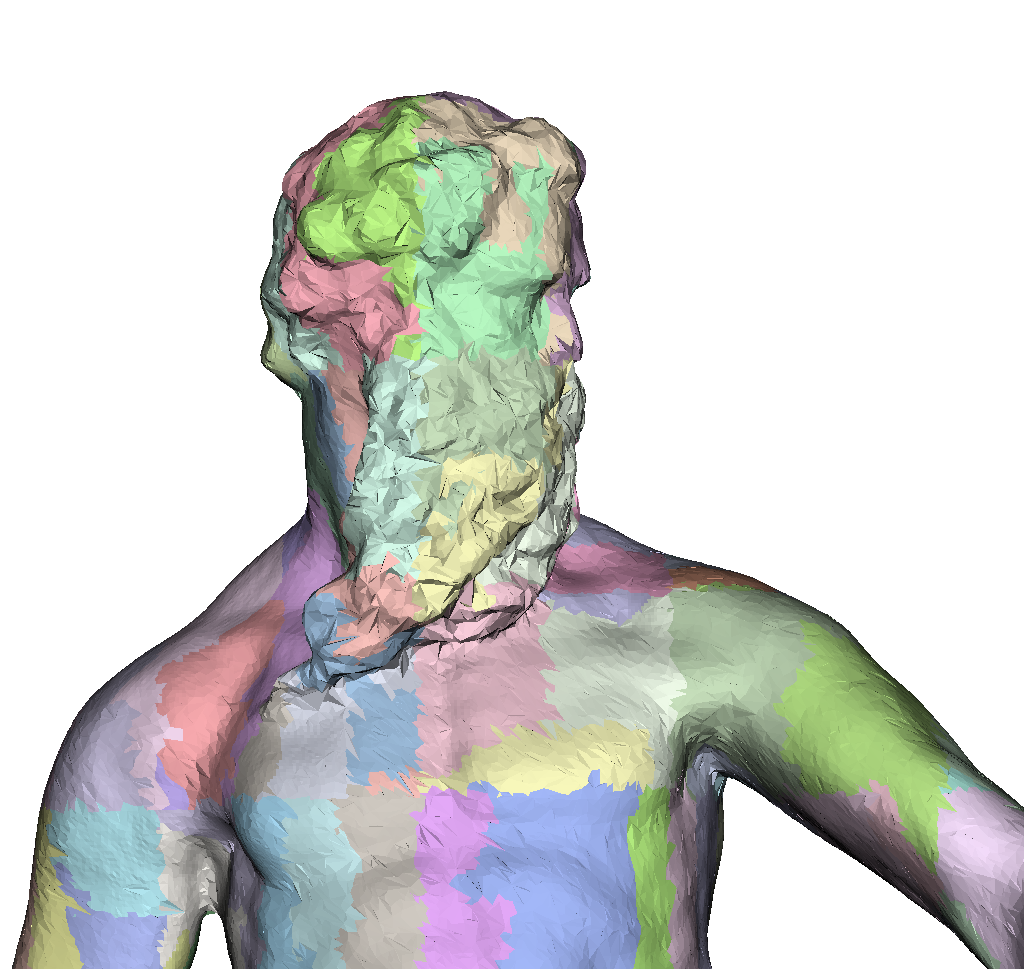}
		}
	}
	\hfil
	\subfloat[99\% simplification]{    
		\resizebox{.49\columnwidth}{!}{
			\includegraphics{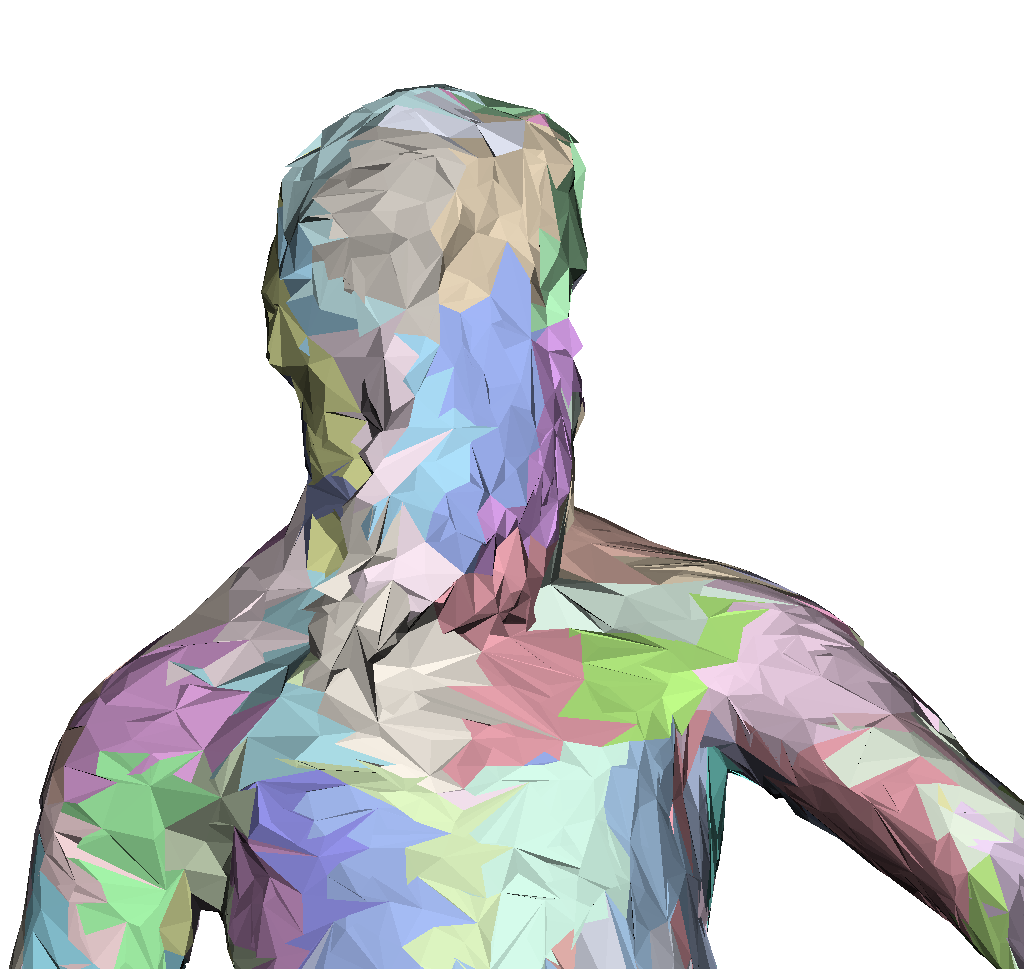}
		}
	}
	\hfil
	\subfloat[99.9\% simplification]{    
		\resizebox{.49\columnwidth}{!}{
			\includegraphics{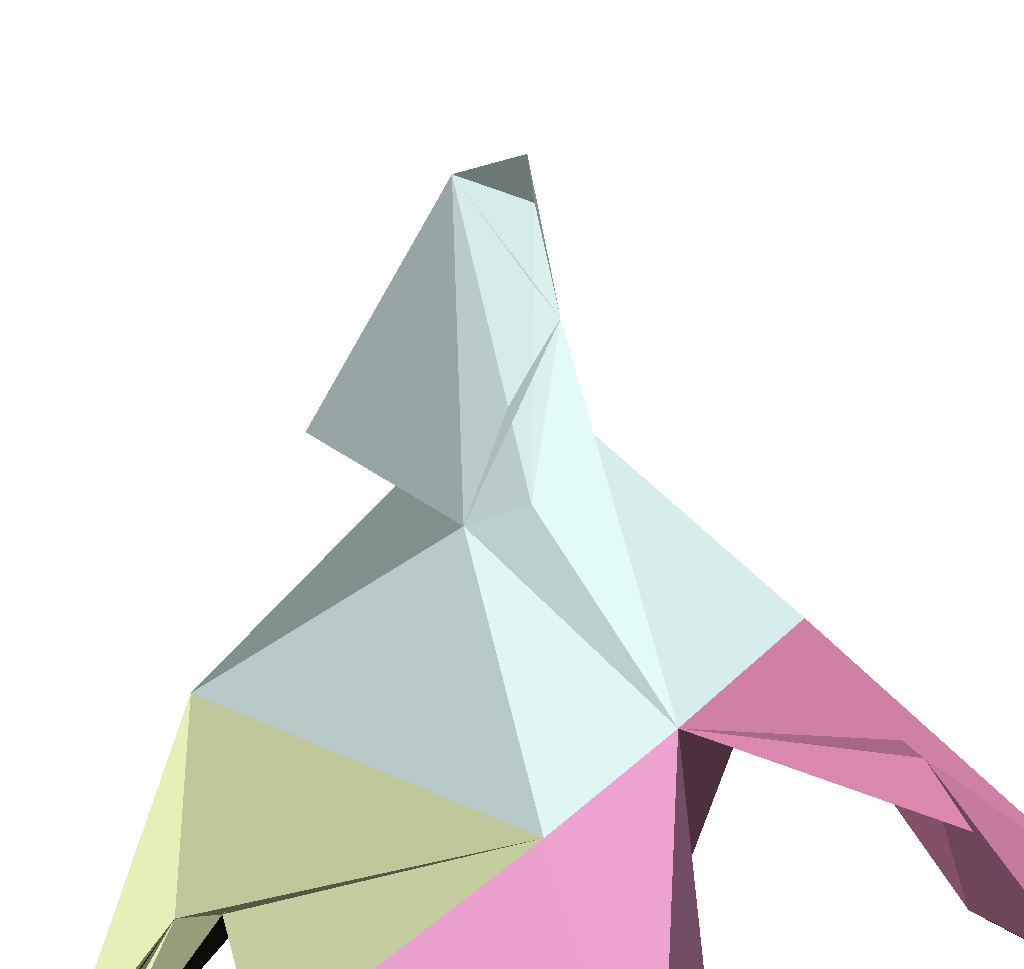}
		}
	}
	\caption{Homology-preserving edge-contractions retain topological invariants of a complex, such as its number of loops,
    even after extreme simplification. 
    (a)-(d) Genus-3 \datasetName{neptune} complex at various simplification percentages. 
    Triangles within the same octree block of the Stellar tree have the same color.
  }
	\label{fig:homology_simplified_neptune}
\end{figure*}    

Stellar decompositions and Stellar trees have been successfully applied in several mesh processing applications.
In this section, we provide a high-level overview of several such applications over large CP complexes with a focus on how
Stellar trees uniquely benefit the application.
%
%
As we will describe, each such application utilizes local topological data structures designed for
the underlying application. Due to the streamed processing approach discussed in Section~\ref{sec:general_strategy}, the storage
requirements for these data structures are proportional to the geometry indexed within a leaf block of the tree
and the generation costs are amortized over all processed cells in the block.

\NOTA{
  \felleComment{for now this section is just a collection of the abstract of the papers from which we have to extract some descriptions}
  \felleComment{current outline: 
      * simplification (tetrahedral mesh edge contraction + homology preserving edge contraction in arbitrary dimension) 
          (PR-star \cite{Weiss2011PR} + CGF papers \cite{Fellegara2020Efficient}) 
      * geometric and topological mesh validation (IMR paper \cite{Fellegara2016Efficient}); 
      * Morse Theory related applications (gradient computation, morphological feature extractions, topological simplification) 
          (EuroVis \cite{Weiss2013primaldual} 
          + Sigspatial IWGS paper \cite{DeFloriani2012spatial} 
          + 2D stellar tree (sigspatial-14 paper) \cite{Fellegara2014Efficient}; 
      * topological data structure generation (old version of this paper + Ph.D. thesis to use as reference \cite{Fellegara2015spatio})}
      * social network analysis (Sigspatial LSBN paper \cite{Fellegara2016Analysis}); 
  }

\subsection{Validation of geometric and topological properties}

  Many popular topological mesh data structures are valid only for a restricted class of complexes due to assumptions they exploit
  in their encodings, such as the cardinality of the adjacency relation among top cells.
  For example, popular edge-based and adjacency-based data structures, 
    such as the \emph{half-edge}~\cite{Mantyla1988Introduction,CGAL,OML15}, 
     \emph{Corner-Table}~\cite{Rossignac20013D}, SOT~\cite{Gurung2009SOT} and IA~\cite{Paoluzzi1993Dimension,Nielson1997Tools},
  require the underlying complex to be pseudo-manifold.
  
  While one can verify such topological conditions using local checks on the \emph{star} or \emph{link} of the vertices of the complex,
  it can be infeasible to reconstruct such relations on large meshes without the aid of an efficient topological data structure.
  On the other hand, global approaches that directly build the required relations do not scale to larger complexes.
  
  In contrast, Stellar trees are ideally suited to verify topological properties of large CP complexes, even in memory-limited environments,
  since each leaf block of the Stellar tree only requires a list of vertices and their incident top cells 
  (i.e., those in the star of the vertices).
  %
  A simple local topological verification operation was utilized in~\cite{Weiss2011PR} 
  to mark boundary vertices of a tetrahedral mesh by checking properties of its link, such as its Euler number.
  This was extended in~\cite{Fellegara2016Efficient} 
  to a full suite of topological checks on a CP complex, implemented using global Stellar tree traversals.
  In particular, a graph traversal of the 1-skeleton was used, in conjunction with a \emph{Union-Find} data structure~\cite{Tarjan1975},
  to count the connected components of a pure simplicial complex.
  A similar traversal of the 1-skeleton of its dual complex (i.e., the graph of the \cDim-adjacency relation) 
  was used to verify the \cDim-connectedness of the complex and whether it is pseudo-manifold.
  Simplified checks for combinatorial manifolds (when applicable) were then performed on the links of each vertex 
  to check that they were locally homeomorphic to (\cDimMinusOne)-spheres (for internal vertices) 
  or to (\cDimMinusOne)-balls (for boundary vertices).
  \NOTA{ Since each block of the tree only needs to be processed a single time, this application did not require/benefit from a cache. }
  \NOTA{ Such custom topological data structures and graph traversals were easy to generate on Stellar trees... }
  \NOTA{ Comparison to IA*? }

\subsection{Topology-preserving simplification}

  One of the earliest applications of the Stellar tree (actually, its predecessor, the PR-star octree~\cite{Weiss2011PR}) 
  was to accelerate a mesh simplification algorithm for tetrahedral meshes based on \emph{edge collapses}.
  An edge collapse is a local topological operation defined in terms of the stars of an edge's two vertices. 
  This operation identifies the pair of vertices along an edge, removes all tetrahedra in the star of that edge
  and updates the mesh connectivity within this local region~\cite{Cignoni2004Selective}.
  Edge collapses are valid when they satisfy the so-called \emph{link condition}~\cite{Dey1999Topology}, 
  which consists of local checks on the links of the edge and its vertices.

  The simplification procedure in~\cite{Weiss2011PR} 
  was implemented as an iterative process that alternated between: 
  (i) streaming through the Stellar tree blocks, where it collapsed eligible candidates, 
  and (ii) rebuilding the Stellar tree's index over the simplified mesh.
  It used discrete distortion~\cite{Mesmoudi2008Discrete} and a quadric error metric~\cite{Garland1997Surface} 
  to organize eligible edges into a priority queue.
  Applying this simplification algorithm to the leaf blocks of a Stellar tree rather than to the entire mesh provides 
  a significant space savings due to the reduced sizes of the edge queues. In many cases, the simplification was 10-50\% faster 
  and required only 0.1\% of the additional memory for auxiliary data structures. Moreover, this speedup was more pronounced as the mesh size increased.

More recently, Stellar trees have been used to perform \emph{homology-preserving} edge-contractions
on general simplicial complexes of arbitrary dimension~\cite{Fellegara2020Efficient}. 
The core edge contraction algorithm 
was implemented using a custom local topological data structure 
over the \emph{top star} of the vertices and edges within each leaf block of the tree, 
similar to the \emph{restricted co-boundary relations} \relation{0,\tDim} and \relation{1,\tDim} (c.f.\ Section~\ref{sec:coboundary_extraction}).
To avoid regenerating these topological relations, it utilized an LRU cache for the expanded leaf blocks as it traversed the tree.
In this mesh simplification application, Stellar trees were applied to datasets with dimensions up to 70 and were able to remove 
more than 90\% of the simplices of the mesh, significantly reducing the dimensionality of the complex
while preserving important topological invariants of the dataset. Compared to existing state-of-the-art data structures 
for edge contractions~\cite{Attali2012Efficient}, Stellar trees were able to simplify complexes using comparable or less runtime
and memory in all cases, and requiring significantly less memory and/or processing time in several cases. 
Notably, in one case, the Stellar tree was able to successfully complete the simplification process in less than 30 minutes,
while~\cite{Attali2012Efficient} did not complete after more than 24 hours.
Figure~\ref{fig:homology_simplified_neptune} shows simplified versions of the genus-3 
\datasetName{neptune} triangle mesh. Each simplified mesh preserves the homology of the complex, such as its number of 
connected components, loops and cavities.
We note that the above application incorporates only topological considerations into its simplification error metric.
Incorporating mesh quality considerations into the error metric could significantly improve the mesh quality~\cite{Garland1997Surface}.

\NOTA{
A similar approach was taken in~\cite{Fellegara2016Analysis} to analyze large geo-localized social networks.
In that work, the social network was represented in terms of its \emph{maximal cliques}, 
i.e., sets of mutually related entities, corresponding to top simplices over the network's flag complex.
The Stellar tree was built over the 2D embedding provided by the geospatial locations of the entities
and simplified using homology-preserving edge-contractions, enabling a study of the network's topological structure 
on a significantly reduced dataset.
}

\subsection{Shape analysis and morphological feature extraction}

\begin{figure}[t]
	\centering
		\resizebox{\columnwidth}{!}{
			\includegraphics{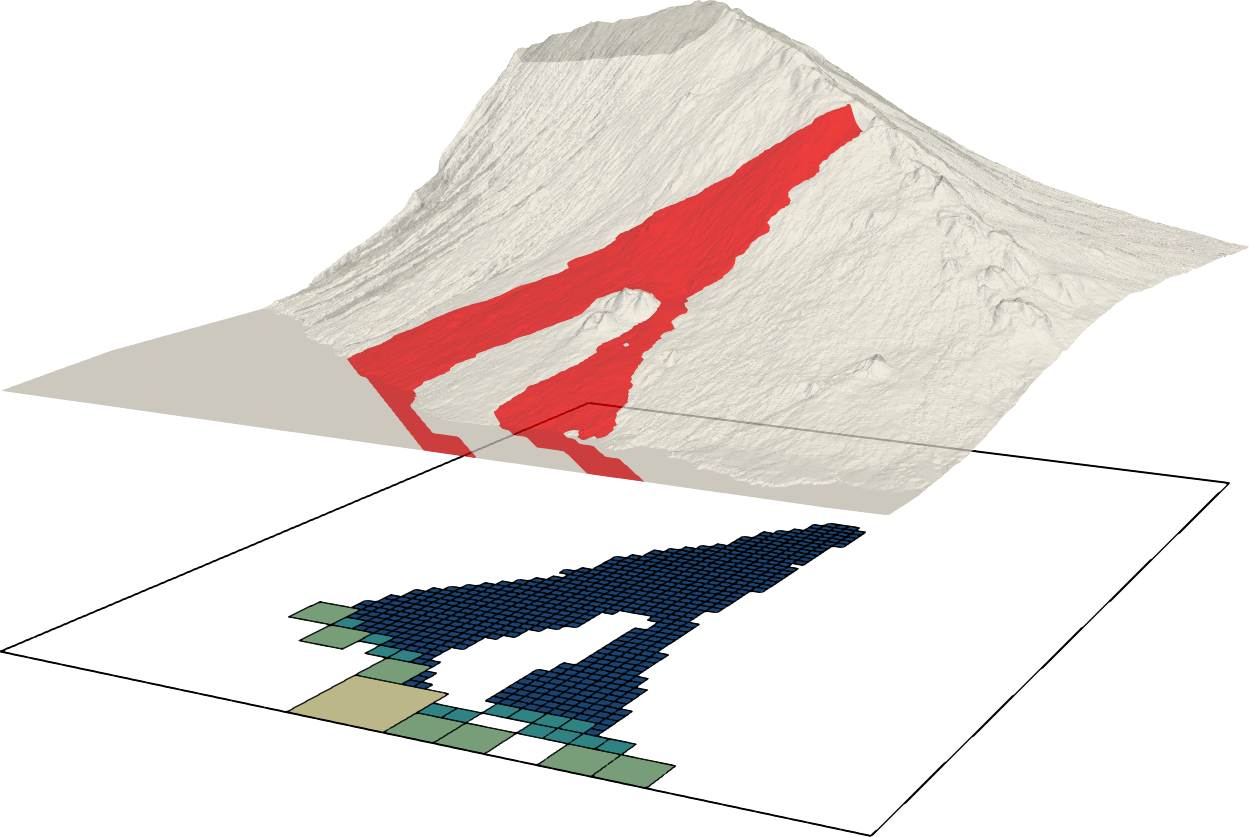}
		}
	\caption{Manifolds of a Morse complex can span vast regions of the domain. 
    This subset of the 4M triangle \datasetName{Maui} terrain dataset highlights the 73K triangles of a single 2-manifold of the Morse complex (shown in red on top) 
    along with the Stellar tree blocks indexing these triangles (rectangles on bottom, colored by quadtree depth).
  }
	\label{fig:maui_desc_2_manifold}
\end{figure} 

While topological validation and simplification operations can be implemented in terms of local operations on the star of a vertex,
shape analysis applications, such as watershed analysis~\cite{Roerdink2000Watershed} and visibility queries on terrain datasets~\cite{Bittner2003Visibility, DeFloriani2003Algorithms} often require algorithms that are seeded locally and span vast interwoven regions of the complex.  
%
This section discusses how Stellar trees have aided in the generation and simplification of the \emph{discrete Morse gradient field} 
and of the associated \emph{Morse} and \emph{Morse-Smale} complexes 
of triangulated terrains~\cite{Fellegara2014Efficient} 
and of tetrahedralized volumetric data~\cite{Weiss2013primaldual}. 

The discrete Morse gradient field~\cite{Forman1998Morse} is composed of \emph{arrows} (ordered pairs) between incident cells of the complex
and can be computed locally using scalar values associated with cells incident in the star of a vertex~\cite{Robins2011Theory}. 
Since the encoding of~\cite{Weiss2013primaldual} compactly encodes
the discrete Morse gradient field as a scalar field on the top simplices of the complex,
the latter can be computed using a local traversal of a Stellar tree's blocks.
Compared to an IA implementation, Stellar trees were able to extract the discrete Morse gradient of scalar fields defined 
over tetrahedral meshes in about half the time (see~\cite{Weiss2013primaldual} for details).
While the experiments in~\cite{Weiss2013primaldual} were performed on \datasetName{vertex-compressed} Stellar trees, 
yielding a 30\% storage savings over the IA data structure, a \datasetName{compressed} Stellar tree encoding would 
likely provide a 50\% total memory savings while maintaining similar performance improvements.

Extracting the Morse complex from a Stellar tree-based encoding is more complicated since it involves traversing the directed acyclic graphs (DAGs) induced 
by the discrete Morse gradient field's arrows.
%
Specifically, in the encoding of~\cite{Weiss2013primaldual}, each $k$-dimensional critical point of the discrete Morse gradient field
corresponds to a $k$-cell of the $d$-dimensional Morse complex.
The disjoint regions of influence of each such critical point, referred to as the \emph{$k$-manifolds} of the Morse complex,
are extracted by traversing the DAG rooted at a given critical point of the discrete Morse gradient field.
Since each such graph traversal can visit the blocks of a Stellar tree multiple times, an LRU cache was used in~\cite{Weiss2013primaldual} to support global extraction algorithms for each $k$-manifold of the Morse complex.

Further, since the extraction algorithm for each dimension's manifolds depends on different topological connectivity relations,
Morse complex extraction benefits from the Stellar tree's ability to generate customized local topological data structures. 
For example, since extracting the 2-manifolds from a tetrahedral complex requires only the $\relation{2,2}$ adjacency relation,
its extraction was optimized by directly starting from $\relation{2,2}$, rather than the $\relation{3,3}$ adjacency relation, as in the IA data structure.

This approach was extended to terrain datasets in~\cite{Fellegara2014Efficient}, 
which also introduced a persistence-based simplification algorithm for noise removal.
Figure~\ref{fig:maui_desc_2_manifold} highlights the 73K triangles in the largest 2-manifold of the Morse complex for the \datasetName{Maui} terrain dataset (in red)
and the blocks of the Stellar tree that were visited when extracting this region (blue-green squares).
While manifold extraction and persistence-based simplification operations were slightly more expensive 
than their IA counterparts for volumetric~\cite{Weiss2013primaldual} and terrain~\cite{Fellegara2014Efficient} datasets, the Stellar tree's vast memory savings and hierarchical encoding open the door to efficient parallel implementations on huge datasets,
which we hope to explore in future work.

\section{Concluding remarks}
\label{sec:stellar_conclusions}


We have introduced the Stellar decomposition as a model for topological data structures over \emph{Canonical-Polytope (CP)} complexes, 
a class of complexes that includes simplicial complexes and certain classes of cell complexes, like quadrilateral and hexahedral meshes.
Stellar decompositions cluster the vertices of a complex into \emph{regions} that contain sufficient information to locally reconstruct the \emph{star} of their vertices.
%
The model is agnostic about the domain of the complex (e.g., manifold, pure, non-manifold)
and we have demonstrated the scalability of this model to large mixed-dimensional datasets in high dimension.

\NOTA{ \outlineComment{Key contribution -- exploit spatial locality through reindexing and SRE compression of lists} }
We introduced the Stellar tree as a concrete realization of the Stellar decomposition model over spatially embedded CP complexes.
Stellar trees couple a spatial index \h\ decomposing the complex's embedding space 
with a simple tuning parameter that limits the number of vertices indexed by a leaf block.

Stellar trees effectively exploit the spatial coherence of a CP complex \sC\ by using the clustering structure of \h\
to reorder the arrays of top cells of \sC\ and to compress the resulting ranges of sequential indexes 
within the lists of vertices and top cells in the leaf blocks of \h.
We have demonstrated over a wide range of datasets that this process typically produces \datasetName{compressed} Stellar trees 
that are only 1-10\% larger than the original indexed base mesh for \sC\ while still retaining sufficient information
to efficiently reconstruct all topological connectivity relations.
The source code for our Stellar tree implementation will be released in the public domain.

In terms of storage size, Stellar trees compare quite favorably with state-of-the-art topological data structures.
They are consistently half the size of their \iastar\ data structure counterparts~\cite{Canino2011IA} 
and one to two orders of magnitude smaller than their Simplex tree counterparts~\cite{Boissonnat2014simplex}.
This is especially notable for high dimensional Vietoris-Rips complexes, a target application for the Simplex tree,
for which Stellar trees have very low overhead.
While Stellar trees support a much broader class of complexes, 
they have similar storage requirements as the dimension-specific SOT data structure~\cite{Gurung2009SOT,Gurung2010SOT}, 
which supports only static pseudo-manifold triangle and tetrahedral complexes.
In future work, it would be interesting to compare the Stellar tree against top-based extensions of the Simplex tree, 
such as the $MST$ and the $SAL$~\cite{Boissonnat2017}, if public-domain implementations become available. 

Despite the simplicity of their leaf block representation, Stellar trees provide a great deal of flexibility to customize
the structure and layout of their \emph{expanded} topological data structures to meet the needs of a given application.
Such data structures are typically constructed by composing several local topological incidence and adjacency relations.
We described efficient algorithms for reconstructing these relations within the subcomplex indexed by the leaf blocks of a Stellar tree
and demonstrated the advantages of this approach compared to similar algorithms on the \iastar\ and Simplex tree data structures.
Stellar trees can also be used as an intermediary representation 
to generate topological data structures in a memory-constrained environment. 
For example, we used Stellar trees to generate
\iastar\ and Simplex tree representations for several of our larger complexes in Section~\ref{sec:storage}
(as we discuss in the Appendix \ref{sec_app:stellar_build_structures}). 
We also provided an overview of several mesh processing applications, 
  ranging from mesh validation, to topology and shape preserving simplification and morphological analysis that have 
  benefited from the Stellar trees representation.

One direction of future work would involve extending the Stellar tree representation to support a broader class of cell complexes. 
For example, it would not be difficult to extend support to indexed polyhedral cell complexes which define their cells in terms of their 
boundary polyhedral faces which are, in turn, defined by oriented lists of vertex indices~\cite{Muigg11}.

  \NOTA{
    A possible future investigation could be the definition of new local encodings for the connectivity within each leaf block. 
    We note that, in our experimental evaluations, the vast majority of a Stellar tree's storage requirements are due to the 
    encoding of the connectivity of the complex. This is especially large in our higher-dimensional complexes, where the connectivity 
    accounts for 99\% of the overall storage, while on lower dimensional complexes, it accounts for around 85\% of the storage. 
    The core idea is to infer this encoding directly from the exploited spatial coherence of the Stellar tree. 
    This will lead to a representation in which each leaf block \B\ fully encodes the sub-complex formed by the vertices 
    and the top cells associated with it, but still with the capability to recover the missing information from the neighbor leaf blocks of \B.
   }
	
\NOTA{ \outlineComment{Future directions -- Stellar tree in parallel/distributed context -- add discussion about relation to distributed mesh data structures.} }
Another avenue for investigation is to extend our processing algorithms for parallel, distributed and/or out-of-core environments, which could be used for applications like multicore homology computation~\cite{Lewis2014Multicore} on point cloud data.
The Stellar tree's compact leaf block representation is already geared towards a parallel execution pattern since each block already has sufficient resources to query the connectivity of its local subcomplex.
%
Preliminary results along this line look promising.
A simple unoptimized OpenMP~\cite{OpenMP} adaptation of boundary and restricted vertex co-boundary queries yielded a 3-4x speedup compared to our serial approach on our 6 core machine.

Finally, while Stellar trees require their underlying complex to be spatially embedded, there is no such restriction on the Stellar decomposition model.
Thus, we plan to investigate Stellar decompositions for \emph{abstract} $CP$ complexes, such as simplicial complexes representing social networks. 
Social network representation and processing poses new challenges in the social big data domain, such as the identification of key-players and \emph{communities} in the dataset, as well as extracting topological properties of the network, like its homology or $k$-connectivity. 
Due to the irregularities of non-spatial datasets, one key challenge would be to define efficient decompositions (i.e., with a low average spanning number $\Chi$) using only the complex's connectivity information.
A preliminary attempt for geolocalized social networks can be found in~\cite{Fellegara2016Analysis}, 
where the social network was represented in terms of its \emph{maximal cliques}, 
i.e., sets of mutually related entities, corresponding to top simplices over the network's flag complex.
The Stellar tree was built over the 2D embedding provided by the geospatial locations of the entities
and simplified using homology-preserving edge-contractions \cite{Fellegara2020Efficient}, 
enabling a study of the network's topological structure on a significantly reduced dataset.

\section*{Acknowledgments}
This  work has been developed while Riccardo Fellegara was with  University of Maryland at College Park, USA.
This work has been partially supported by the US National Science Foundation under grant number IIS-1910766 
and	by the University of Maryland under the 2017-2018 BSOS Dean Research Initiative Program. 
It has also been performed under the auspices of the U.S. Department of Energy by Lawrence Livermore National Laboratory under Contract DE-AC52-07NA27344, and of the German Aerospace Center (DLR) under Grant DLR-SC-2467209.
%
Datasets are courtesy of 
the \emph{Volvis} repository (\datasetName{bonsai}, \datasetName{f16} and \datasetName{foot})~\cite{Volvis},
the \emph{Volume Library} (\datasetName{vismale})~\cite{VolLib}, 
\emph{CMU Unstructured Mesh Suite} (\datasetName{san fernando})~\cite{CMUMeshSuite},
\emph{Aim@Shape} repository (\datasetName{lucy}, \datasetName{statuette} and \datasetName{neptune})~\cite{AIM@shape},
\emph{Virtual Terrain Project} ($VTP$) (\datasetName{maui})~\cite{Virtual}, 
and Claudio Silva (\datasetName{fighter2}).

\bibliographystyle{apalike}
\bibliography{main_bibliography}

\appendix
\section*{Appendix}
\label{sec:appendix}	
\begin{figure*}[t]
	\centering
	\subfloat[Base mesh \sC := (\sCV, \sCT)]{
		\resizebox{.25\textwidth}{!}{
			\includegraphics{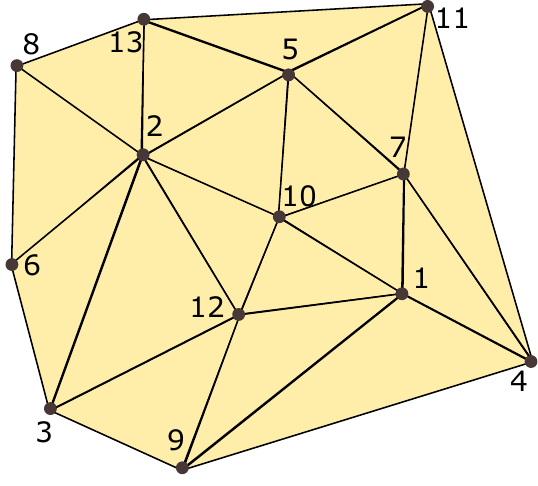}
		}
		\label{fig:generation_base_mesh}
	}
	\hfil
	\subfloat[Vertices inserted into \D] { 
		\resizebox{.25\textwidth}{!}{
			\includegraphics{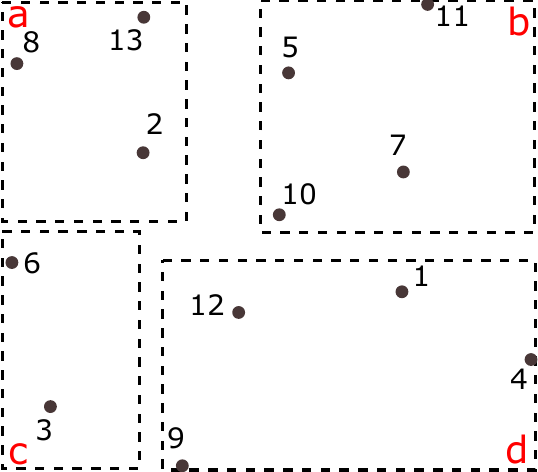}
		}
		\label{fig:generation_insert_verts}
	}
	\hfil
	\subfloat[Vertices reindexed]{
		\resizebox{.25\textwidth}{!}{
			\includegraphics{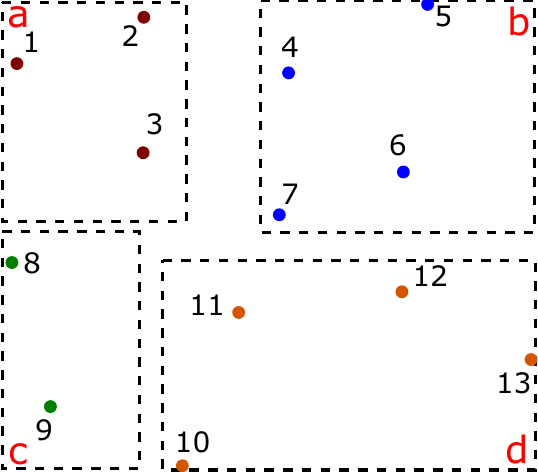}
		}
		\label{fig:generation_reindex_verts}
	}
	\caption{Generating a partition \D\ over the vertices of a triangle mesh (a). 
		After inserting the vertices (b), we reindex \sCV\ according to the regions of \D\ (c).} 
	\label{fig:stellar_dec_vertices_reindexing}
\end{figure*}

\section{Generating a Stellar decomposition}
\label{sec_app:stellar_dec_generation}

In this section, we describe how to generate a \datasetName{compressed} Stellar decomposition from an indexed CP complex \sC\
%
and a given partition \D\ on the vertices of \sC. This process consists of three main phases:
\begin{enumerate}
	\item reindex the vertices of \sC\ following a traversal of the regions of \D\ and SRE-compress the \vR\ arrays;
	\item insert the \topcpcells\ of \sC\ into \D;  
	\item reindex the \topcpcells\ of \sC\ based on locality within common regions of \D\ and SRE-compress the regions \tR\ arrays. 
\end{enumerate}

As it can be noted, the generation process ignores how the partitioning on the vertices is obtained, since this step is defined by the data structure instantiating a Stellar decomposition.
The reindexing of the vertices follows a traversal of the regions of \D\ in such a way that 
all vertices mapped to a region have a contiguous range of indices in the reindexed global vertex array \sCV\ (as detailed in~\ref{sec:verts_reordering}).
Figure~\ref{fig:stellar_dec_vertices_reindexing} illustrates a reindexing of the vertices of a triangle mesh in the plane into a decomposition
defind by four rectangular regions (dashed lines).

We then insert each \ktopcp\ \simplex, with index \tIndex\ in \sCTk, into all the regions of \D\ that index its vertices. 
This is done by iterating through the vertices of \simplex\  and inserting \tIndex\ into the \tR\ array
of each region \R\ whose vertex map \PhiMapVert(\R) contains at least one of these vertices.
As such, each \ktopcp\ \simplex\ appears in at least one and at most $|\relation{\tDim,0}(\simplex)|$ regions of \D.
Due to the vertex reindexing of step 1, this operation is extremely efficient. 
Determining if a vertex of a given cell lies in a block requires only a range comparison on its index \vIndex.

Finally, we reindex the \topcp\ arrays \sCT\ to better exploit the locality induced by the vertex-based partitioning and compress the local \tR\ arrays using a sequential range encoding over this new index.
The reindexing and the compression of the \topcpcells\ is obtained following a traversal of the regions of \D\ in such a way that all \topcpcells\ mapped from the same set of regions have a contiguous range of indices in the reindexed arrays \sCT.
This last step is detailed in~\ref{sec:tops_reordering} and in~\ref{sec_app:tops_reordering_long}.
As we demonstrate in the paper, this compression yields significant storage savings in a wide range of mesh datasets.  

\begin{algorithm}[t]	
	\caption{ $\AlgoName{compress\_and\_reindex\_vertices}(\D,\sC)$ }
	\label{alg:main_vertices_reordering}				
	\begin{algorithmic}[1]			
		\Input {\D\ is the decomposition defined on the vertices of \sC}
		\Input {\sC\ is the CP complex}
		\Variable { $v\_perm$ is an array containing new vertex indices} 
		\Variable {$current$ references the current vertex id}
		\Variable {\vstart\ and \vend\ are the range of vertex indices in region \R}

		\State{current $\leftarrow$ 0}
		
		\LineComment {\textbf{Step 1}: Generate and apply new vertex index ranges}
		\ForAll{regions \R\ \textbf{in} \D}
			\State { \vstart\ $\leftarrow$ current } \SideComment{\vstart\ is the first vertex index in \R}
			\ForAll {vertices $\vertex$ \textbf{in} \PhiMapVert(\R) (with index \vIndex\ in \sCV)} 
				\State {v\_perm[\vIndex] $\leftarrow$ current}
				\State {current $\leftarrow$ current $+ 1$}
			\EndFor
			\State { \vend\ $\leftarrow$ current } \SideComment{\vend\ is the first vertex index outside \R}
		\EndFor
				
		\LineComment{\textbf{Step 2}: Update \relation{k,0} relation for all \ktopcpcells\ in \sC}
		\ForAll {\topcpcells\ \simplex\ \textbf{in} \sCT} 
			\For {j $\leftarrow$ 0 \textbf{to} $|\relation{k,0}(\simplex)|$} 
				\LineComment{the $j$-th entry of \relation{k,0}(\simplex) has value \vIndex}
				\LineComment{v\_perm[\vIndex] contains the new index of vertex \vertex}
				\State{\relation{k,0}(\simplex)[j] $\leftarrow$ v\_perm[\vIndex]}
			\EndFor
		\EndFor  
		
		\LineComment {\textbf{Step 3}: Update the vertex array in \sCV} 
		\LineComment {(see Algorithm \ref{alg:update_array} in \ref{sec_app:tops_reordering_long})}
		\State { \AlgoName{permute\_array}(\sCV,v\_perm) }
	\end{algorithmic}
\end{algorithm}

\subsection{Reindexing and compressing the vertices}
\label{sec:verts_reordering}

After generating the partition \D\ and the vertex map \PhiMapVert\ for the Stellar decomposition, 
we reindex the vertex array \sCV\ to better exploit the coherence induced by \D.
At the end of this process, each region of \D\ has a consecutive range of indices within the global vertex array \sCV, and thus it trivially compresses under SRE to two values per block.
We denote the starting and ending vertex indices as \vstart\ and \vend, respectively.

\NOTA{
  \felleComment{since step 1 is way simpler for a stellar decomposition (we just have regions.. no more leaves and internal blocks), I
  integrated it within Algorithm \ref{alg:main_vertices_reordering}}
}
This reindexing procedure is organized into three major steps, as outlined in Algorithm~\ref{alg:main_vertices_reordering}.
The first step performs a traversal of the regions, which generates new indices for the vertices in \sC.
For a region \R, it generates a contiguous range of indices for the vertices in \R.
%
For example, in Figure~\ref{fig:stellar_dec_vertices_reindexing}, after executing step 1 on region \textit{b}, we have $\vstart = 4$ and $\vend= 7$. 

The new indexes are then incorporated into mesh \sC\ by updating the vertex indices in \relation{\tDim,0} relations for all \ktopcells\ in \sCTk\
(see step 2 of Algorithm~\ref{alg:main_vertices_reordering}) and then permuting the vertices 
(detailed in step 3 of Algorithm~\ref{alg:update_array} in \ref{sec_app:tops_reordering_long}).  
These updates take place in memory without requiring any extra storage.

%
%

\subsection{Reindexing and compressing the \topcpcells}
\label{sec:tops_reordering}
\NOTA{\felleComment{here I removed/rephrased any reference to \emph{space}.. no more spatially coherent, but just coherent}}

\begin{algorithm}[t]		
	\caption{ $\AlgoName{compress\_and\_reindex\_cells}(\D,\sC)$ }
	\label{alg:main_tops_reordering}				
	\begin{algorithmic}[1]			
		\medskip
		\Input {\D\ is the decomposition defined on the vertices of \sC}
		\Input {\sC\ is the CP complex}
		\Variable {$M$ is an associative array mapping an integer identifier to each unique tuple of regions from \D}
		\Variable {$I$ is an array associated with the unique tuples of regions in $M$}
		\Variable {$t\_position$ is an array associated with the \topcpcells}
		
		\LineComment {\textbf{Step 1}: find unique tuples of regions and their counts} 
		\LineComment {(see Algorithm \ref{alg:get_region_top_association} in \ref{sec_app:tops_reordering_long})}
		\State {\AlgoName{extract\_tuples}(\D,\sC,M,I,t\_position) } 
		
		\LineComment {\textbf{Step 2}: find new position indices for \topcpcells}
		\LineComment {(see Algorithm \ref{alg:get_tops_reorderd_indexes} in \ref{sec_app:tops_reordering_long})}
		\State {\AlgoName{extract\_cell\_indices}(I,t\_position) } 

		\LineComment {\textbf{Step 3}: reorder and SRE compress the \topcpcells\ arrays}
		\LineComment {(see Algorithm \ref{alg:compress_cells_representation} in \ref{sec_app:tops_reordering_long})}	
		\State {\AlgoName{compress\_cells}(\D,t\_position) } 
		
		\LineComment {\textbf{Step 4}: update the \topcpcells\ array in \sC} 
		\LineComment {(see Algorithm \ref{alg:update_array} in \ref{sec_app:tops_reordering_long})}
		\State {\AlgoName{permute\_array}(\sCT,t\_position) } 
	\end{algorithmic}
\end{algorithm}

After inserting the \topcpcells\ of \sCT\ into \D, 
we reorder the \topcpcells\ array \sCT\ based on the partitioning
and apply SRE compaction to the region arrays to generate our \datasetName{compressed} encoding.
%
This reindexing exploits the coherence of \topcpcells\ that are indexed by the same set of regions,
translating the proximity in \D\ into \emph{index} proximity in \sCT.
This procedure is organized into four main phases, as shown in Algorithm~\ref{alg:main_tops_reordering}.
A detailed description can be found in \ref{sec_app:tops_reordering_long}.

\begin{figure}[t]
	\centering
	\subfloat[Original triangles]{
		\resizebox{.22\textwidth}{!}{
			\includegraphics{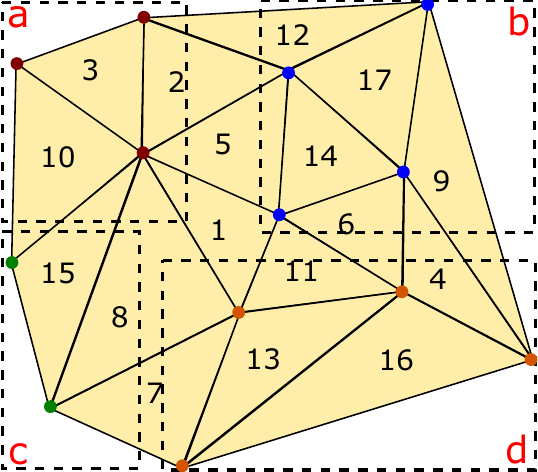}
		}
	}
	\hfil
	\subfloat[Reindexed triangles]{
		\resizebox{.22\textwidth}{!}{
			\includegraphics{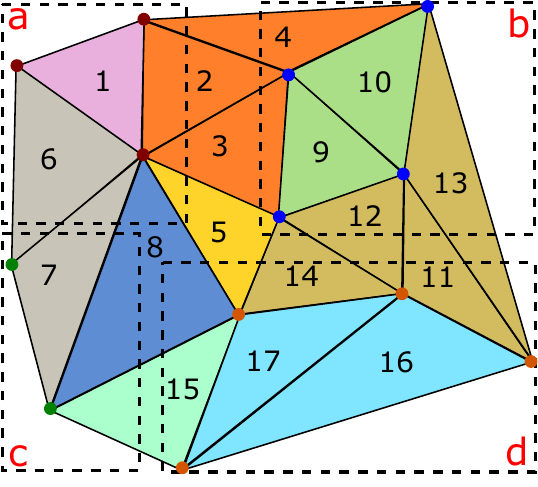}
		}
	}
	\caption{Top cell indices before (a) and after (b) tuple-based reindexing. }
	\label{fig:reindexing}
\end{figure}

The \AlgoName{extract\_tuples} procedure (see Algorithm \ref{alg:get_region_top_association} in \ref{sec_app:tops_reordering_long}), 
traverses the regions of \D\ to find the tuple of regions $\tuple=(\R_1,\dots,\R_n)$ in \D\ that index a \topcp\ \simplex. 
Inverting this relation provides the list of top cells from \sC\ mapped to each such tuple of regions. 
As we iterate through regions, we ensure that each \topcp\ in the complex is processed by only one region \R,
by skipping the \topcpcells\ whose minimum vertex index \vIndex\ is not in $\PhiMapVert(\R)$.
For example in Figure~\ref{fig:reindexing}(a), triangle 5 is indexed by region $a$ and $b$, and, thus, its tuple is $\tuple = (a,b)$. The complete list of triangles in tuple $(a,b)$ is $\{2,5,12\}$.

We use this inverted relation in \AlgoName{extract\_cell\_indices} (see Algorithm~\ref{alg:get_tops_reorderd_indexes} in \ref{sec_app:tops_reordering_long}),  
to generate a new coherent order for the \topcpcells\ of \sCT. 
Specifically, the prefix sum of the tuple cell counts 
provides the starting index for cells in that group.
For example, when considered in lexicographic order, the first three region tuples, $(a)$, $(a,b)$ and $(a,b,d)$ in Figure~\ref{fig:reindexing}(b),
with 1, 3 and 1 triangles, respectively, get starting indices $1$, $2$ and $5$.
We then assign increasing indices to the \topcpcells\ of each group. 
Thus, e.g., the three triangles belonging to tuple $(a,b)$ get indices $\{2,3,4\}$ after this reindexing.

Finally, in \AlgoName{compress\_cells} and \AlgoName{permute\_array}
(Algorithm~\ref{alg:compress_cells_representation} and~\ref{alg:update_array} in \ref{sec_app:tops_reordering_long}), 
we reorder and SRE-compact the \tR\ region arrays 
and the global \topcpcells\ array \sCT.

\section{Algorithm details: reindexing and compressing the \topcpcells}
\label{sec_app:tops_reordering_long}

\begin{figure*}[t]
	\centering
	\subfloat[Original triangles]{
		\resizebox{.25\textwidth}{!}{
			\includegraphics{imgs/tri_reindexing_orig_regions}
		}
	}
	\hfil
	\subfloat[After Step 1]{
		\resizebox{.2\textwidth}{!}{
			\includegraphics{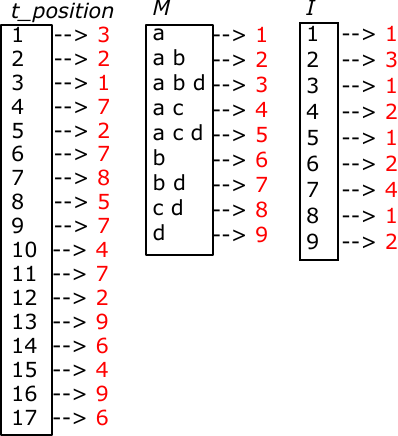}
		}
	}
	\hfil
	\subfloat[After Step 2]{
		\resizebox{.13\textwidth}{!}{
			\includegraphics{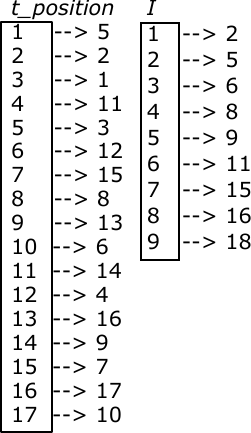}
		}
	}
	\hfil
	\subfloat[Reindexed triangles]{
		\resizebox{.25\textwidth}{!}{
			\includegraphics{imgs/tri_reindexing_new_regions}
		}
	}
	\caption{Top cell reindexing.
		(a) initial tree with four leaf blocks $a, b, c, d$
		(b,c) auxiliary data structures after Steps 1 and 2
		(d) reindexed tree.} 
	\label{fig:reindexing_verbose}
\end{figure*}

This appendix provides details for reindexing the top CP cells during the Stellar tree generation algorithm (Algorithm~\ref{alg:main_tops_reordering}) as outlined
in~\ref{sec:tops_reordering}.
The reindexing exploits the coherence of \topcpcells\ that are indexed by the same set of regions
by translating the proximity in \D\ into \emph{index} proximity in \sCT\ and depends on three auxiliary data structures:
\begin{itemize}
	\item an associative array, \mapM, which maps an (integer) identifier to each unique tuple of regions;
	\item an array of integers, \arrI, having the same number of entries as \mapM.
				Initially, it is used to track the number of \topcpcells\ associated with each tuple of regions.
				In a successive phase, it tracks the next index for a \topcp\ in a tuple;
	\item an array of integers, \arrTPos, of size $|\sCT|$. 
				Initially, it is used to associate \topcpcells\ with their region tuple identifier.
				In a successive phase, it is used to store the new coherent indices for the \topcpcells. \\
\end{itemize}
%
%
\NOTA{
	A 2D example is shown in Figure \ref{fig:reindexing_verbose}. 
	In Figure \ref{fig:reindexing_verbose}(a) it is shown the original tree \h\ to which are added the triangles in the leaf blocks keeping the original sorting of \sCT.
	Then, Figure \ref{fig:reindexing_verbose}(b) shows an illustration of the technique, with the two main structures that are conceptually defined during the exploiting of the spatial coherence of the triangles:
	\begin{enumerate}
		\item on the left, it is shown the array which keeps, for each triangle in \sCT, the array of leaf blocks tuple that index it;
		\item while on the right, it is shown the structure that encodes the association between the leaves tuples and the triangles indexed in the same tuple.
	\end{enumerate}
	Finally, in Figure \ref{fig:reindexing_verbose}(c), it is shown the final tree, in which it has been exploited the spatial coherence of the triangles.
}

\begin{algorithm}[tb]			
	\caption{ $\AlgoName{extract\_tuples}(\D,\sC,M,I,t\_position)$}
	\label{alg:get_region_top_association}				
	\begin{algorithmic}[1]			
		\medskip
		\Input {\D\ is the decomposition defined on the vertices of \sC}
		\Input {\sC\ is the CP complex}
		\Input {$M$ is a map associating a unique identifier to each tuple}
		\Input {$I$ is an array that tracks \topcpcells\ in each tuple}
		\Input {$t\_position$ is an array linking \topcpcells\ to tuples}
		
		\ForAll{regions \R\ \textbf{in} \D}
			\ForAll {$\simplex$ \textbf{in} \PhiMapTop(\R) (with index \tIndex\ in \sCT)}
				\LineComment{extract the minimum vertex index of \simplex}	
				\State {\vIndex\ $\leftarrow$ \AlgoName{get\_min\_vertex\_index}(\simplex)}	
				\LineComment{we visit \simplex\ if $\R$ indexes \vIndex}
				\If{\vIndex\ \textbf{in} \PhiMapVert(\R)}
					\LineComment{extract the tuple \tupleR\ of regions indexing \simplex}
					\State {\tupleR\ $\leftarrow$ \AlgoName{extract\_tuple}(\D,\simplex)}
					\LineComment{get the tuple key in $M$ (if not present insert it)}
					\State{key $\leftarrow$ M[\tupleR]}					
					\LineComment{increment the counter of this tuple in I}
					\State{I[key] $\leftarrow$ I[key]$+1$}					
					\LineComment{associate the tuple key to the \tIndex\ entry in t\_position}
					\State{t\_position[\tIndex] $\leftarrow$ key} 
				\EndIf
			\EndFor	
		\EndFor	
		
	\end{algorithmic}
\end{algorithm}
%
%
\begin{algorithm}[tb]			
	\caption{ $\AlgoName{extract\_cell\_indices}(I,t\_position)$ }
	\label{alg:get_tops_reorderd_indexes}				
	\begin{algorithmic}[1]			
		\medskip
		\Input {$I$ is an array associated with the unique leaf tuples in $M$}
		\Input {$t\_position$ contains tuple index for top cell}
		\Ensure {$t\_position$ contains new top simplex position indexes}
		
		\Variable {$c$ is the counter variable of the current position index}
		
		\LineComment{convert the cell counts in I into the starting indexes for \topcpcells\ grouped by the same tuple}
		\ForAll{key \textbf{in} I}
			\State {tmp $\leftarrow$ I[key]} \SideComment{I[key] contains cells count for tuple key}
			\State {I[key] $\leftarrow$ c}
			\State {c $\leftarrow$ c $+$ tmp}
		\EndFor
		
		\LineComment{assign to each \topcp\ its new position index in \sCT}
		\ForAll{$\simplex$ \textbf{in} \sCT\ (with index \tIndex\ in \sCT)}
			\State {key $\leftarrow$ t\_position[\tIndex] }
			\State {t\_position[\tIndex] = I[key]} \SideComment{set new position index for \simplex}
			\State {I[key] $\leftarrow$ I[key]$+1$}
		\EndFor
	\end{algorithmic}
\end{algorithm}
%
\begin{algorithm}[tb]			
	\caption{ $\AlgoName{compress\_cells}(\D,t\_position)$}
	\label{alg:compress_cells_representation}				
	\begin{algorithmic}[1]			
		\medskip
		\Input {\D\ is the decomposition defined on the vertices of \sC}
		\Input {$t\_position$ contains the new \topcp\ position indices}
		
		\ForAll{regions \R\ \textbf{in} \D}		
			\State {\tR\_aux\ $\leftarrow$ \PhiMapTop(\R)} \SideComment{copy the \topcpcells\ array of \R}	
			\State { \PhiMapTop(\R) $\leftarrow \emptyset$ } \SideComment{reset that array}	
			
			\LineComment {update the indices in \tR\_aux with those from t\_position}
			\For {id $\leftarrow$ 0 \textbf{to} $|$\tR\_aux$|$}
				\State {\tR\_aux[id] $\leftarrow$ t\_position[\tR\_aux[id]]}	
			\EndFor
			
			\State {\AlgoName{sort}(\tR\_aux) }

			\State {start\_id $\leftarrow$ \tR\_aux.\AlgoName{first}()}
			\State {counter $\leftarrow$ 0}
			
			\For {id $\leftarrow$ 1 \textbf{to} $|$\tR\_aux$|$}
				\LineComment{if we find consecutive indexes}
				\If{ \tR\_aux[id]+1 =  \tR\_aux[id+1] } 
					\State {counter $\leftarrow$ counter$+1$}
				\Else 				
					\If{ counter $> 1$ } \SideComment{found a run of indices}
						\LineComment{create a run in \tR\ of \R}
						\State{\AlgoName{create\_sre\_run}(\R,start\_id,counter)}
					\Else \SideComment{simply add the \topcp\ index in \tR}
						\State{\R.\AlgoName{add\_top}(start\_id)}				
					\EndIf
					\LineComment{reset the two auxiliary variable}				
					\State {start\_id $\leftarrow$ \tR\_aux.\AlgoName{next}()}
					\State {counter $\leftarrow$ 0}				
				\EndIf
			\EndFor
		\EndFor
	\end{algorithmic}
\end{algorithm}

\begin{algorithm}[tb]			
	\caption{ $\AlgoName{permute\_array}(array,permutation)$ }
	\label{alg:update_array}				
	\begin{algorithmic}[1]			
		\medskip
		\Input {$array$ is the simplex array to update}
		\Input {$permutation$ is the array containing the new position indices}
		\For {id $\leftarrow$ 0 \textbf{to} $|$array$|$}
			\LineComment{current vertex is in correct position or updated already}		
			\If {permutation[id] $=$ id}
				\State {permutation[id] $\leftarrow\ -1$} \SideComment{mark $id$ as updated}
			\Else \SideComment{id is not updated already}
				\LineComment{iteratively update vertices positions}
				\While {permutation[id] $\neq$ id}	
					\LineComment{swap id and permutation[id] entries in array}
					\State {\AlgoName{swap}(array,id,permutation[id])}
					\LineComment{mark $id$ as updated in $permutation$}
					\LineComment{then, get the $id$ of the next vertex to update}
					\State {id $\leftarrow$ \AlgoName{mark\_and\_get\_next}(id,permutation)}
				\EndWhile
				\State {$permutation[id]$ $\leftarrow$ -1} \SideComment{mark $id$ as updated}
			\EndIf
		\EndFor
	\end{algorithmic}
\end{algorithm}

The remainder of this appendix summarizes the four major steps of Algorithm~\ref{alg:main_tops_reordering}.
Figure~\ref{fig:reindexing_verbose} illustrates this reorganization process over a triangle mesh.
\NOTA{
	\paragraph*{Extract the leaves and \topcpcells\ association}
	\kennyComment{Summary:
		Iterate through leaves of tree.  
		\\ $\forall$ \topcp\ \simplex\ in \tB\ of \B\ w/ $min(\relation{\tDim,0}(\simplex)) \in \PhiMapVert(\B)$:
		\\ -- Find its tuple\_of\_blocks, 
		\\ -- uniqueID = M[tuple\_of\_blocks] (get or create new key) 
		\\ -- I[uniqueID]++
		\\ -- t\_positions( \tIndex) = uniqueID	
	}
}
\paragraph{\AlgoName{extract\_tuples}}
In Algorithm~\ref{alg:get_region_top_association}, we generate map \mapM,
count the number of \topcpcells\ associated with each tuple of regions in array \arrI\
and initialize the \arrTPos\ array entries with its tuple identifier:

\begin{itemize}
	\item 
	for each region \R\ in \D, we visit the \topcpcells\ \simplex\ in \PhiMapTop(\R) whose minimum vertex index \vIndex\ (row 4)
	is indexed in \R. This ensures that each \topcp\ is processed only once.
	Regions of \D\ are uniquely indexed by the index of their starting vertex \vstart;
	
	\item 
	for each such \topcp\ \simplex\ with index \tIndex, we traverse the partitioning on the vertices to find the tuple of regions from \D\ that index \simplex\ (row 5 function \AlgoName{extract\_tuple}).
	We then look up its unique identifier $key$ in \mapM\ (or create a new one and insert it into \mapM) (row 6).  
	We then increment the count for this tuple, 
	and associate \simplex\ with this tuple 
	(rows 7 and 8).
\end{itemize}	
At the end of the traversal of \D, each entry of \arrTPos\ contains the identifier of the tuple of regions indexing its corresponding top cell and \arrI\ contains the number of \topcpcells\ indexed by each tuple. \mapM\ is no longer needed and we can discard it.

The content of auxiliary data structures, after this step, is illustrated in Figure \ref{fig:reindexing_verbose}(b).
For example, triangle $5$ is indexed by regions $a$ and $b$, whose key in \mapM\ is $2$. This tuple contains two triangles other than $5$, as indicated by the corresponding counter in \arrI.

\NOTA{
	\paragraph*{Extract the spatial coherence of the \topcpcells}
	\kennyComment{
		Summary:
		Use I and t\_positions to find a reordering of \sCT.
		\\ -- No longer need M, clear its memory
		\\ (1) I := \texttt{prefixSum} of I
		\\ (2) Forall \simplex: t\_positions[t] = I[t\_positions[t]]++
		\\ --  No longer need I, clear its memory
	}
}
\paragraph{\AlgoName{extract\_cell\_indices}}
In Algorithm \ref{alg:get_tops_reorderd_indexes}, 
we use the \arrI\ and \arrTPos\ arrays to find the updated index for each \topcp\ in \sCT, which is computed in place in \arrTPos.

	First, we convert the cell counts in array \arrI\ into starting indexes for the \topcpcells\ grouped by the same set of regions, by taking the \emph{prefix sum} of array \arrI\ 
	(rows 1 to 4).

	Then, we use array \arrI\ to update \arrTPos\ array by iterating over the \topcpcells, and replacing the tuple identifier in \arrTPos\ with the next available index from \arrI\ and increment the counter in \arrI\ 
	(rows 5 to 8).
At this point, \arrTPos\ is a permutation array that encodes a more coherent ordering for the \topcpcells\ and \arrI\ is no longer needed.

The content of auxiliary data structures after this step, is shown in Figure \ref{fig:reindexing_verbose}(c).
At the end, each entry of \arrI\ contains the first index of the next tuple, while \arrTPos\ the new position for the $i$-th triangle.

\NOTA{
	\paragraph*{Compress the tree representation}
	\kennyComment{Summary: Use t\_positions to reorder \tB\ arrays and SRE compress them.
		Use t\_positions to reorder \sCT\ arrays.}
}

\paragraph{\AlgoName{compress\_tree\_cells}}
In Algorithm \ref{alg:compress_cells_representation}, we apply this order to the arrays \tR\ of \topcpcells\ of each region \R\ and compact the \tR\ arrays using the SRE compression.
This procedure iteratively visits all regions of a Stellar decomposition.
Within each region \R, an auxiliary array, called $\tR\_aux$, is used, encoding, initially, a copy of the array of \topcpcells\ position indices encoded by \R\ (row 2). Then, these indices are updated with the coherent ones from $t\_permutation$ (rows 4 and 5), and, finally, by sorting this array we have sequential indices in consecutive position of $\tR\_aux$ (row 6).

Next, we identify consecutive index runs by iterating over $\tR\_aux$ array (rows 7 to 18).
In this phase, we use two auxiliary variables, a counter, encoding the size of the current run, and a variable, called $start\_id$, encoding the starting index of the current run. If we find two consecutive indices, we simply increment $counter$. Otherwise, we check if we have a run (row 13), or if we have to simply add the index in $start\_id$ to \tR\ array of \R\ (row 16). If we have to encode a run in \tR\ (procedure \AlgoName{create\_sre\_run}, row 17), we apply the strategy, described in Section~4.2.2, for encoding it.  

\paragraph{\AlgoName{permute\_array}}
Finally, in Algorithm \ref{alg:update_array}, we update the global \topcpcells\ array \sCT.
This is done by iteratively swap the entries in \sCT\ (rows 5 to 8), applying the new coherent indices encoded in $permutation$ array. 
This procedure does in place updates and, thus, does not require any additional auxiliary data structure.

\NOTA{
	Recall that, we assume that all the blocks of \h\ are \emph{half-open} blocks unless a block \B\ is incident 
	in the block representing the ambient space \aSpace, as in this case we consider that face of \B\ as \emph{closed}.
}

\NOTA{
	We insert the vertices sequentially, over which we start a visit of \h\ to insert a vertex \vertex\ in a leaf block \B\ containing it.
	We recall that we do not explicitly encode within each block its domain, but we compute it at runtime, keeping track of the split planes. 
	We assume that the root block \hR\ completely covers the complex domain and
	represents it as a \emph{closed block}. 
}

\NOTA{
	We insert each \top\ \simplex\ at once and for each boundary vertex \vertex\ of a \ktop\ \simplex, we find the leaf block \B\ of \hL\ that indexes \vertex\ 
	and then, we add \simplex\ to \top\ array of \B\ (i.e., we insert \tIndex\ index in \tB\ array).
}

\NOTA{
	\begin{enumerate}
		\item first, we navigate the tree to compress it and to get the spatially coherent ordering of the vertices (see Algorithm \ref{alg:get_vertices_ordering});
		\item then, for each \simplex\ in \sCT, we update its boundary relation \relation{k,0} (see rows 4 to 6 of Algorithm \ref{alg:main_vertices_reordering});
		\item finally, we update the \sCV\ array according to the new position ordering on the vertices.
	\end{enumerate}
	As auxiliary data structure for this procedure, we need an array of integer references, that we call $v\_permutation$. 
	This array contains exactly $|\sCV|$ entries, and at each entry $i$, it is associated the $i$-th vertex \vertex\ in \sCV. 
	This entry in $v\_permutation$ contains the new \emph{spatially coherent} position of \vertex\ in \sCV. 
	Thus, the extra storage overhead is exactly $|\sCV|$.
	$v\_permutation$ is the output of Algorithm \ref{alg:get_vertices_ordering}, and it is required as input during the update 
	of the \topcp\ boundary relations (rows 4 to 6 of Algorithm \ref{alg:main_vertices_reordering}) and during the update of \sCV\ array.
	%
	Algorithm \ref{alg:get_vertices_ordering} is a recursive procedure in which we visit all the blocks in \h. 
	If we are in an internal block (see rows 1 to 5 of the Algorithm), we recursively visit all the children, while if we are in a leaf block \B\ 
	(see rows 7 to 12 of the Algorithm), we visit the vertices array \PhiMapVert(b), we set up a consecutive indexes run in \B\ 
	and we update the corresponding entries in $v\_permutation$.
}

\NOTA{
	After this stage, each leaf block contains a contiguous range of vertices and each internal block contains a continuous range 
	of vertices equal to the union of the ranges contained in its descendant.
	Moreover, we do not require extra structures to encode this range, thanks to the \emph{sequential run-length} compression, and, thus, 
	in each vertex array, we require just two entries that have been initialized in order to represent the unique vertices run into the leaf block. 
	We denote the extreme vertices of this run as \vstart\ and \vend.
	Then, we proceed with the updating of boundary relation \relation{k,0} on all \ktops\ in \sCT\ (see rows 4 to 6 in Algorithm \ref{alg:main_vertices_reordering}). 
	For each vertex \vertex\ (with index \vIndex\ in \sCV) in \relation{k,0} of a \ktop\ \simplex, we get its spatial coherent position 
	in \sCV\ from $v\_permutation$, by accessing the \vIndex\ entry. Once we get this position we update the entry, associated to \vertex, 
	in \relation{k,0}(\simplex) with the value in $v\_permutation[\vIndex]$.
	Finally, we update the \sCV\ array accordingly with the spatially coherent ordering (in Algorithm \datasetName{update\_array}). 
	During the algorithm, we iteratively swap the index positions, updating at each swap operation  a vertex \vertex, 
	for which we gather its new spatially coherent position from $v\_permutation[\vIndex]$. This algorithm does exactly $|\sCV|$ swaps without requiring any extra storage.
}

\NOTA{
	In order to update $M$ and $t\_position$, we need to extract the tuple of leaf blocks indexing \simplex.
	If \simplex\ is completely indexed into \B, then we have its tuple already available. Otherwise, we have to visit the tree to find the leaf blocks that index the other vertices of \simplex.
	Once we get the tuple of leaf blocks indexing \simplex, we have to check if the tuple is already into $M$, and if it is not present we insert it. Conversely, if it is already into $M$, we simply increment the counter that keeps track of the \topcpcells\ indexed into that tuple.
}

\section{Generating topological data structures}
\label{sec_app:stellar_build_structures}

\NOTA{ Motivate: 
		Generation of existing state of the art topological data structures as proxy for mesh processing applications.
		This localized generation algorithm is useful in its own right due to the memory savings: the auxiliary storage is proportional to the complexity in a leaf block.
}
\NOTA{
In this section, we want to show that the Stellar tree has the same expressive power as the topological data structures, that it can adapt to algorithms defined for a particular topological data structure, by building a local topological data structure only when needed, thus, requiring a fraction of the memory of a global topological data structure. 
Moreover, we want to show that the Stellar tree can be used also as a tool to generate topological data structures. 
}

As a proxy for more complicated mesh processing workflows, this section describes how Stellar trees can be used to generate an existing topological data structure over CP complexes: the \iastar\ data structure. 
The \iastar\ data structure is the most compact data structure in the class of connectivity-based representations 
since it encodes only the vertices and the top cells of a CP complex \sC, as well as a subset of topological relations connecting these cells. 
As with other topological data structures, the \iastar\ over \sC\ is typically generated directly 
from the indexed representation of \sC\ by extracting its adjacency and co-boundary relations. 
However, this direct approach could present issues when scaling to large complexes on commodity hardware due to the high storage requirements 
for auxiliary data structures needed to reconstruct these relations.

%
%
This application demonstrates the versatility of the Stellar tree representation 
and exercises many of the operations necessary for other mesh processing tasks.
We define customized topological relations and auxiliary data structures as we stream through the leaf blocks of the tree and take either a \emph{global} approach, to reconstruct the full topological data structure, 
or a \emph{local} approach, which reconstructs coherent subsets of the full data structure restricted to the portion of the complex indexed within each leaf block.
In the former case, Stellar trees enable generating the global topological data structures 
using a fraction of the memory as would be required to directly generate them from an indexed representation.
In the latter case, the local approach can be used to adapt local regions of the Stellar tree's underlying complex 
to algorithms defined for existing topological data structures.

In the following, we present a local generation algorithm over a single leaf block of the Stellar tree,
and compare the local and global generation algorithms against a direct approach 
that generates the data structure from the original indexed mesh representation. 
We do this within the Stellar tree framework by setting the bucketing threshold to infinity,
since $\kv = \infty$ produces a tree that indexes the entire complex \sC\ in its root block. 


%
Recall from Section~8.2 of the paper 
that the \iastar\ data structure is an adjacency-based topological data structure defined over non-manifold \cDim-dimensional CP complexes that gracefully degrades to the IA representation over manifold complexes. 
The IA data structure is defined over pseudo-manifolds, and, thus, each $(\cDimMinusOne)$-cell can be incident in at most two top CP \cDim-cells.
We first describe how to generate the IA data structure from the Stellar tree, and then extend this to the \iastar\ data structure.


The IA data structure encodes 
the following topological relations: 
\begin{inparaenum}[(i)]
	\item boundary relation $\relation{\cDim,0}(\sigma)$,	 
	\item partial co-boundary relation $\partialrelation{0,\cDim}(\vertex)$ for each vertex $\vertex$, consisting of one arbitrarily selected top CP \cDim-cell in the star of \vertex, and 
	\item adjacency relation $\relation{\cDim,\cDim}(\sigma)$, for each top CP \cDim-cell $\simplex$.
\end{inparaenum}
If $\simplex_1$ is adjacent to $\simplex_2$ through $(\cDimMinusOne)$-cell $\tau$, and $\tau$ is the $i$-th face of $\simplex_1$, then $\simplex_2$ will be in position $i$ in the ordered list of $\relation{\cDim,\cDim}(\simplex_1)$.

Since the Stellar tree explicitly encodes the $\relation{\cDim,0}$ relations for all top CP \cDim-cells, the generation of a \emph{local} IA data structure consists of extracting \partialrelation{0,\cDim}(\vertex), for each \vertex\ in \PhiMapVert(\R), and $\relation{\cDim,\cDim}$(\simplex), for each top CP \cDim-cell \simplex\ in \PhiMapTop(\R).
For vertices in  \PhiMapVert(\R), the former is computed by iterating over the top CP \cDim-cells in \PhiMapTop(\R), 
and selecting the first \topcp\ incident in \vertex\ that we find.

\begin{algorithm}[t]			
	\caption{ $\AlgoName{extract\_\relation{\cDim,\cDim}\_manifold}(\R,\sC)$ }
	\label{alg:top-tops}
	\begin{algorithmic}[1]			
		\medskip
		\Input {$\R$ is a leaf block in \h}
		\Input {$\sC$ is the CP complex indexed by $\h$}
		\Variable {$\cDim\_1\_cell\_top$ encodes \relation{\cDimMinusOne,\cDim} for the (\cDimMinusOne)-cells in \R}
		%
		\ForAll {top CP \cDim-cells $\simplex$ \textbf{in} $\PhiMapTop(\R)$} 
			\ForAll {(\cDimMinusOne)-cells $\tau$ \textbf{in} \relation{\cDim,\cDimMinusOne}($\simplex$)}
				\State{add \simplex\ to $\cDimMinusOne\_cell\_top[\tau]$}
			\EndFor
		\EndFor
		\ForAll {(\cDimMinusOne)-cells $\tau$ \textbf{in} $\cDim\_1\_cell\_top$} 
			\If { $| \cDim\_1\_cell\_top[\tau] | = 2$ }
				\State{$\{\simplex_1,\simplex_2\} \leftarrow \cDim\_1\_cell\_top[\tau]$}
				\State{set $\simplex_1$ as adjacent \cDim-cell for $\simplex_2$ in $\tau$}
				\State{set $\simplex_2$ as adjacent \cDim-cell for $\simplex_1$ in $\tau$}
			\Else \SideComment{$|\cDim\_1\_cell\_top[\tau]| = 1$}
				\State{mark cell $\tau$ of \simplex\ as a boundary cell }
			\EndIf
		\EndFor 
	\end{algorithmic}
\end{algorithm}

\begin{figure}[t]
	\centering
	\resizebox{.5\columnwidth}{!}{
		\includegraphics{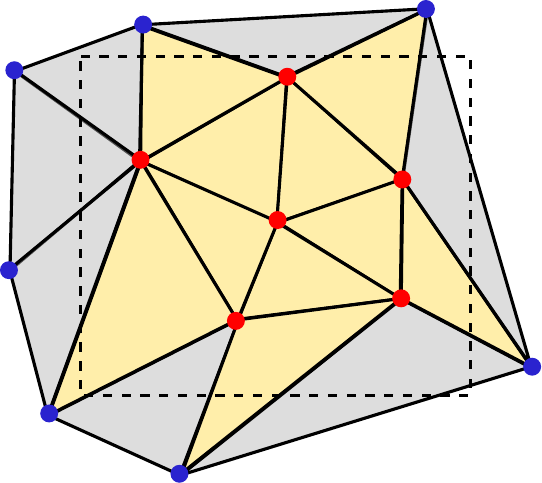}
	}
	\caption{ \emph{Local} adjacency reconstruction finds adjacencies across faces 
		with a vertex in the leaf block \R\ (dashed). 
		For yellow triangles, all edges have a vertex in \R, while some edges of gray triangles do not.
	} 
	\label{fig:adj_extraction_case}
\end{figure}

Algorithm~\ref{alg:top-tops} provides a description of a \emph{local} strategy for extracting \relation{\cDim,\cDim}(\simplex) 
relations within block \R\ of the tree.
Note that it finds only the adjacencies across (\cDim-1)-faces that have at least one vertex in \PhiMapVert(\R).
While we can locally reconstruct the full adjacency relation for top CP \cDim-cells with \cDim\ vertices in \PhiMapVert(\R), a top CP \cDim-cell \simplex\ with fewer vertices in \PhiMapVert(\R) might be missing at least one adjacency.
For example, in Figure~\ref{fig:adj_extraction_case}, we can completely reconstruct the adjacency relations of the triangles 
having two vertices in \R\ (in yellow), while we can only partially reconstruct the adjacencies of triangles having just one vertex in \R\ (in gray). 
 Adjacencies on the edges opposite to the vertices in red cannot be reconstructed inside \R\ for gray triangles.
%

The algorithm first iterates on the top CP \cDim-cells in \PhiMapTop(\R) (rows 1--3). 
Given a top CP \cDim-cell \simplex, we cycle over the \cDim-tuples of the vertices of \simplex, where each \cDim-tuple defines a (\cDimMinusOne)-cell on the boundary of \simplex. The auxiliary data structure $d\_1\_cell\_top$ encodes, for each \cDim-tuple $\tau$, the top \cDim-cells sharing $\tau$, 
corresponding to the \relation{\cDimMinusOne,\cDim} relation of $\tau$.
Then, the algorithm iterates over  $d\_1\_cell\_top$ to initialize adjacency relations \relation{\cDim,\cDim}. 
Given a (\cDimMinusOne)-cell $\tau$, if $\tau$ has two \cDim-cells in its co-boundary (row 5), 
namely $\simplex_1$ and $\simplex_2$, we set $\simplex_1$ and $\simplex_2$ as adjacent along $\tau$ (rows 7--8).
Due to its local nature, the Stellar tree adjacency reconstruction provides considerable storage savings compared to its global counterpart: 
the storage requirements are proportional to the number of top CP \cDim-cells in \R, 
rather than those in \sCT.

Extending this algorithm to generate a \emph{global} IA data structure requires only a few modifications. Aside from encoding the auxiliary data structures at a global level, the other major difference with respect to the local approach is that, within each leaf block \R, \relation{\cDimMinusOne,\cDim} relations are extracted only for those (\cDimMinusOne)-cells $\tau$ for which the two top CP \cDim-cells sharing $\tau$ have not been already initialized.
The \iastar\ data structure extends the IA data structure to arbitrary non-manifold CP \tDim-complexes, with $0< \tDim\leq \cDim$.
Recall that, in addition to the relations stored in the IA data structure,  it encodes: 
\begin{inparaenum}[(i)]
	\item adjacency relation $\relation{\tDim,\tDim}(\sigma)$, for each \ktopcp\ $\simplex$;
	\item co-boundary relation \relation{0,1}(\vertex) restricted to the top 1-cells, for each vertex \vertex;
	\item \emph{augmented} partial co-boundary relation (\partialrelation{0,\tDim}(\vertex)), $1 < \tDim \leq \cDim$, for each vertex $\vertex$, consisting of one arbitrarily selected \ktopcp\ from each \emph{\tDim-cluster} in the star of \vertex, where a \tDim-cluster is a (\tDimMinusOne)-connected component of the star of \vertex\ restricted to its top CP $k$-cells; and
	\item co-boundary relation $\relation{\tDimMinusOne,\tDim}(\tau)$, for each non-manifold (\tDimMinusOne)-cell $\tau$ bounding a \ktopcp. 
\end{inparaenum}

Extracting \relation{\tDim,\tDim} relations, when $\tDim<\cDim$, and \relation{\tDimMinusOne,\tDim} relations for non-manifold (\tDimMinusOne)-cells is performed by a suitable extension of Algorithm \ref{alg:top-tops}.
%
%
Augmented partial co-boundary relation \partialrelation{0,\tDim}(\vertex), for $\tDim > 1$, is computed  by extracting  
the restricted star of \vertex\ (Algorithm~3 in the paper) and by using \relation{\tDim,\tDim} relation 
for the \topcpcells\ in the star of \vertex\ to identify the (\tDimMinusOne)-connected components incident in \vertex.
\relation{0,1}(\vertex) is initialized by iterating over the top 1-cells in the restricted star of \vertex.

\paragraph*{Experimental results} 
\label{sec:adjtiming}

In Table~\ref{tab:timings_adj_ds_gen}, we compare the time, and storage requirements to generate an IA  or  \iastar\ data structure, depending on whether the complex has a manifold or non-manifold domain, using the Stellar tree or directly extracting it from the indexed representation.
%
For each dataset, we compare the Stellar trees generated by using thresholds \ks\ and \kl\ and by using a local and a global algorithm against the direct approach on the original indexed representation of the complex.
For the manifold (\emph{triangular}, \emph{quadrilateral}, \emph{tetrahedral} and \emph{hexahedral}) and pure (\emph{probabilistic}) datasets, 
where all top cells have dimension \cDim, we used Algorithm~\ref{alg:top-tops} to compute the adjacencies.

\begin{table}[tp]
	\centering
	\caption{Generation times (seconds) and storage (number of references) for the \iastar\ data structure
			from Stellar trees (\ks\ and \kl) and the direct approach ($dir.$) on the original indexed representation of the complex.
      With the exception of \datasetName{v-rips} complexes, the \iastar\ is equivalent to the IA representation on these datasets.
	}
	{
	\resizebox{0.95\columnwidth}{!}{
		\begin{tabular}{cccrr|rrr}
			\toprule
			\multicolumn{1}{c}{\multirow{3}{*}{\textbf{Data}}} & \multicolumn{1}{c}{\multirow{3}{*}{\textbf{}}}  &
			\multicolumn{1}{c}{\multirow{3}{*}{\textbf{\kv}}} &
			\multicolumn{2}{c}{\multirow{3}{*}{\textbf{Time}}} & \multicolumn{3}{c}{\textbf{Storage}} \\
			\cmidrule(lr){6-8}
			& & & &
			& \multicolumn{2}{c}{\textbf{IA/IA$^*$}} & \multicolumn{1}{c}{\textbf{aux.}} \\
			\cmidrule(lr){4-5} 
			\cmidrule(lr){6-7} 
			& & & \multicolumn{1}{c}{\textbf{local}} & \multicolumn{1}{c}{\textbf{global}}
			& \multicolumn{1}{c}{\textbf{local}} & \multicolumn{1}{c}{\textbf{global}} & \multicolumn{1}{c}{\textbf{d.s.}} \\
			\midrule
 
\multirow{3}{*}{\datasetName{neptune}} & \multirow{10}{*}{\begin{sideways}\datasetName{tri.}\end{sideways}} & $\ks$ & 6.88	&	5.69	
 & 0.36K & \multirow{3}{*}{6.01M} & 0.70K \\
 &  &  $\kl$ & 6.51	&	5.84	
  & 1.65K &  & 3.24K \\
 &  &  $dir.$ &  & 9.69 &  & 
 & 12.0M \\
  \cmidrule(lr){1-1} \cmidrule(lr){3-8} 
 
\multirow{3}{*}{\datasetName{statuette}} &  &  $\ks$ & 17.5	&	14.5	
 & 0.38K & \multirow{3}{*}{15.0M} & 0.72K \\
 &  &  $\kl$ & 17.0	&	14.9	
  & 1.62K &  & 3.22K \\
 &  &  $dir.$ &  & 20.7 &  & & 
 30.0M \\
 \cmidrule(lr){1-1} \cmidrule(lr){3-8}
 
\multirow{3}{*}{\datasetName{lucy}} &  &  $\ks$ & 47.4	&	39.6	
 & 0.42K & \multirow{3}{*}{42.0M} &  0.82K \\
 &  &  $\kl$ & 49.0	&	40.7	
  & 1.64K &  & 3.28K \\
 &  &  $dir.$ &  & 50.6 &  & 
 &  84.1M \\
 \midrule

\multirow{3}{*}{\datasetName{neptune}} & \multirow{10}{*}{\begin{sideways}\datasetName{quad.}\end{sideways}} &  $\ks$ & 36.6	&	31.6	
 & 0.27K & \multirow{3}{*}{24.0M} & 0.52K \\
 &  &  $\kl$ & 35.0	&	31.3	
  & 1.70K &  & 3.37K \\
 &  &  $dir.$ &  & 44.1 &  & 
   & 48.1M \\
 \cmidrule(lr){1-1} \cmidrule(lr){3-8}
 
\multirow{3}{*}{\datasetName{statuette}} &  &  $\ks$ & 92.2	&	78.9	
 & 0.26K & \multirow{3}{*}{60.0M} & 0.50K \\
 &  &  $\kl$ & 90.4	&	79.4	
  & 1.74K &  & 3.38K \\
 &  &  $dir.$ &  & 102 & & 
 & 120M \\
 \cmidrule(lr){1-1} \cmidrule(lr){3-8}
 
\multirow{3}{*}{\datasetName{lucy}} &  &  $\ks$ & 250	&	218	
 & 0.27K & \multirow{3}{*}{168M} & 0.53K \\
 &  &  $\kl$ & 250	&	221	
  & 1.74K &  & 3.40K \\
 &  &  $dir.$ &  & 252 &  & 
&  336M \\
 \midrule

  \multirow{3}{*}{\datasetName{bonsai}} & \multirow{10}{*}{\begin{sideways}\datasetName{tetra.}\end{sideways}} &  $\ks$ & 56.5	&	38.5
   & 3.20K & \multirow{3}{*}{28.6M} & 10.2K \\
  &  &  $\kl$ & 49.4	&	38.8	
   & 6.29K &  & 20.6K \\
  &  &  $dir.$ &  & 60.0 & & 
 &  97.7M \\
  \cmidrule(lr){1-1} \cmidrule(lr){3-8}
  
 \multirow{3}{*}{\datasetName{vismale}} &  &  $\ks$ & 61.2	&	42.0	
  & 3.22K & \multirow{3}{*}{31.1M} & 10.4K \\
  &  &  $\kl$ & 53.3	&	43.0	
   & 5.52K &  & 17.4K \\
  &  &  $dir.$ &  & 66.3 &  & 
 &  106M \\
  \cmidrule(lr){1-1} \cmidrule(lr){3-8}
  
 \multirow{3}{*}{\datasetName{foot}} &  &  $\ks$ & 72.6	&	47.2
  & 3.21K & \multirow{3}{*}{34.5M} & 10.3K \\
  &  &  $\kl$ & 58.7	&	47.1
   & 6.42K &  & 21.0K \\
  &  &  $dir.$ &  & 72.7 & & 
  &  118M \\
  \midrule

\multirow{3}{*}{\datasetName{f16}} & \multirow{10}{*}{\begin{sideways}\datasetName{hexa.}\end{sideways}} &  $\ks$ & 152	&	102	
 & 0.32K & \multirow{3}{*}{53.3M} & 0.83K \\
 &  &  $\kl$ & 129	&	103	
  & 2.38K &  & 6.42K \\
 &  &  $dir.$ &  & 237 & &
  &  152M \\
 \cmidrule(lr){1-1} \cmidrule(lr){3-8}
 
\multirow{3}{*}{\datasetName{san fern}} &  &  $\ks$ & 380	&	217	
 & 0.38K & \multirow{3}{*}{117M} & 1.05K \\
 &  &  $\kl$ & 273	&	219	
  & 2.64K &  & 7.31K \\
 &  &  $dir.$ &  & 285 & & 
 &  336M \\
 \cmidrule(lr){1-1} \cmidrule(lr){3-8}
 
\multirow{3}{*}{\datasetName{vismale}} &  &  $\ks$ & 844	&	459	
 & 0.23K & \multirow{2}{*}{261M} & 0.59K \\
 &  &  $\kl$ & 591	&	477	
  & 2.13K &  & 5.82K \\
 &  &  $dir.$ &  & OOM & & -- & --  \\ 
\midrule
\midrule

\multirow{3}{*}{\datasetName{5D}} & \multirow{10}{*}{\begin{sideways}\datasetName{prob.}\end{sideways}} & $\ks$ & 209	&	77.6
	 & 15.3K & \multirow{3}{*}{26.9M} & 84.2K \\
 &  &  $\kl$ & 148	&	75.0
  & 95.3K & & 535K \\
 &  &  $dir.$ &  & 108 & & 
  & 159M \\
 \cmidrule(lr){1-1} \cmidrule(lr){3-8}
 
\multirow{3}{*}{\datasetName{7D}} &  & $\ks$ & 4.84K & 1.63K 
& 1.05M & \multirow{2}{*}{258M} & 7.66M \\
 &  &  $\kl$ & 3.89K & 1.53K 
 & 4.30M & & 32.5M \\
 &  &  $dir.$ &  & OOM &  & -- & -- \\
 \cmidrule(lr){1-1} \cmidrule(lr){3-8}
 
\multirow{3}{*}{\datasetName{40D}} &  & $\ks$ & 30.1K & 24.3K 
& 1.36M & \multirow{2}{*}{16.7M} & 55.3M \\
&  &  $\kl$ & 28.3K & 22.9K 
& 5.04M & & 205M \\
 &  &  $dir.$ &  & OOM &  & -- & -- \\

\midrule

\multirow{3}{*}{\datasetName{vismale 7D}} & \multirow{10}{*}{\begin{sideways}\datasetName{v-rips}\end{sideways}} & $\ks$ & 45.8	&	42.1 & 2.24K & \multirow{3}{*}{11.0M} & 4.59K \\
 &  &  $\kl$ & 47.3	& 42.4  & 3.36K & & 6.29K \\
 &  &  $dir.$ &  & 43.3 & & & 35.2M \\
 \cmidrule(lr){1-1} \cmidrule(lr){3-8}
 
\multirow{3}{*}{\datasetName{foot 10D}} &  & $\ks$ & 694 & 595
& 13.6K & \multirow{3}{*}{68.9M} & 88.8K \\
 &  &  $\kl$ & 528 & 558 & 17.4K & & 115K \\
 &  &  $dir.$ &  & 899 &  &  & 552M \\
 \cmidrule(lr){1-1} \cmidrule(lr){3-8}
 
\multirow{3}{*}{\datasetName{lucy 34D}} &  & $\ks$ & 804 & 763 & 5.44K & \multirow{2}{*}{55.1M} & 23.2K \\
&  &  $\kl$ & 688 & 615 & 12.6K & & 58.3K \\
 &  &  $dir.$ &  & OOM &  & -- & -- \\
%
%
%
%
%
 \bottomrule

	 \end{tabular}
	}
	}
 \label{tab:timings_adj_ds_gen}
\end{table}

When comparing execution times, we find that the global Stellar tree approach is about 25\% faster than the direct approach in most cases.
However, due in part to the redundant lookups in the adjacency calculation, 
the local approach is slightly slower than the global approach, but still 10\% faster than the direct approach in most cases. 
For example, it is almost twice as fast on \emph{F16}, on par on \emph{Lucy} and slower on the \emph{5D probabilistic} dataset.
Considering the effects of the bucket threshold \kv, we observe little discernible difference 
on the global Stellar tree approach.
However, a larger bucketing threshold (\kl) yielded up to a 25\% speedup in the local approach on our larger datasets, compared to its smaller (\ks) counterpart.

Lastly, we consider the storage requirements for generating the IA / \iastar\ data structure.
For both the local and global Stellar tree tree approaches, the auxiliary storage requirements are limited to the complexity of each leaf block,
requiring only a few KBs of auxiliary storage for the manifold and non-manifold datasets, and a few MBs for the pure (\emph{probabilistic}) datasets.
In contrast, the direct approach requires hundreds of MBs for the medium sized datasets. We were not able to generate the
\iastar\ data structures using the direct approach on our largest datasets, 
which ran out of memory (\emph{OOM}) on our workstation, despite its 64 GB of available RAM.

\end{document}